\documentclass[%
 reprint,
 amsmath,amssymb,
 aps,
]{revtex4-2}

\usepackage{newtxtext,newtxmath}
\usepackage[]{hyperref}
\usepackage{etoolbox}
\hypersetup{
    colorlinks=true, 
    linkcolor=blue, 
    filecolor=blue, 
    urlcolor=blue, 
    citecolor=blue,
}

\usepackage{graphicx}% Include figure files
\usepackage{dcolumn}% Align table columns on decimal point
\usepackage{bm}% bold math
\usepackage{float}% force figures for placement
\newcommand{\snn}{\ensuremath{\sqrt{s_{\mathrm{NN}}}}}
\newcommand{\tpt}{\ensuremath{p_{\mathrm{T}}}}
\newcommand{\substy}[2]{$#1_\text{#2}$}
\newcommand{\dnch}[1]{d\ensuremath{N_{\mathrm{ch}}}/d\ensuremath{#1}}
\newcommand{\partpm}[1]{$#1^{\pm}$}
\newcommand{\partpr}{\ensuremath{\text{p}(\overline{\text{p}})}}
\newcommand{\gev}{GeV/\textit{c}}
\newcommand{\ptcorr}{\ensuremath{\sqrt{\langle \langle \Delta p_{\mathrm{T},i}\Delta p_{\mathrm{T},j} \rangle \rangle}\mathrm{/} \langle \langle p_{\mathrm{T}} \rangle \rangle}}
\newcommand{\vn}[1]{\ensuremath{v_{\mathrm{#1}}}}
\newcommand{\vnn}[2]{\ensuremath{v_{\mathrm{#1}} \{#2\}}}

\begin{document}

\preprint{APS/123-QED}

%%%%%%%%%%%%%%%%%%%%%%%%%%%%%%%%%%%%%%%%%%%%%%%%%%%%%%%%%%%%%%%%%%%%%%%%%%%%%%%%%%%%%%%%
%%%%%%%%%%%%%%%%%%%%%%%%%%%%%%%%%%%%%%%%%%%%%%%%%%%%%%%%%%%%%%%%%%%%%%%%%%%%%%%%%%%%%%%%
%%%%%%%%%%%%%%%%%%%%%%%%%%%%%%%%%%%%%%%%%%%%%%%%%%%%%%%%%%%%%%%%%%%%%%%%%%%%%%%%%%%%%%%%
%%%%%%%%%%%%%%%%%%%%%%%%%%%%%%%%%%%%%%%%%%%%%%%%%%%%%%%%%%%%%%%%%%%%%%%%%%%%%%%%%%%%%%%%
%%%%%%%%%%%%%%%%%%%%%%%%%%%%%%%%%%%%%%%%%%%%%%%%%%%%%%%%%%%%%%%%%%%%%%%%%%%%%%%%%%%%%%%%
%%%%%%%%%%%%%%%%%%%%%%%%%%%%%%%%%%%%%%%%%%%%%%%%%%%%%%%%%%%%%%%%%%%%%%%%%%%%%%%%%%%%%%%%
%%%%%%%%%%%%%%%%%%%%%%%%%%%%%%%%%%%%%%%%%%%%%%%%%%%%%%%%%%%%%%%%%%%%%%%%%%%%%%%%%%%%%%%%
%%%%%%%%%%%%%%%%%%%%%%%%%%%%%%%%%%%%%%%%%%%%%%%%%%%%%%%%%%%%%%%%%%%%%%%%%%%%%%%%%%%%%%%%
\title{Scaling of soft QGP signatures in relativistic lead, xenon and oxygen collisions in EPOS4}
%%%%%%%%%%%%%%%%%%%%%%%%%%%%%%%%%%%%%%%%%%%%%%%%%%%%%%%%%%%%%%%%%%%%%%%%%%%%%%%%%%%%%%%%
%%%%%%%%%%%%%%%%%%%%%%%%%%%%%%%%%%%%%%%%%%%%%%%%%%%%%%%%%%%%%%%%%%%%%%%%%%%%%%%%%%%%%%%%
%%%%%%%%%%%%%%%%%%%%%%%%%%%%%%%%%%%%%%%%%%%%%%%%%%%%%%%%%%%%%%%%%%%%%%%%%%%%%%%%%%%%%%%%
%%%%%%%%%%%%%%%%%%%%%%%%%%%%%%%%%%%%%%%%%%%%%%%%%%%%%%%%%%%%%%%%%%%%%%%%%%%%%%%%%%%%%%%%
%%%%%%%%%%%%%%%%%%%%%%%%%%%%%%%%%%%%%%%%%%%%%%%%%%%%%%%%%%%%%%%%%%%%%%%%%%%%%%%%%%%%%%%%
%%%%%%%%%%%%%%%%%%%%%%%%%%%%%%%%%%%%%%%%%%%%%%%%%%%%%%%%%%%%%%%%%%%%%%%%%%%%%%%%%%%%%%%%
%%%%%%%%%%%%%%%%%%%%%%%%%%%%%%%%%%%%%%%%%%%%%%%%%%%%%%%%%%%%%%%%%%%%%%%%%%%%%%%%%%%%%%%%
%%%%%%%%%%%%%%%%%%%%%%%%%%%%%%%%%%%%%%%%%%%%%%%%%%%%%%%%%%%%%%%%%%%%%%%%%%%%%%%%%%%%%%%%

%%%%%%%%%%%%%%%%%%%%%%%%%%%%%%%%%%%%%%%%%%%%%%%%%%%%%%%%%%%%%%%%%%%%%%%%%%%%%%%%%%%%%%%%
%%%%%%%%%%%%%%%%%%%%%%%%%%%%%%%%%%%%%%%%%%%%%%%%%%%%%%%%%%%%%%%%%%%%%%%%%%%%%%%%%%%%%%%%
%%%%%%%%%%%%%%%%%%%%%%%%%%%%%%%%%%%%%%%%%%%%%%%%%%%%%%%%%%%%%%%%%%%%%%%%%%%%%%%%%%%%%%%%
%%%%%%%%%%%%%%%%%%%%%%%%%%%%%%%%%%%%%%%%%%%%%%%%%%%%%%%%%%%%%%%%%%%%%%%%%%%%%%%%%%%%%%%%
%%%%%%%%%%%%%%%%%%%%%%%%%%%%%%%%%%%%%%%%%%%%%%%%%%%%%%%%%%%%%%%%%%%%%%%%%%%%%%%%%%%%%%%%
%%%%%%%%%%%%%%%%%%%%%%%%%%%%%%%%%%%%%%%%%%%%%%%%%%%%%%%%%%%%%%%%%%%%%%%%%%%%%%%%%%%%%%%%
%%%%%%%%%%%%%%%%%%%%%%%%%%%%%%%%%%%%%%%%%%%%%%%%%%%%%%%%%%%%%%%%%%%%%%%%%%%%%%%%%%%%%%%%
%%%%%%%%%%%%%%%%%%%%%%%%%%%%%%%%%%%%%%%%%%%%%%%%%%%%%%%%%%%%%%%%%%%%%%%%%%%%%%%%%%%%%%%%
\author{Salman Khurshid Malik}
 \email{salmankm@keemail.me}
\author{Fakhar Ul Haider}
 %\email{fakharulhaider@gmail.com}
\author{Pratibha Bhagat}
 \email{pratibha.bhagat0401@gmail.com}
\author{Anju Bhasin}
% \email{anju.bhasin@cern.ch}
\author{Ramni Gupta}
% \email{ramni.gupta@cern.ch}
\affiliation{%
 Department of Physics, University of Jammu, J\&K, India (180006)\\
}%

%%%%%%%%%%%%%%%%%%%%%%%%%%%%%%%%%%%%%%%%%%%%%%%%%%%%%%%%%%%%%%%%%%%%%%%%%%%%%%%%%%%%%%%%
%%%%%%%%%%%%%%%%%%%%%%%%%%%%%%%%%%%%%%%%%%%%%%%%%%%%%%%%%%%%%%%%%%%%%%%%%%%%%%%%%%%%%%%%
%%%%%%%%%%%%%%%%%%%%%%%%%%%%%%%%%%%%%%%%%%%%%%%%%%%%%%%%%%%%%%%%%%%%%%%%%%%%%%%%%%%%%%%%
%%%%%%%%%%%%%%%%%%%%%%%%%%%%%%%%%%%%%%%%%%%%%%%%%%%%%%%%%%%%%%%%%%%%%%%%%%%%%%%%%%%%%%%%
%%%%%%%%%%%%%%%%%%%%%%%%%%%%%%%%%%%%%%%%%%%%%%%%%%%%%%%%%%%%%%%%%%%%%%%%%%%%%%%%%%%%%%%%
%%%%%%%%%%%%%%%%%%%%%%%%%%%%%%%%%%%%%%%%%%%%%%%%%%%%%%%%%%%%%%%%%%%%%%%%%%%%%%%%%%%%%%%%
%%%%%%%%%%%%%%%%%%%%%%%%%%%%%%%%%%%%%%%%%%%%%%%%%%%%%%%%%%%%%%%%%%%%%%%%%%%%%%%%%%%%%%%%
%%%%%%%%%%%%%%%%%%%%%%%%%%%%%%%%%%%%%%%%%%%%%%%%%%%%%%%%%%%%%%%%%%%%%%%%%%%%%%%%%%%%%%%%

\date{\today}

%%%%%%%%%%%%%%%%%%%%%%%%%%%%%%%%%%%%%%%%%%%%%%%%%%%%%%%%%%%%%%%%%%%%%%%%%%%%%%%%%%%%%%%%
%%%%%%%%%%%%%%%%%%%%%%%%%%%%%%%%%%%%%%%%%%%%%%%%%%%%%%%%%%%%%%%%%%%%%%%%%%%%%%%%%%%%%%%%
%%%%%%%%%%%%%%%%%%%%%%%%%%%%%%%%%%%%%%%%%%%%%%%%%%%%%%%%%%%%%%%%%%%%%%%%%%%%%%%%%%%%%%%%
%%%%%%%%%%%%%%%%%%%%%%%%%%%%%%%%%%%%%%%%%%%%%%%%%%%%%%%%%%%%%%%%%%%%%%%%%%%%%%%%%%%%%%%%
%%%%%%%%%%%%%%%%%%%%%%%%%%%%%%%%%%%%%%%%%%%%%%%%%%%%%%%%%%%%%%%%%%%%%%%%%%%%%%%%%%%%%%%%
%%%%%%%%%%%%%%%%%%%%%%%%%%%%%%%%%%%%%%%%%%%%%%%%%%%%%%%%%%%%%%%%%%%%%%%%%%%%%%%%%%%%%%%%
%%%%%%%%%%%%%%%%%%%%%%%%%%%%%%%%%%%%%%%%%%%%%%%%%%%%%%%%%%%%%%%%%%%%%%%%%%%%%%%%%%%%%%%%
%%%%%%%%%%%%%%%%%%%%%%%%%%%%%%%%%%%%%%%%%%%%%%%%%%%%%%%%%%%%%%%%%%%%%%%%%%%%%%%%%%%%%%%%
%%%%%%%%%%%%%%%%%%%%%%%%%%%%      ABSTRACT     %%%%%%%%%%%%%%%%%%%%%%%%%%%%%%%%%%%%%%%%%

\begin{abstract}
The collective expansion and hydrodynamic evolution in heavy-ion collisions is well-established. However, whether femtometer-scale droplets of QGP are produced in small systems at high energies remains a fundamental open question. Analysis of Pb--Pb collisions at \snn = 5.02 TeV, Xe--Xe at \snn = 5.44 TeV and O--O collisions at \snn = 5.36 TeV using EPOS4 is reported to make predictions and postdictions. The results are compared with ALICE data for Pb--Pb and Xe--Xe collisions. Charged particle multiplicity (\dnch{\eta}), transverse-momentum (\tpt) spectra for pions (\partpm{\pi}), kaons (\partpm{K}), and protons (\partpr) are studied. The inclusion of hadronic afterburner, UrQMD (Ultra-relativistic Quantum Molecular Dynamics) is found to be necessary to correctly describe baryon yields. \tpt-fluctuations are also studied via normalized \tpt~correlator, \ptcorr. Lastly, anisotropic flow harmonics (\vnn{2}{2}, \vnn{3}{2}) are computed as a function of \tpt~ and centrality. Since EPOS4 has not been extensively studied for flow observables, this study thereby provides a non‑trivial assessment of its collective dynamics. The results are compared with experiment wherever data is available. Taken together, this study provides a unified description of soft observables from Pb--Pb through Xe--Xe down to O--O and offer quantitative guidance on how such collisions may inform of the properties of the QGP.
% \begin{description}
% \item[Usage]
% Secondary publications and information retrieval purposes.
% \item[Structure]
% You may use the \texttt{description} environment to structure your abstract;
% use the optional argument of the \verb+\item+ command to give the category of each item. 
% \end{description}
\end{abstract}

\maketitle

%\tableofcontents

%%%%%%%%%%%%%%%%%%%%%%%%%%%%%%%%%%%%%%%%%%%%%%%%%%%%%%%%%%%%%%%%%%%%%%%%%%%%%%%%%%%%%%%%
%%%%%%%%%%%%%%%%%%%%%%%%%%%%%%%%%%%%%%%%%%%%%%%%%%%%%%%%%%%%%%%%%%%%%%%%%%%%%%%%%%%%%%%%
%%%%%%%%%%%%%%%%%%%%%%%%%%%%%%%%%%%%%%%%%%%%%%%%%%%%%%%%%%%%%%%%%%%%%%%%%%%%%%%%%%%%%%%%
%%%%%%%%%%%%%%%%%%%%%%%%%%%%%%%%%%%%%%%%%%%%%%%%%%%%%%%%%%%%%%%%%%%%%%%%%%%%%%%%%%%%%%%%
%%%%%%%%%%%%%%%%%%%%%%%%%%%%%%%%%%%%%%%%%%%%%%%%%%%%%%%%%%%%%%%%%%%%%%%%%%%%%%%%%%%%%%%%
%%%%%%%%%%%%%%%%%%%%%%%%%%%%%%%%%%%%%%%%%%%%%%%%%%%%%%%%%%%%%%%%%%%%%%%%%%%%%%%%%%%%%%%%
%%%%%%%%%%%%%%%%%%%%%%%%%%%%%%%%%%%%%%%%%%%%%%%%%%%%%%%%%%%%%%%%%%%%%%%%%%%%%%%%%%%%%%%%
%%%%%%%%%%%%%%%%%%%%%%%%%%%%%%%%%%%%%%%%%%%%%%%%%%%%%%%%%%%%%%%%%%%%%%%%%%%%%%%%%%%%%%%%
%%%%%%%%%%%%%%%%%%%%%%%%%%%%      INTRODUCTION     %%%%%%%%%%%%%%%%%%%%%%%%%%%%%%%%%%%%%
%%%%%%%%%%%%%%%%%%%%%%%%%%%%%%%%%%%%%%%%%%%%%%%%%%%%%%%%%%%%%%%%%%%%%%%%%%%%%%%%%%%%%%%%
%%%%%%%%%%%%%%%%%%%%%%%%%%%%%%%%%%%%%%%%%%%%%%%%%%%%%%%%%%%%%%%%%%%%%%%%%%%%%%%%%%%%%%%%
%%%%%%%%%%%%%%%%%%%%%%%%%%%%%%%%%%%%%%%%%%%%%%%%%%%%%%%%%%%%%%%%%%%%%%%%%%%%%%%%%%%%%%%%
%%%%%%%%%%%%%%%%%%%%%%%%%%%%%%%%%%%%%%%%%%%%%%%%%%%%%%%%%%%%%%%%%%%%%%%%%%%%%%%%%%%%%%%%
%%%%%%%%%%%%%%%%%%%%%%%%%%%%%%%%%%%%%%%%%%%%%%%%%%%%%%%%%%%%%%%%%%%%%%%%%%%%%%%%%%%%%%%%
%%%%%%%%%%%%%%%%%%%%%%%%%%%%%%%%%%%%%%%%%%%%%%%%%%%%%%%%%%%%%%%%%%%%%%%%%%%%%%%%%%%%%%%%
%%%%%%%%%%%%%%%%%%%%%%%%%%%%%%%%%%%%%%%%%%%%%%%%%%%%%%%%%%%%%%%%%%%%%%%%%%%%%%%%%%%%%%%%
%%%%%%%%%%%%%%%%%%%%%%%%%%%%%%%%%%%%%%%%%%%%%%%%%%%%%%%%%%%%%%%%%%%%%%%%%%%%%%%%%%%%%%%%
%\section{\label{sec:level1}Introduction:\protect\\}

\section{Introduction}

% The field of high-energy physics is fundamentally driven by the quest to understand the nature of strongly interacting matter at extreme temperatures and densities. According to Quantum Chromodynamics (QCD)~\cite{Barber:1979yr}, the theory governing the strong interaction, ordinary nuclear matter is expected to transform into a deconfined state of quarks and gluons~\cite{Gell-Mann:1961omu} when subjected to sufficiently intense thermal or baryonic conditions. This state, known as the quark--gluon plasma (QGP)~\cite{Shuryak:1978ij}, represents a phase in which color charges are no longer confined within hadrons but propagate over extended volumes. The transition temperature is predicted by ab initio lattice QCD calculations to be on the order of 150–160 MeV~\cite{ZABRODIN1994407}. 

Ultra-relativistic heavy-ion collisions at the Large Hadron Collider (LHC)~\cite{CERN-Brochure-2017-002-Eng} generate short-lived but extremely hot and dense fireballs of deconfined quarks and gluons, known as the quark-gluon plasma (QGP)~\cite{Shuryak:1978ij}. The deconfined phase persists for only a few fm/$c$ prior to cooling and confinement into hadrons. Due to its transient nature, properties of the QGP must be inferred indirectly from the particles measured after kinetic freeze-out~\cite{XU2017290}. These final-state hadrons retain the evolution of the collision, carrying signatures of the early interactions~\cite{Harris:2023tti}, the development of initial-state fluctuations, the subsequent collective hydrodynamic expansion and the complex dynamics of hadronization~\cite{Huang:2019tgz}. The combination of a number of complementary observables enables the characterization of QGP dynamics.

Within the soft sector~\cite{Vernet:2009xp} dominated by low transverse-momentum particles and governed by the bulk evolution of the medium, special emphasis is placed on particle yields, transverse-momentum spectra~\cite{Mohanty:2003yn}, event-by-event fluctuations~\cite{Jeon:2003gk} and anisotropic flow coefficients~\cite{Oldenburg:2004ri}. These observables are directly shaped by the collective expansion and transport properties of the system and therefore provide sensitive probes of its thermodynamic state and dynamical evolution. Particle production captures entropy and energy deposition,~\tpt-spectra encode radial flow. Fluctuations link initial inhomogeneities to medium response, while anisotropic flow quantifies the hydrodynamic conversion of spatial to momentum-space anisotropies. A description of these signatures within a theoretical framework is essential for understanding formation of deconfined matter and its fundamental properties.

Particle production constitutes the most fundamental observable in relativistic nuclear collisions. The yields and spectra of produced particles~\cite{Preghenella:2011np} are intimately connected to the initial energy density~\cite{FOWLER1988657}, subsequent expansion dynamics and freeze-out conditions of the system. In particular, charged-particle multiplicity densities ~\cite{Zampolli:2013ewn} serves as a proxy for entropy production and provide a natural measure of collision activity, enabling meaningful comparisons across different systems and energies. The transverse momentum (\tpt) distributions~\cite{Cleymans:2012ya} contains additional information about the collective radial expansion of the medium. The characteristic flattening of low-\tpt~spectra relative to elementary collisions provides a measure of radial flow~\cite{Capella:2006be} driven by strong pressure gradients in the expanding QGP. The mass dependence of spectral shapes further constrains the strength of collective expansion and connects microscopic particle production to macroscopic medium properties. When examined across different collision systems, particle production observables allow one to probe how bulk medium properties evolve with system size. Simple geometric or participant-scaling expectations may break down if the underlying particle production mechanisms change or if non-equilibrium effects become significant. Therefore, studying the scaling behaviour of particle yields provides insight into whether entropy production and collective expansion exhibit universal features across systems of varying size.

Event-by-event fluctuations~\cite{Jeon:2003gk} provide insight into the underlying dynamics beyond average quantities. The event-by-event fluctuations in particle production originate from a convolution of initial-state geometry variations, the particulars of energy deposition, and the dynamics inherent to the system's evolution. Prominent among these are transverse-momentum fluctuations and two-particle $p_{T}$ correlations~\cite{Heinz:1999rw}, which are helpful at detecting fluctuations in temperature, local density, and correlations originating from the early stages of the collision. \tpt~correlations offer a window into the interplay between initial-state fluctuations to intermediate energy deposition and final-state interactions. In large systems, these correlations are significantly influenced by collective expansion and by contrast in smaller systems, they are expected to retain a stronger imprint of partonic dynamics and non-equilibrium effects. A systematic investigation of how \tpt~fluctuations evolve with system size therefore provides essential constraints on the onset of collectivity and the degree of thermalization achieved in the medium.

While particle yields and momentum correlations characterize bulk properties and fluctuations, flow observables directly probe the medium's dynamic response and transport coefficients. The azimuthal anisotropy of particle emission arises from the conversion of initial spatial asymmetries into momentum-space anisotropies via intense final-state interactions, a process highly sensitive to the shear viscosity to entropy density ratio (\( \eta/s \)). Among the various flow phenomena, radial flow \cite{PhysRevC.57.1891} and elliptic flow are the two most prominent classes. Elliptic flow (\(v_2\))~\cite{Danielewicz:1994nb, Braun-Munzinger:2007edi}, the second Fourier component of the azimuthal distribution, originates from the almond-shaped overlap geometry in non-central collisions. The resulting anisotropic pressure gradients convert this spatial eccentricity into momentum anisotropy. Radial flow, initially deduced from transverse momentum spectra, characterizes the isotropic transverse expansion at freeze-out and in non-central collisions, includes an anisotropic component reflecting the difference between in-plane and out-of-plane expansion velocities. Their interplay produces the characteristic mass splitting of elliptic flow, where heavier particles exhibit smaller \(v_2\) values a well understood hydrodynamic signature. The anisotropic component of radial flow~\cite{Tang_2013, Huovinen:2006jp, Santos_2014} holds particular significance for dissipative effects, because shear viscosity couples to anisotropic velocity gradients, its determination is essential for quantifying the QGP’s viscosity. The successful description of these observables by viscous hydrodynamics establishes the QGP as a nearly perfect fluid with exceptionally low \( \eta/s \). Extending flow studies to smaller collision systems tests the limits of collectivity. As system size decreases, medium lifetime shrinks and initial-state fluctuations become increasingly important. The central question is whether observed anisotropies in small systems continue to reflect hydrodynamic response or become dominated by non-collective correlations, with profound implications for understanding thermalization and the universal nature of collectivity.

In this work, a systematic investigation of the scaling, or continuous system-size evolution,  of particle yields, transverse-momentum (\tpt) fluctuations and anisotropic flow harmonics in Pb--Pb, Xe--Xe and O--O collisions with EPOS4 is performed. Pb--Pb collisions serve as the reference system, while Xe--Xe interactions probe an intermediate nuclear size. Predictions for the O--O system are provided to explore the behaviuor of collective phenomena toward smaller collision systems. The analysis is carried out using two configurations of EPOS4, one including the hadronic afterburner based on UrQMD (Ultra-relativistic Quantum Molecular Dynamics)~\cite{Bleicher:1999xi} and one without the afterburner stage. The UrQMD afterburner accounts for late-stage hadronic rescattering and resonance interactions, which can significantly influence final-state particle spectra and flow observables and calculations without the afterburner isolate the contributions from the initial conditions and hydrodynamic evolution, thereby allowing a controlled assessment of the role of hadronic re-scattering effects.

This paper is organized as follows: Section II provides a concise overview of the EPOS4 model. Section III presents the results for transverse-momentum spectra, event-by-event \tpt\ fluctuations and anisotropic flow harmonics across the three collision systems. Section IV summarizes the main findings.

%%%%%%%%%%%%%%%%%%%%%%%%%%%%%%%%%%%%%%%%%%%%%%%%%%%%%%%%%%%%%%%%%%%%%%%%%%%%%%%%%%%%%%%%
%%%%%%%%%%%%%%%%%%%%%%%%%%%%%%%%%%%%%%%%%%%%%%%%%%%%%%%%%%%%%%%%%%%%%%%%%%%%%%%%%%%%%%%%
%%%%%%%%%%%%%%%%%%%%%%%%%%%%%%%%%%%%%%%%%%%%%%%%%%%%%%%%%%%%%%%%%%%%%%%%%%%%%%%%%%%%%%%%
%%%%%%%%%%%%%%%%%%%%%%%%%%%%%%%%%%%%%%%%%%%%%%%%%%%%%%%%%%%%%%%%%%%%%%%%%%%%%%%%%%%%%%%%
%%%%%%%%%%%%%%%%%%%%%%%%%%%%%%%%%%%%%%%%%%%%%%%%%%%%%%%%%%%%%%%%%%%%%%%%%%%%%%%%%%%%%%%%
%%%%%%%%%%%%%%%%%%%%%%%%%%%%%%%%%%%%%%%%%%%%%%%%%%%%%%%%%%%%%%%%%%%%%%%%%%%%%%%%%%%%%%%%
%%%%%%%%%%%%%%%%%%%%%%%%%%%%%%%%%%%%%%%%%%%%%%%%%%%%%%%%%%%%%%%%%%%%%%%%%%%%%%%%%%%%%%%%
%%%%%%%%%%%%%%%%%%%%%%%%%%%%%%%%%%%%%%%%%%%%%%%%%%%%%%%%%%%%%%%%%%%%%%%%%%%%%%%%%%%%%%%%
%%%%%%%%%%%%%%%%%%%%%%%%%%%%      EPOS4         %%%%%%%%%%%%%%%%%%%%%%%%%%%%%%%%%%%%%
%%%%%%%%%%%%%%%%%%%%%%%%%%%%%%%%%%%%%%%%%%%%%%%%%%%%%%%%%%%%%%%%%%%%%%%%%%%%%%%%%%%%%%%%
%%%%%%%%%%%%%%%%%%%%%%%%%%%%%%%%%%%%%%%%%%%%%%%%%%%%%%%%%%%%%%%%%%%%%%%%%%%%%%%%%%%%%%%%
%%%%%%%%%%%%%%%%%%%%%%%%%%%%%%%%%%%%%%%%%%%%%%%%%%%%%%%%%%%%%%%%%%%%%%%%%%%%%%%%%%%%%%%%
%%%%%%%%%%%%%%%%%%%%%%%%%%%%%%%%%%%%%%%%%%%%%%%%%%%%%%%%%%%%%%%%%%%%%%%%%%%%%%%%%%%%%%%%
%%%%%%%%%%%%%%%%%%%%%%%%%%%%%%%%%%%%%%%%%%%%%%%%%%%%%%%%%%%%%%%%%%%%%%%%%%%%%%%%%%%%%%%%
%%%%%%%%%%%%%%%%%%%%%%%%%%%%%%%%%%%%%%%%%%%%%%%%%%%%%%%%%%%%%%%%%%%%%%%%%%%%%%%%%%%%%%%%
%%%%%%%%%%%%%%%%%%%%%%%%%%%%%%%%%%%%%%%%%%%%%%%%%%%%%%%%%%%%%%%%%%%%%%%%%%%%%%%%%%%%%%%%
%%%%%%%%%%%%%%%%%%%%%%%%%%%%%%%%%%%%%%%%%%%%%%%%%%%%%%%%%%%%%%%%%%%%%%%%%%%%%%%%%%%%%%%%
\section{\label{sec:epos4}EPOS4 Model\protect\\}

EPOS~\cite{Werner:2023fne} is a Monte Carlo event generator designed to simulate particle production mechanisms across different collision environments, from small systems such as proton--proton to large nucleus--nucleus interactions. The model provides a unified description connecting initial particle production, collective medium evolution, and final-state hadronic interactions on an event-by-event basis. The theoretical foundation of EPOS4 is based on a Gribov--Regge multiple-scattering formalism supplemented with perturbative QCD concepts~\cite{Drescher:2000ha}. Within this framework, the Abramovskii--Gribov--Kancheli (AGK)~\cite{Koplik:1975iu} cutting rules establish a correspondence between multiple scattering amplitudes and observable particle production, leading to simple geometric scaling behaviour of inclusive yields, such as binary scaling in nucleus--nucleus collisions. 

In ultra-relativistic collisions, particle production arises from multiple partonic interactions occurring in parallel~\cite{Werner:2023mod}. EPOS4 represents an advancement over EPOS3~\cite{PIEROG2009102}, where particle production was described in terms of cut Pomerons and remnant excitations, with energy shared among different scatterings in an effective manner. However, the hadronization step in EPOS3 followed an essentially grand-canonical approach, such that energy--momentum and conserved quantum numbers were only approximately conserved at the event level.  In contrast EPOS4~\cite{Werner:2025yse} uses a microcanonical approach for the exact conservation of energy and quantum numbers during hadronization. It also features a rigorous parallel treatment of primary scatterings that ensures strict energy–momentum conservation across all sub-collisions within a single event~\cite{Werner:2023mod}. Such an improvement is particularly important at LHC energies, where numerous partonic interactions occur simultaneously~\cite{PhysRevC.108.064903}.

EPOS4 uses core-corona~\cite{Werner:2023jps} separation as a universal scaling tool, by distinguishing between the dense 'core' and the dilute 'corona,' the model creates a single, unified description that remains accurate across all system sizes and geometries. At the onset of a collision, energy is deposited into longitudinal color strings. EPOS4 filters initial string segments based on their local surroundings, segments in regions exceeding a critical density threshold where energy loss prevents their escape are absorbed into a collective, expanding core and those segments that exist in low density zones or possess high enough transverse momentum ($p_T$) to bypass significant energy loss escape the medium to form the corona via string fragmentation. While the corona contributes to the total multiplicity via independent string fragmentation, the core is rapidly thermalized. This dense region serves as the starting point for 3D viscous hydrodynamics~\cite{Werner:2010aa}, which models the collective expansion and flow of the produced matter. The hydrodynamic module operates event-by-event and incorporates realistic spatial and temporal fluctuations inherited from the initial state. These density variations create strong anisotropic pressure gradients that drive the collective expansion of the system and lead to the formation of final-state flow harmonics. The evolution of the medium is governed by an equation of state consistent with lattice QCD results and by specified transport coefficients, most notably the shear viscosity-to-entropy density ratio, $\eta/s$~\cite{Kovtun:2004de}. In EPOS4, $\eta/s$ is chosen to be small (typically $\eta/s \approx 0.08$), reflecting the near-perfect fluid behavior observed in heavy-ion collisions and enabling the development of strong collective flow.

The hydrodynamic evolution proceeds until the local energy density drops below a predefined switching criterion corresponding to the hadronization hypersurface. At this stage, the hydrodynamically evolved core is converted into hadrons using a microcanonical statistical prescription. In this approach, the local energy--momentum flow across the hypersurface is used to construct an effective invariant mass, which subsequently decays into hadrons according to microcanonical phase space. This procedure ensures the exact conservation of energy, momentum, and conserved quantum numbers such as baryon number, electric charge, and strangeness. Such a treatment is particularly relevant for smaller systems, where limited phase space and conservation constraints can significantly influence particle production, especially for strange hadrons.

After hadronization, all produced particles undergo a microscopic hadronic cascade simulated using the UrQMD model. This afterburner accounts for elastic and inelastic hadronic re-scattering as well as resonance decays, modeling the non-equilibrium evolution from chemical to kinetic freeze-out~\cite{Bass:1998ca}. The inclusion of UrQMD allows for a realistic description of late-stage modifications to particle yields, transverse-momentum spectra, and flow observables.

%The hydrodynamic evolution proceeds until the local energy density drops below a predefined switching criterion corresponding to the hadronization hypersurface. At this stage, the fluid is converted into hadrons using a microcanonical statistical prescription that ensures exact conservation of baryon number, electric charge, and strangeness. This treatment is particularly important for smaller systems, where limited phase space and conservation constraints can significantly influence particle production, especially for strange hadrons. After hadronization, all produced particles undergo a microscopic hadronic cascade simulated using the UrQMD model. This afterburner accounts for elastic and inelastic hadronic re-scattering as well as resonance decays, modeling the non-equilibrium evolution from chemical to kinetic freeze-out~\cite{Bass:1998ca}. The inclusion of UrQMD allows for a realistic description of late-stage modifications to particle yields, transverse-momentum spectra, and flow observables.

In the present study, calculations are performed both with and without the UrQMD afterburner. This controlled comparison enables the isolation of hadronic rescattering effects and provides insight into how late-stage interactions influence bulk particle production, transverse-momentum correlations, and anisotropic flow across different collision systems.

\begin{figure*}[htbp!]
  \centering

% --------- Pb-Pb ---------
  \begin{minipage}{0.48\textwidth}
    \centering
    \includegraphics[width=\linewidth]{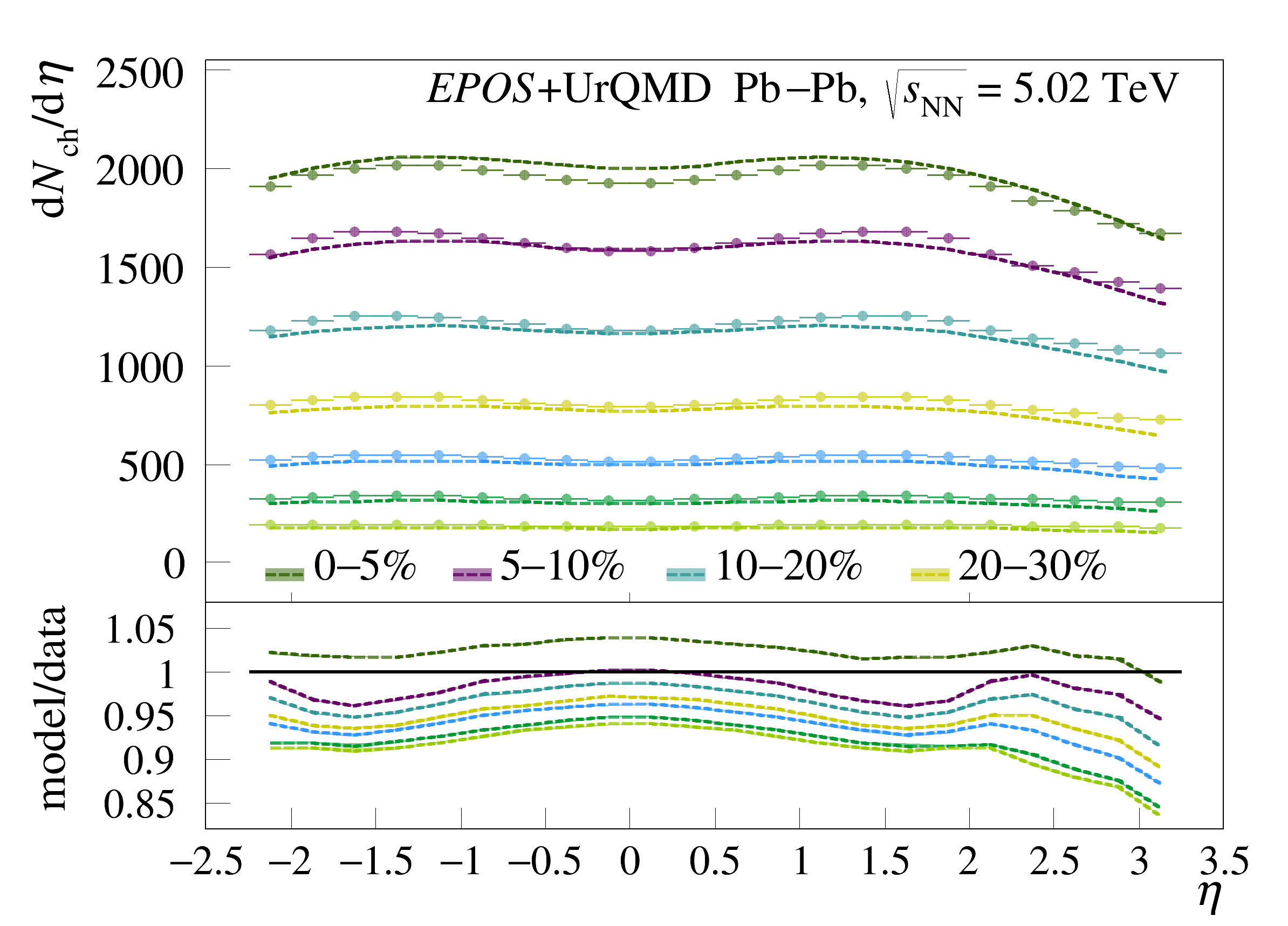}
  \end{minipage}
  \hfill
  \begin{minipage}{0.48\textwidth}
    \centering
    \includegraphics[width=\linewidth]{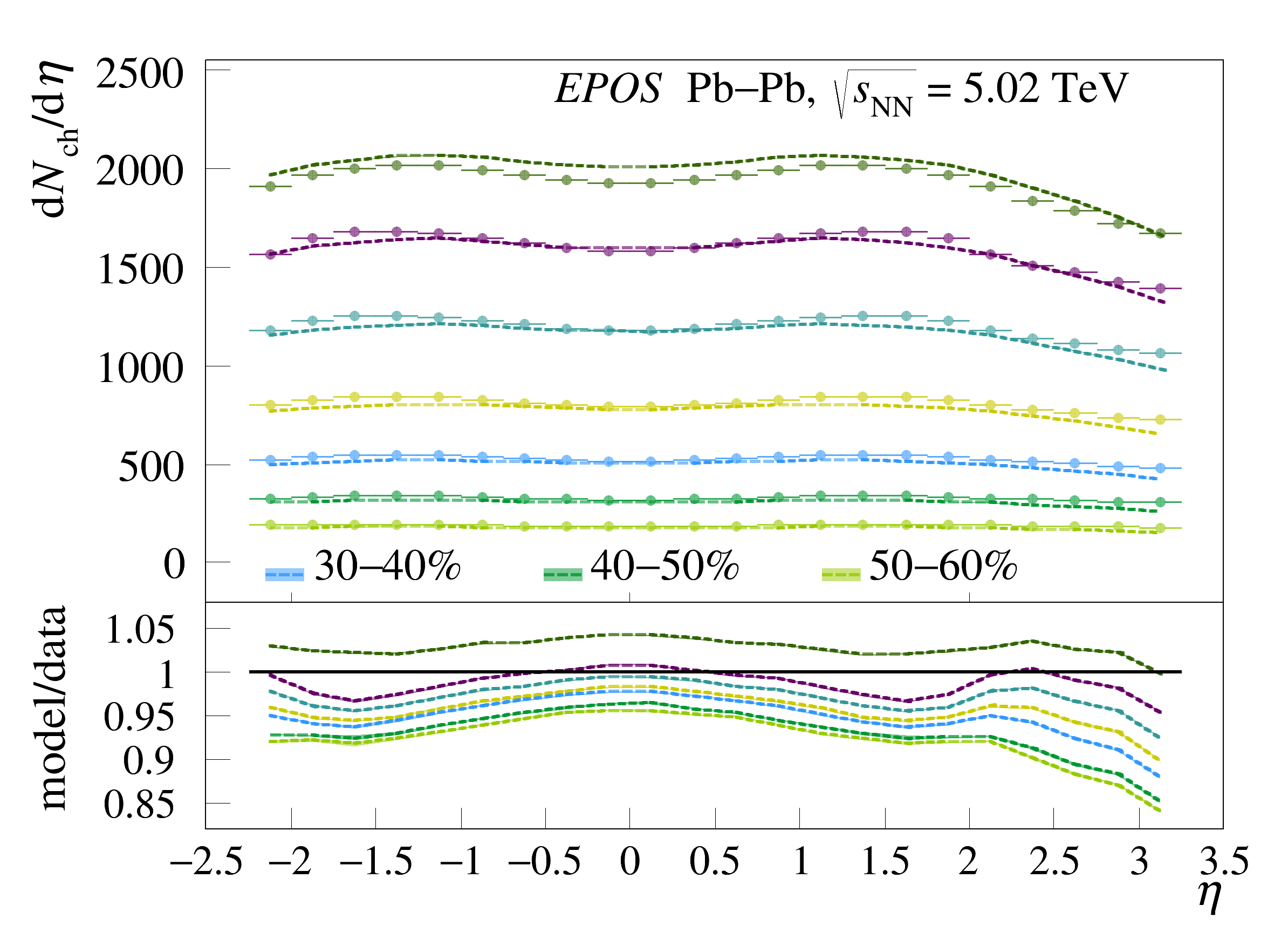}
  \end{minipage}

% --------- Xe-Xe ---------
  \begin{minipage}{0.48\textwidth}
    \centering
    \includegraphics[width=\linewidth]{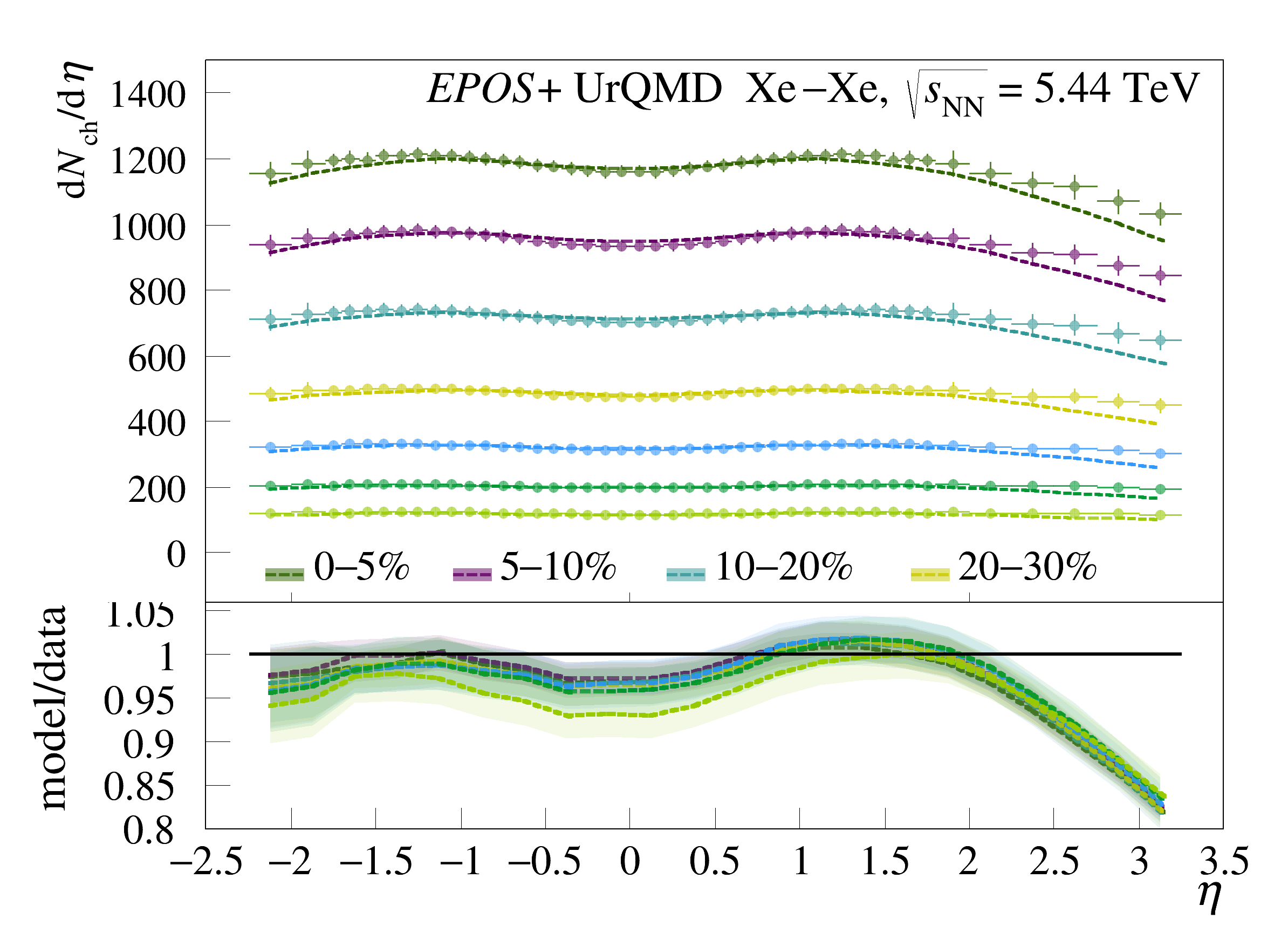}
  \end{minipage}
  \hfill
  \begin{minipage}{0.48\textwidth}
    \centering
    \includegraphics[width=\linewidth]{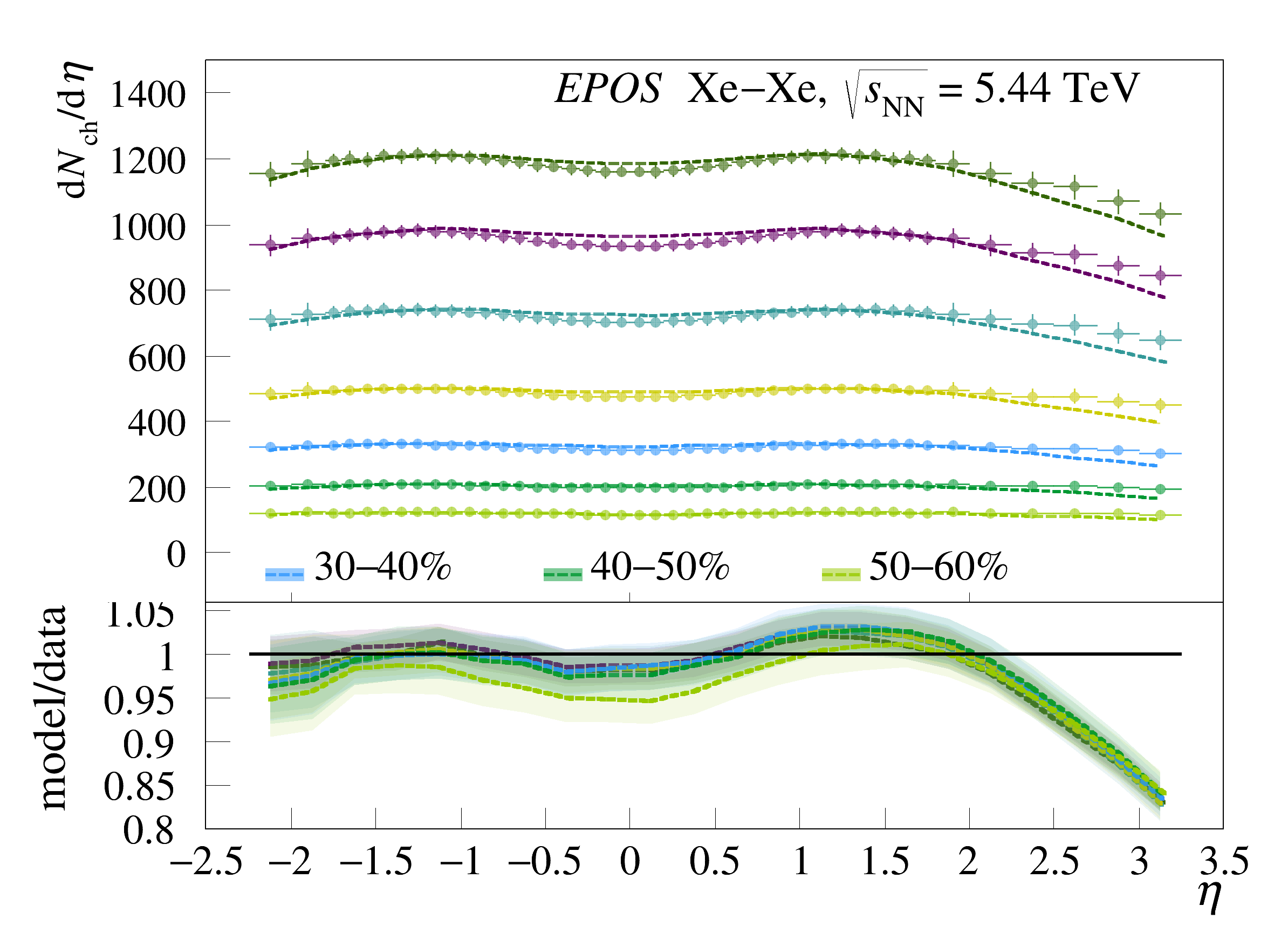}
  \end{minipage}

  \caption{Charged-particle pseudorapidity density in Pb--Pb collisions at \snn = 5.02 TeV (upper row) and Xe--Xe collisions at \snn = 5.44 TeV (lower row). The left panel shows EPOS4 results with the UrQMD hadronic afterburner, while the right panel shows results without hadronic rescattering. Lines of different colours indicate the EPOS4 predictions for various centrality classes, and the markers represent the corresponding ALICE data~\cite{ALICE:2016fbt, ALICE:2018cpu}.}
  \label{fig:dnchdeta}
\end{figure*}

\begin{figure}[htbp!]
  \centering

% --------- O-O ---------
  \begin{minipage}{0.48\textwidth}
    \centering
    \includegraphics[width=\linewidth]{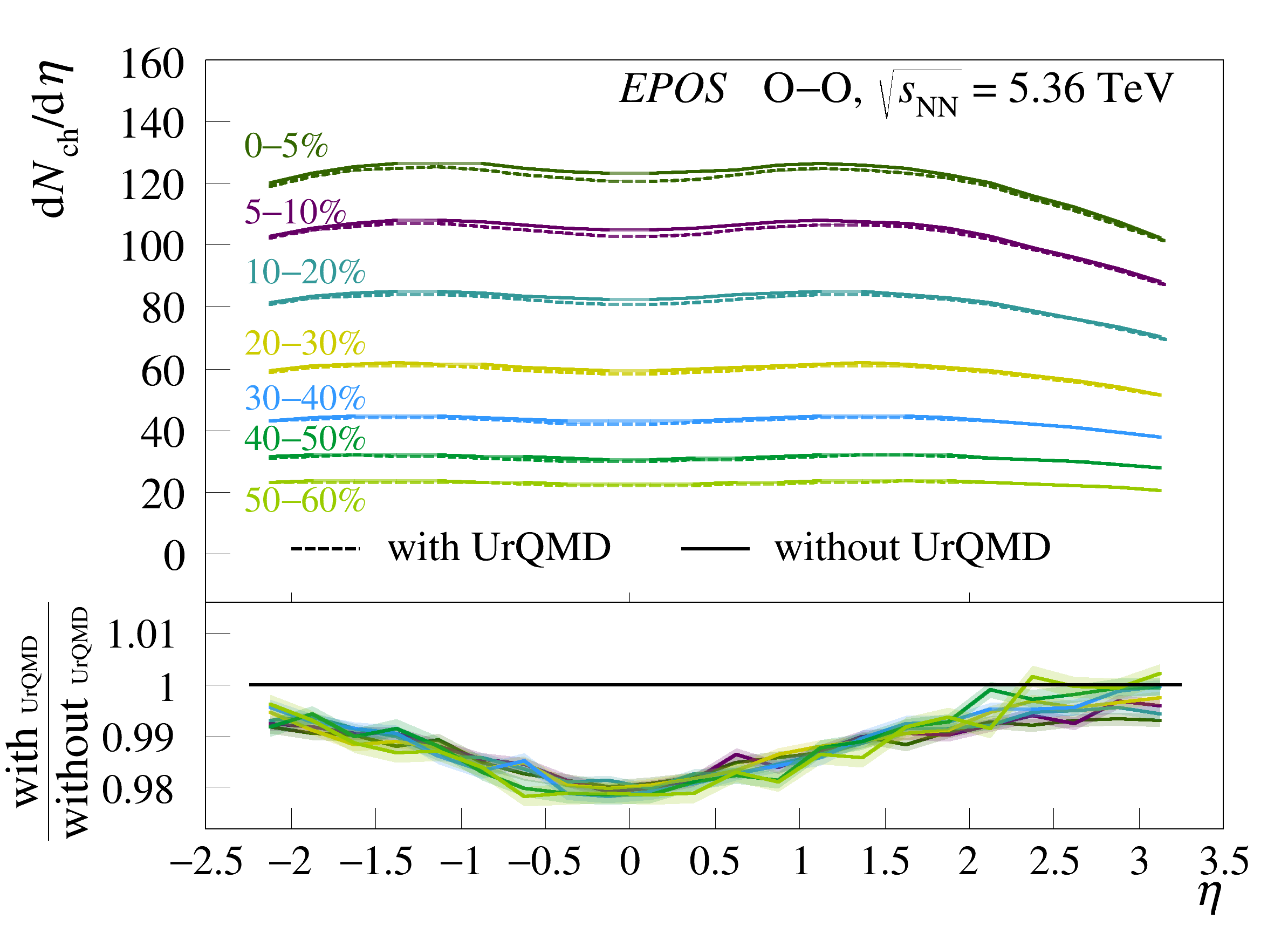}
  \end{minipage}
  \caption{Charged-particle pseudorapidity density in O--O collisions at \snn = 5.36 TeV. EPOS4 predictions including the UrQMD hadronic afterburner (dashed lines) and without hadronic rescattering (solid lines) are shown for several centrality classes.}
  \label{fig:dnchdeta1}
\end{figure}

%%%%%%%%%%%%%%%%%%%%%%%%%%%%%%%%%%%%%%%%%%%%%%%%%%%%%%%%%%%%%%%%%%%%%%%%%%%%%%%%%%%%%%%%
%%%%%%%%%%%%%%%%%%%%%%%%%%%%%%%%%%%%%%%%%%%%%%%%%%%%%%%%%%%%%%%%%%%%%%%%%%%%%%%%%%%%%%%%
%%%%%%%%%%%%%%%%%%%%%%%%%%%%%%%%%%%%%%%%%%%%%%%%%%%%%%%%%%%%%%%%%%%%%%%%%%%%%%%%%%%%%%%%
%%%%%%%%%%%%%%%%%%%%%%%%%%%%%%%%%%%%%%%%%%%%%%%%%%%%%%%%%%%%%%%%%%%%%%%%%%%%%%%%%%%%%%%%
%%%%%%%%%%%%%%%%%%%%%%%%%%%%%%%%%%%%%%%%%%%%%%%%%%%%%%%%%%%%%%%%%%%%%%%%%%%%%%%%%%%%%%%%
%%%%%%%%%%%%%%%%%%%%%%%%%%%%%%%%%%%%%%%%%%%%%%%%%%%%%%%%%%%%%%%%%%%%%%%%%%%%%%%%%%%%%%%%
%%%%%%%%%%%%%%%%%%%%%%%%%%%%%%%%%%%%%%%%%%%%%%%%%%%%%%%%%%%%%%%%%%%%%%%%%%%%%%%%%%%%%%%%
%%%%%%%%%%%%%%%%%%%%%%%%%%%%%%%%%%%%%%%%%%%%%%%%%%%%%%%%%%%%%%%%%%%%%%%%%%%%%%%%%%%%%%%%
%%%%%%%%%%%%%%%%%%%%%%%%%%%%       RESULTS       %%%%%%%%%%%%%%%%%%%%%%%%%%%%%%%%%%%%%%%
%%%%%%%%%%%%%%%%%%%%%%%%%%%%%%%%%%%%%%%%%%%%%%%%%%%%%%%%%%%%%%%%%%%%%%%%%%%%%%%%%%%%%%%%
%%%%%%%%%%%%%%%%%%%%%%%%%%%%%%%%%%%%%%%%%%%%%%%%%%%%%%%%%%%%%%%%%%%%%%%%%%%%%%%%%%%%%%%%
%%%%%%%%%%%%%%%%%%%%%%%%%%%%%%%%%%%%%%%%%%%%%%%%%%%%%%%%%%%%%%%%%%%%%%%%%%%%%%%%%%%%%%%%
%%%%%%%%%%%%%%%%%%%%%%%%%%%%%%%%%%%%%%%%%%%%%%%%%%%%%%%%%%%%%%%%%%%%%%%%%%%%%%%%%%%%%%%%
%%%%%%%%%%%%%%%%%%%%%%%%%%%%%%%%%%%%%%%%%%%%%%%%%%%%%%%%%%%%%%%%%%%%%%%%%%%%%%%%%%%%%%%%
%%%%%%%%%%%%%%%%%%%%%%%%%%%%%%%%%%%%%%%%%%%%%%%%%%%%%%%%%%%%%%%%%%%%%%%%%%%%%%%%%%%%%%%%
%%%%%%%%%%%%%%%%%%%%%%%%%%%%%%%%%%%%%%%%%%%%%%%%%%%%%%%%%%%%%%%%%%%%%%%%%%%%%%%%%%%%%%%%
%%%%%%%%%%%%%%%%%%%%%%%%%%%%%%%%%%%%%%%%%%%%%%%%%%%%%%%%%%%%%%%%%%%%%%%%%%%%%%%%%%%%%%%%
\section{Results}
In this work, soft-QGP observables in Pb--Pb and Xe--Xe collisions are analyzed with EPOS4, and the same setup is used to obtain predictions for O--O collisions. A dataset sample of $~$0.5M events for Pb--Pb (\snn = 5.02 TeV) and Xe--Xe (\snn = 5.44 TeV) each was generated. Along with this, a larger sample of$~$1M events for O--O was generated at \snn = 5.36 TeV. Calculations are performed both with and without the UrQMD afterburner. This comparison enables the isolation of hadronic rescattering effects and helps understand how late-stage interactions influence bulk particle production, transverse-momentum correlations, and anisotropic flow across different collision systems. The following sections detail the system-size evolution of bulk particle production, transverse-momentum fluctuations, and anisotropic flow. The analysis is structured as follows. First, the centrality determination is discussed. This is followed by an examination of transverse momentum spectra followed by discussion of transverse momentum fluctuations. Finally, the anisotropic flow coefficients are analyzed to assess collective behaviour across different system sizes.

\begin{figure*}[htbp!]
  \centering

% --------- Pb-Pb ---------
  \begin{minipage}{0.48\textwidth}
    \centering
    \includegraphics[width=\linewidth]{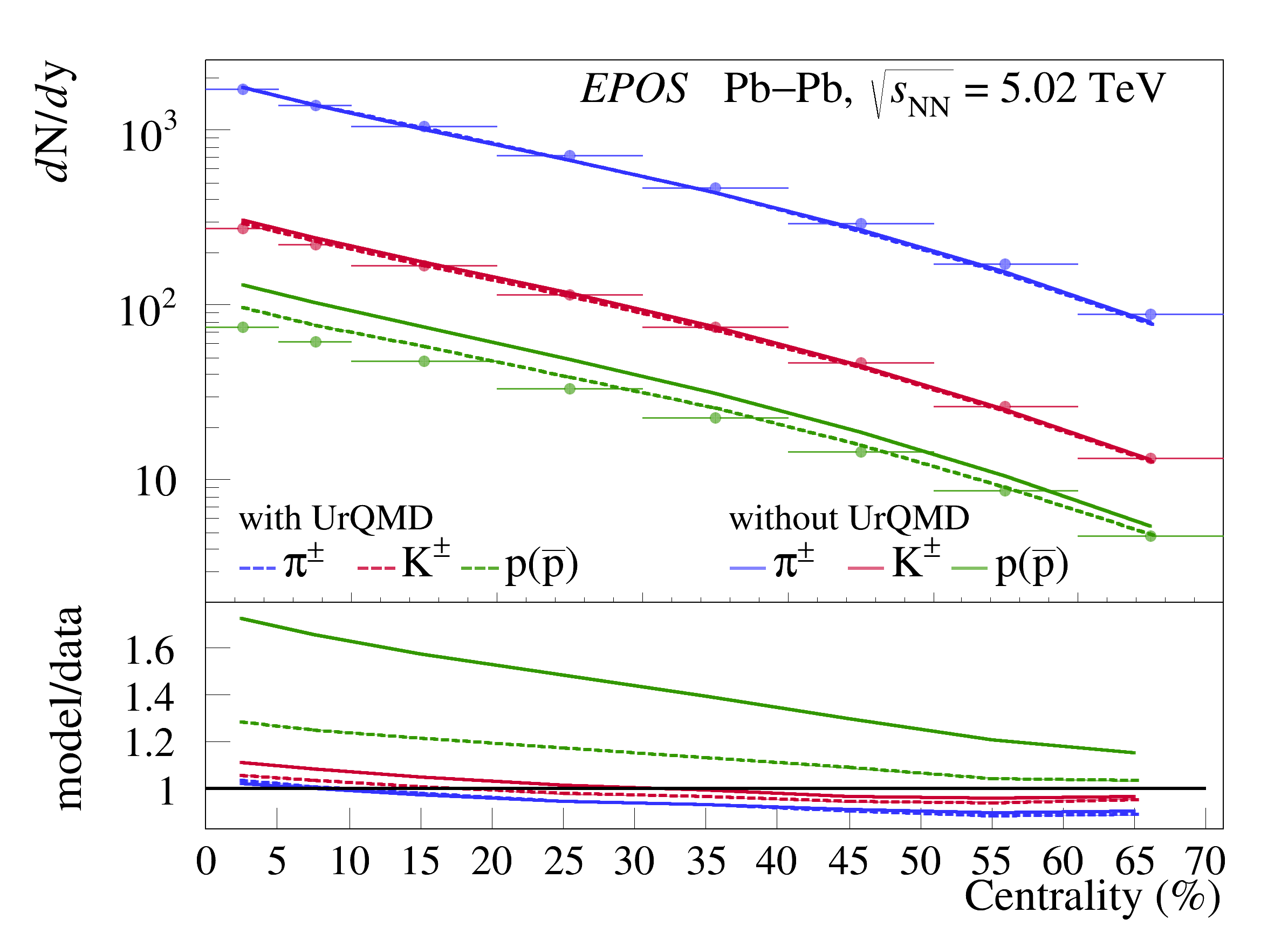}
  \end{minipage}
  \hfill
  \begin{minipage}{0.48\textwidth}
    \centering
    \includegraphics[width=\linewidth]{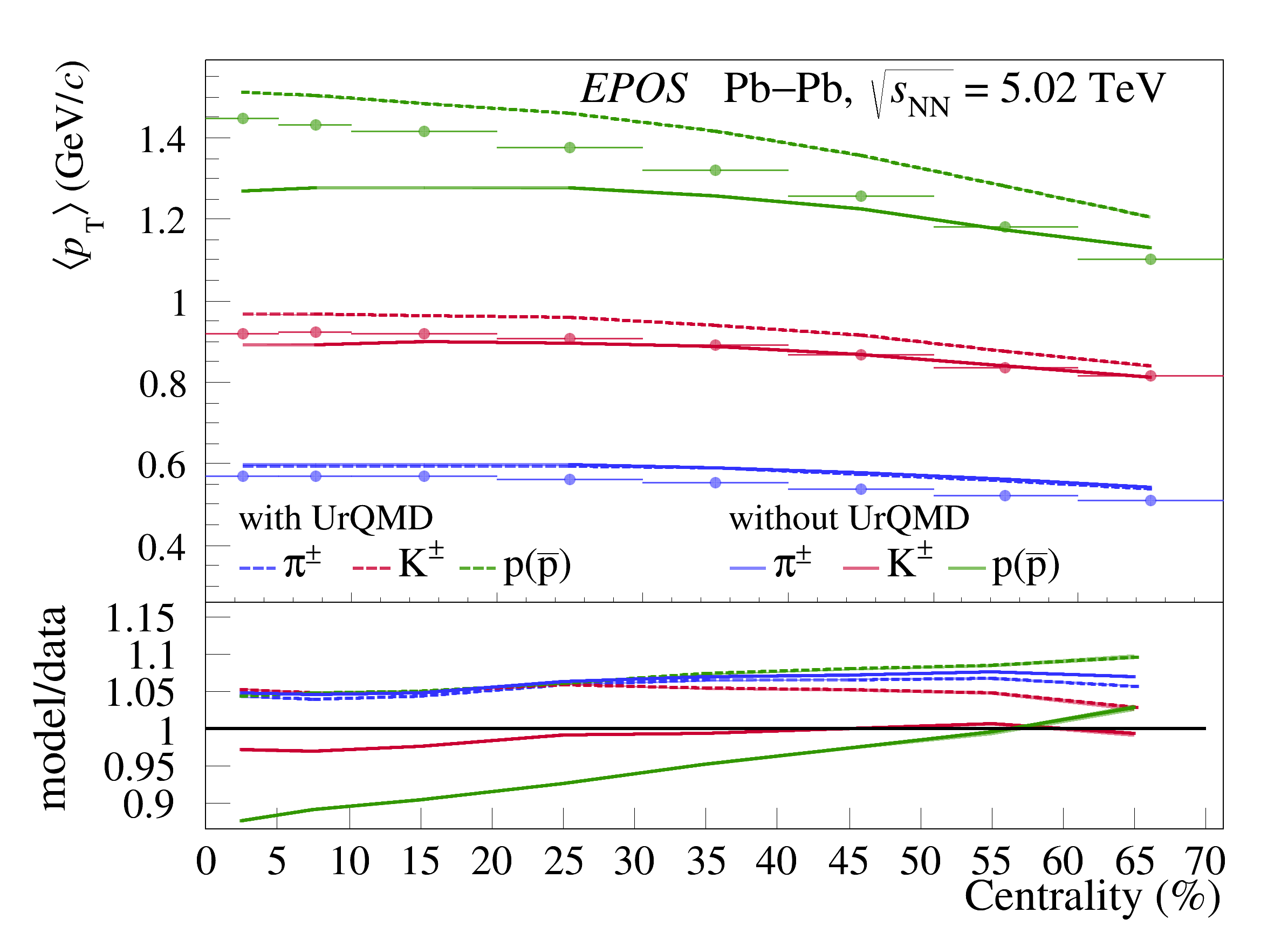}
  \end{minipage}

% --------- Xe-Xe ---------
   \begin{minipage}{0.48\textwidth}
    \centering
    \includegraphics[width=\linewidth]{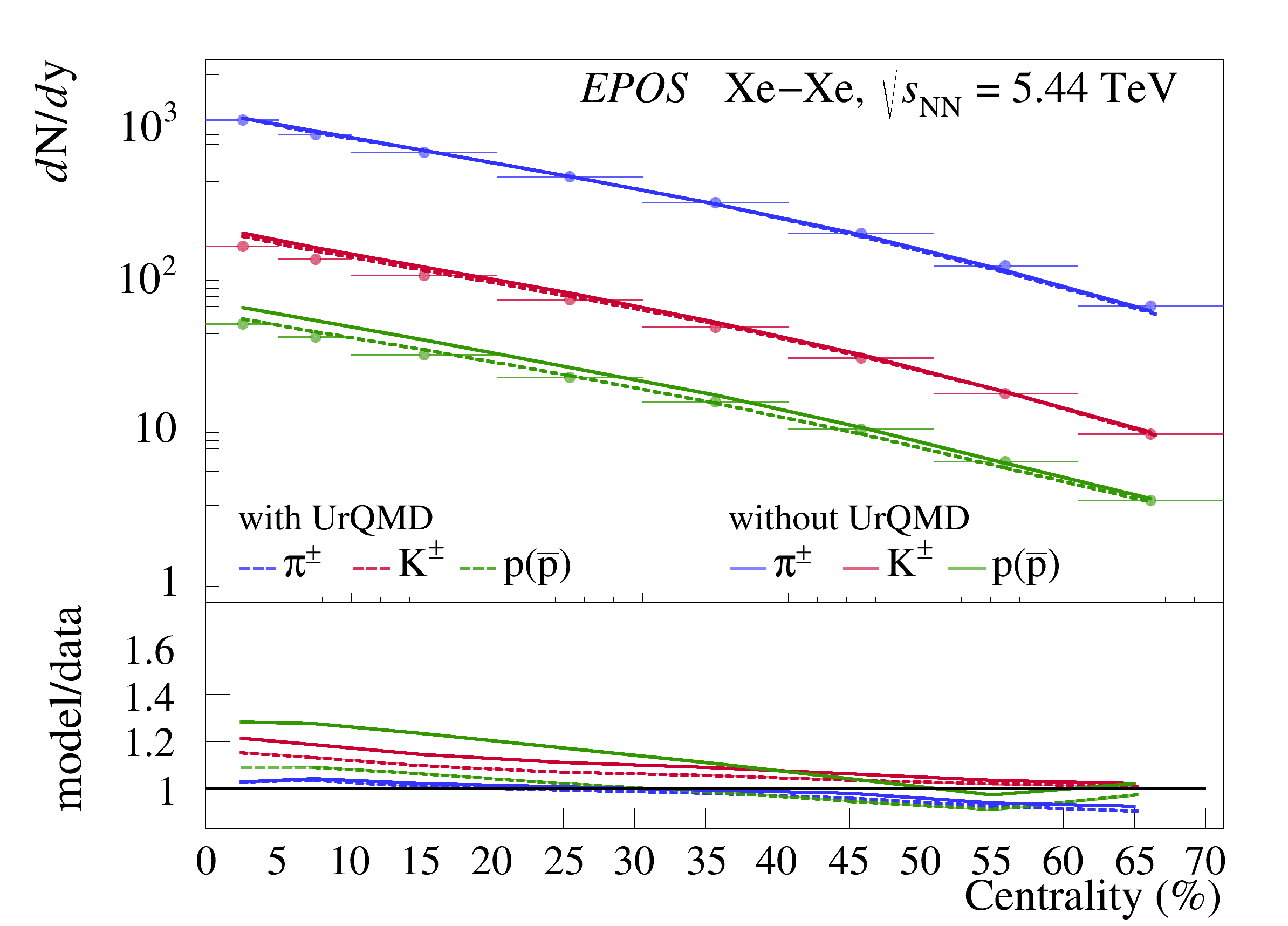}
  \end{minipage}
  \hfill
  \begin{minipage}{0.48\textwidth}
    \centering
    \includegraphics[width=\linewidth]{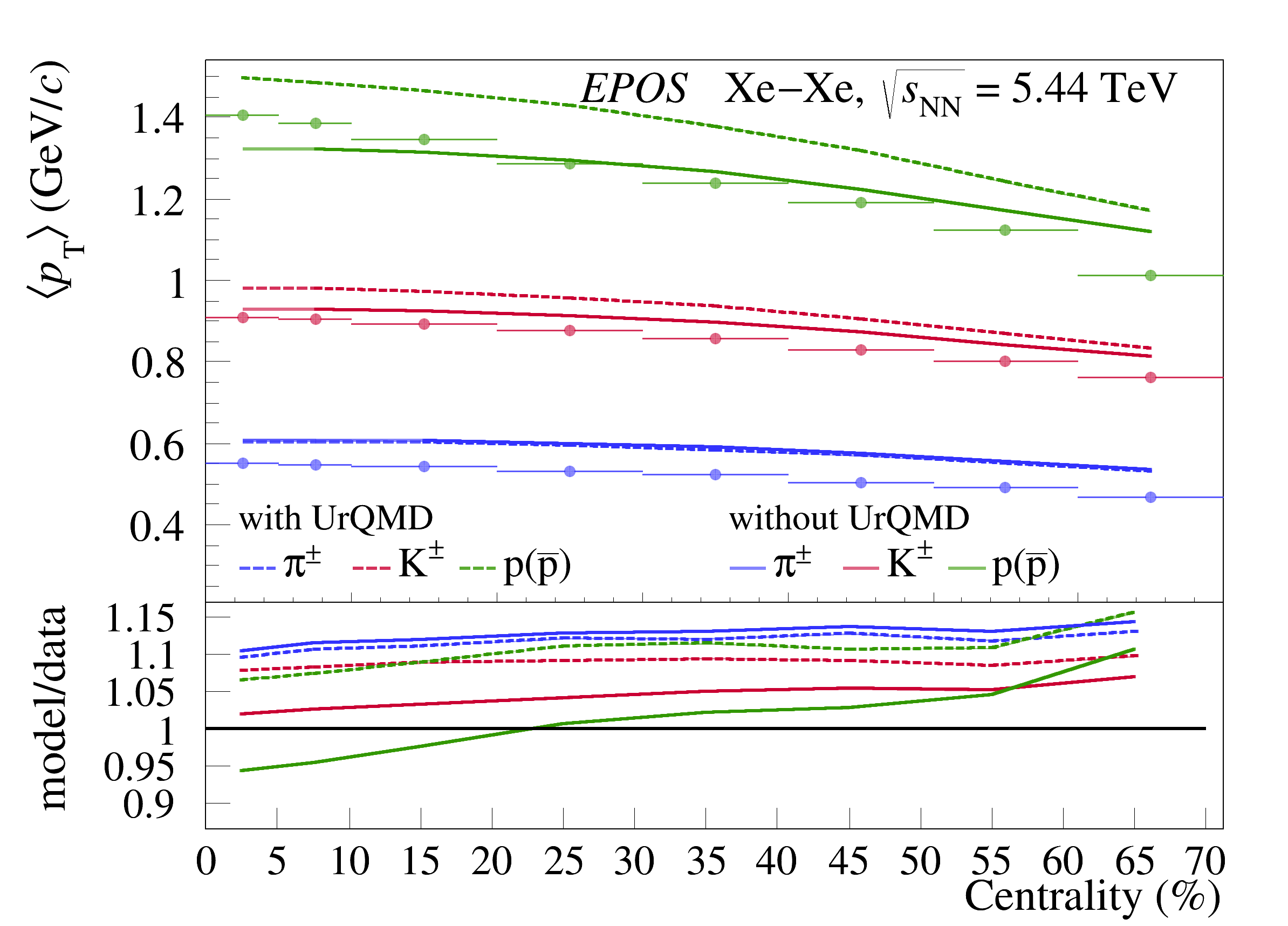}
  \end{minipage}

% --------- O-O ---------
   \begin{minipage}{0.48\textwidth}
    \centering
    \includegraphics[width=\linewidth]{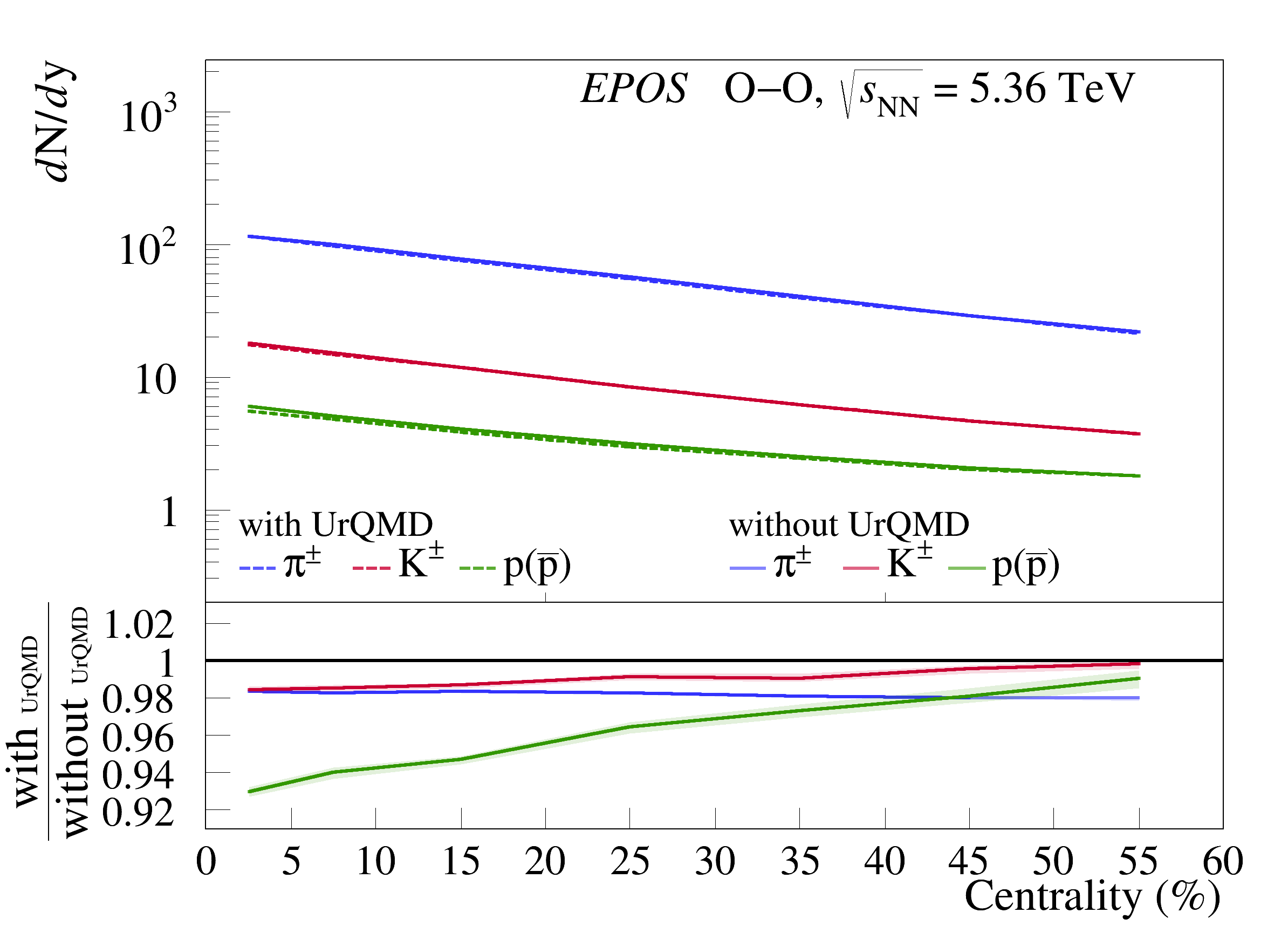}
  \end{minipage}
  \hfill
  \begin{minipage}{0.48\textwidth}
    \centering
    \includegraphics[width=\linewidth]{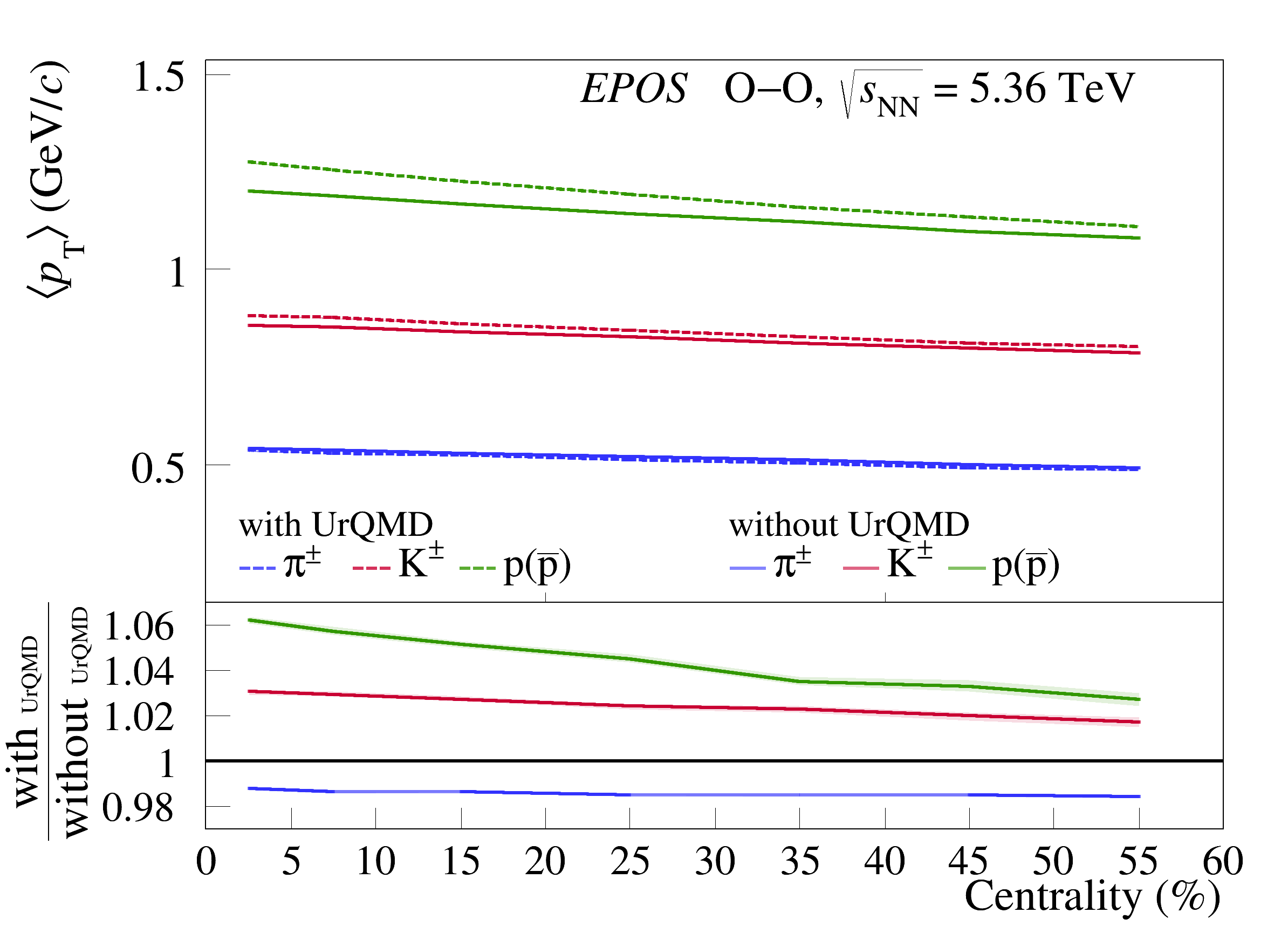}
  \end{minipage}

  \caption{Identified particle yields, $dN/dy$ (left) and mean transverse momentum, $\langle \tpt \rangle$ (right) of pions, kaons, and protons as a function of centrality in Pb--Pb collisions at \snn = 5.02 TeV. In each panel, EPOS4 calculations with and without the UrQMD hadronic afterburner are shown for several centrality classes, together with ALICE data points~\cite{ALICE:2019hno, ALICE:2021lsv} plotted as markers including statistical uncertainties.}
  \label{fig:dndy_meanpt}
\end{figure*}

%%%%%%%%%%%%%%%%%%%%%%%%%%%%%%%%%%%%%%%%%%%%%%%%%%%%%%%%%%%%%%%%%%%%%%%%%%%%%%%%%%%%%%%%
%%%%%%%%%%%%%%%%%%%%%%%%%%%%%%%%%%%%%%%%%%%%%%%%%%%%%%%%%%%%%%%%%%%%%%%%%%%%%%%%%%%%%%%%
%%%%%%%%%%%%%%%%%%%%%%%%%%%%%%%%%%%%%%%%%%%%%%%%%%%%%%%%%%%%%%%%%%%%%%%%%%%%%%%%%%%%%%%%
%%%%%%%%%%%%%%%%%%%%%%%%%%  CENTRALITY DETERMINA %%%%%%%%%%%%%%%%%%%%%%%%%%%%%%%%%%%%%%%
%%%%%%%%%%%%%%%%%%%%%%%%%%%%%%%%%%%%%%%%%%%%%%%%%%%%%%%%%%%%%%%%%%%%%%%%%%%%%%%%%%%%%%%%
%%%%%%%%%%%%%%%%%%%%%%%%%%%%%%%%%%%%%%%%%%%%%%%%%%%%%%%%%%%%%%%%%%%%%%%%%%%%%%%%%%%%%%%%

%%%%%%%%%%%%%%%%%%%%%%%%%%%%%%%%%%%%%%%%%%%%%%%%%%%%%%%%%%%%%%%%%%%%%%%%%%%%%%%%%%%%%%%%
\subsection{\label{sec:centralitydet}Centrality determination and Bulk production}
Centrality classes in EPOS4 are defined directly via the collision impact parameter ($b$). For Pb--Pb, the centrality classes are mapped to specific impact parameter intervals using published values~\cite{Adam:2113256} corresponding to the experimental geometric cross-section percentiles. Because published experimental values do not yet exist for Xe--Xe and O--O collisions, we determined its centrality classes by directly slicing the simulated impact parameter distribution into the corresponding geometric percentiles. This geometric approach serves as a proxy for the experimental determination of centrality, which typically relies on the amplitude of the signal in forward detectors, such as the V0 scintillator arrays in ALICE~\cite{ALICE:2013axi}. In smaller systems or peripheral heavy-ion collisions, where relative multiplicity fluctuations are large, such a selection can correspond to a narrower and more idealized event sample than an experimental selection based on final-state particle multiplicities. For the results presented here, the impact parameter intervals are chosen to correspond to the standard centrality classes (0--5\%, 5--10\%, 10--20\%, 20--30\%, 30--40\%, 40--50\%, 50--60\%) used in Pb--Pb and Xe--Xe measurements.

The charged-particle pseudorapidity density, \dnch{\eta}  serves as a fundamental observable for characterizing the global properties of the system and the total entropy production. Figure~\ref{fig:dnchdeta} (upper portion) presents the \dnch{\eta} distributions for Pb--Pb collisions at various centrality classes. The EPOS4 calculations including the UrQMD hadronic afterburner (left panel) show excellent agreement with the ALICE data~\cite{ALICE:2016fbt} throughout the pseudorapidity range. The model reproduces both the magnitude of the multiplicity density in the central plateau region and the fall-off at forward and backward rapidities associated with limiting fragmentation. The results for the intermediate system, Xe--Xe collisions are given in the lower portion of Fig.~\ref{fig:dnchdeta}. Similar to the Pb--Pb case, the model accurately describes the shape of the pseudorapidity distributions and the centrality dependence. The distribution for the smaller system, O--O is shown in Fig.~\ref{fig:dnchdeta1}. The distribution exhibits a shape similar to the larger systems but with significantly reduced overall multiplicity. The impact of the hadronic afterburner on bulk production is assessed by comparing the results with and without UrQMD. In Fig.~\ref{fig:dnchdeta}, the difference between the dashed lines (with UrQMD) and solid lines (without UrQMD) is generally small for the total charged-particle density. The ratio in Fig.~\ref{fig:dnchdeta} shows afterburner effect, with the ratio of with to without UrQMD remaining below unity across all centrality region, indicating a modest overall suppression of charged-particle multiplicity due to hadronic rescattering.

While the total charged multiplicity is relatively robust against hadronic rescattering, the yields of identified particles show more sensitivity. Figure~\ref{fig:dndy_meanpt} (left panel) presents the \tpt-integrated yields (\dnch{y}) for pions (\partpm{\pi}), kaons (\partpm{K}), and protons (\partpr) as a function of centrality in Pb--Pb collisions at 5.02 TeV. The model provides a reasonable description of the pion and kaon yields. For protons, a distinct effect of the hadronic afterburner is observed. The ratio of calculations with UrQMD to those without (bottom panel of Fig.~\ref{fig:dndy_meanpt}) falls below unity for protons, particularly in central collisions. This suppression arises from baryon-antibaryon annihilation processes within the dense hadronic phase simulated by UrQMD~\cite{Steinheimer:2017vju}. These rescattering effects are necessary to bring the proton yields into better agreement with the experimental data, indicating the late hadronic stage plays an important role in shaping the baryon production. The mean transverse momentum, $\langle \tpt \rangle$, shown in the right panel of Fig.~\ref{fig:dndy_meanpt}, exhibits a characteristic mass ordering, with protons having the highest $\langle \tpt \rangle$, followed by kaons and pions. This ordering is a signature of radial flow generated when the core is going through hydrodynamic expansion~\cite{Werner:2023jps}. The UrQMD afterburner further modifies the spectra; for heavier particles it leads to a moderate hardening, reflecting additional radial flow generated during the hadronic cascade and the depletion of low-\tpt protons through annihilation processes.

\begin{figure*}[htbp!]
  \centering

  % --------- with UrQMD ---------
  \begin{minipage}{0.325\textwidth}
    \centering
    \includegraphics[width=\linewidth]{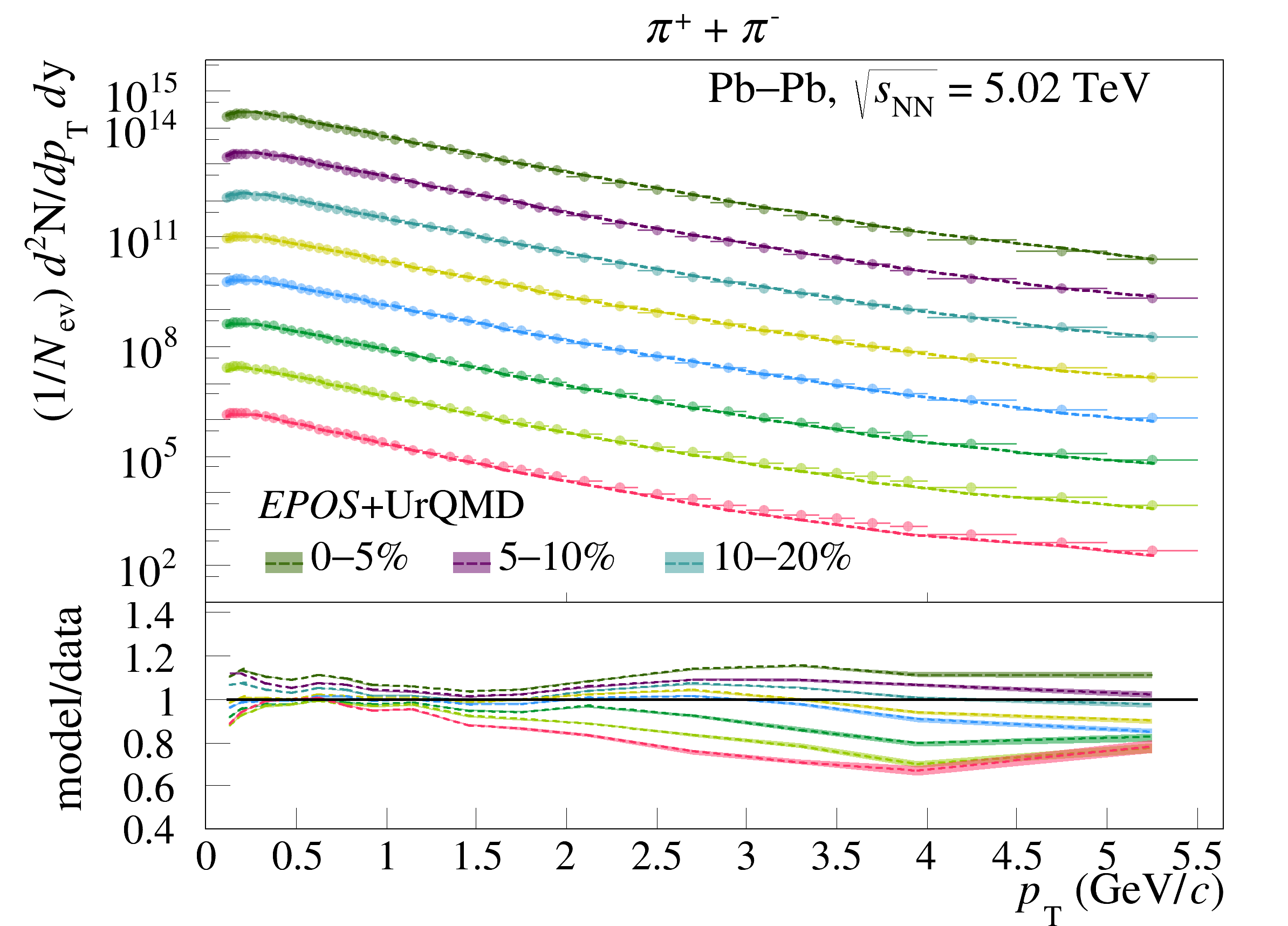}
    % \subcaption{Pb--Pb $\pi$}   % optional, if using subcaption
  \end{minipage}
  \hfill
  \begin{minipage}{0.325\textwidth}
    \centering
    \includegraphics[width=\linewidth]{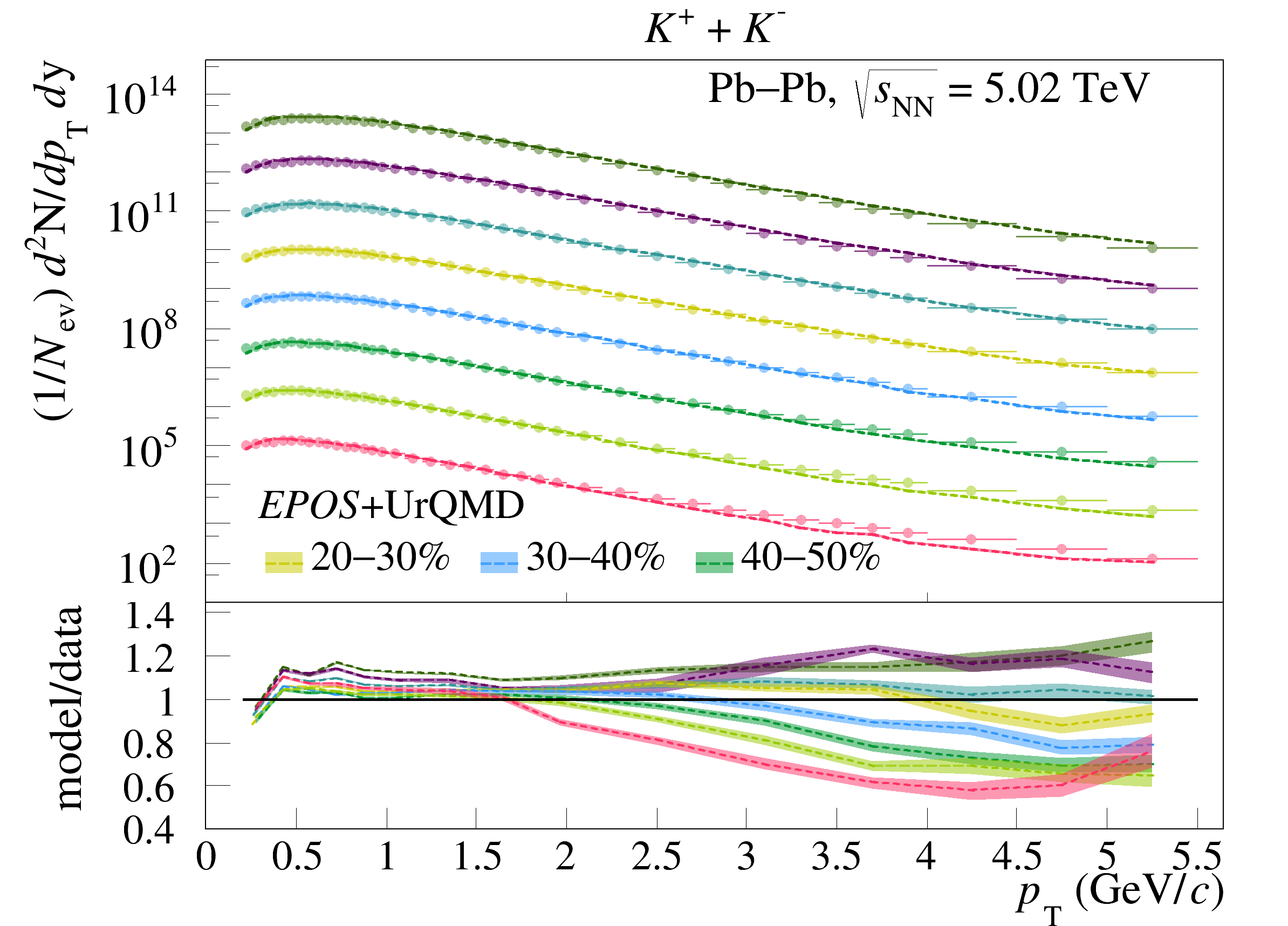}
    % \subcaption{Pb--Pb K}
  \end{minipage}
  \hfill
  \begin{minipage}{0.325\textwidth}
    \centering
    \includegraphics[width=\linewidth]{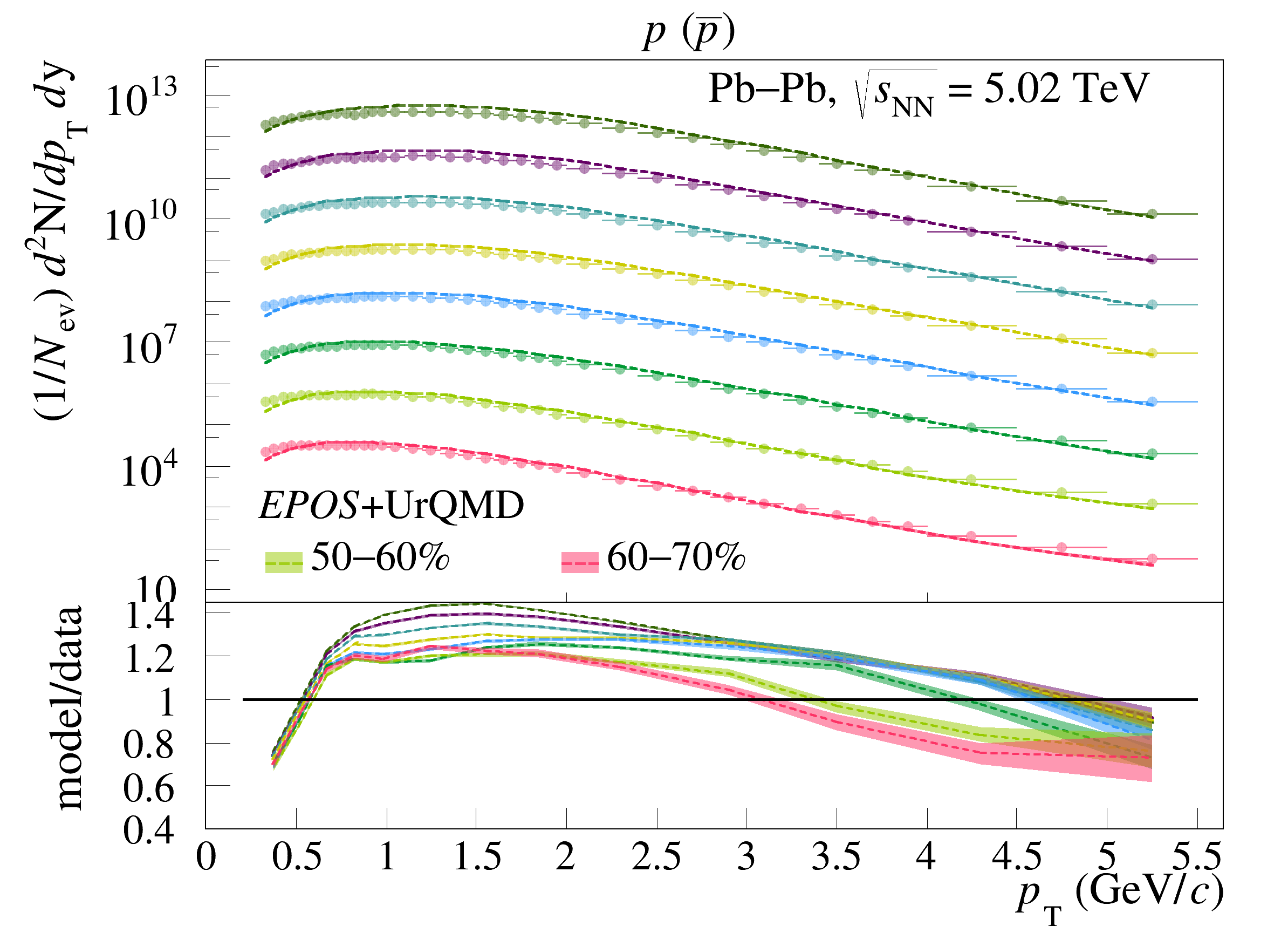}
    % \subcaption{Pb--Pb p}
  \end{minipage}

  %\vspace{-1.5em} % vertical gap between rows

  % --------- without UrQMD ---------
  \begin{minipage}{0.325\textwidth}
    \centering
    \includegraphics[width=\linewidth]{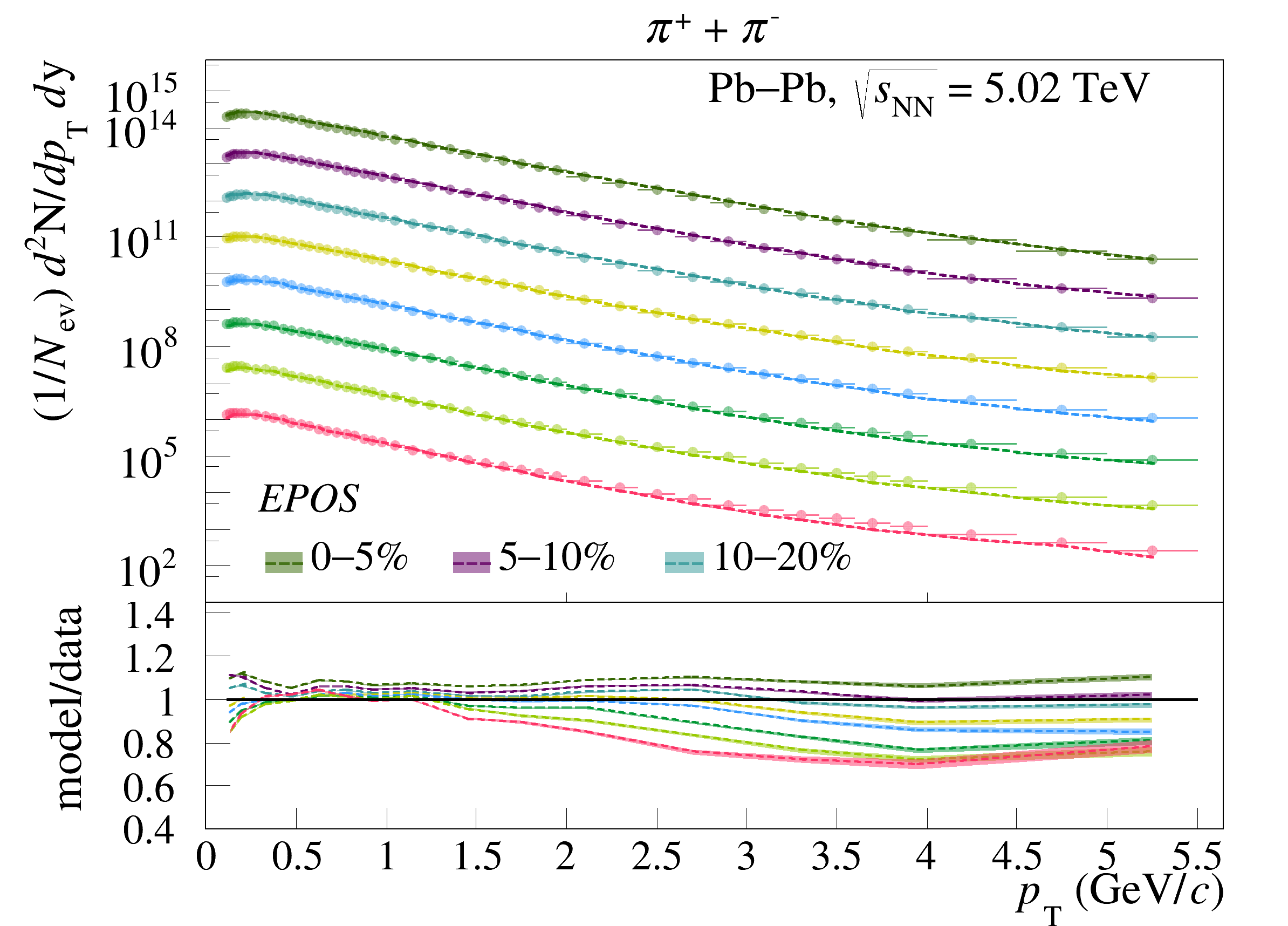}
    % \subcaption{Pb--Pb $\pi$}
  \end{minipage}
  \hfill
  \begin{minipage}{0.325\textwidth}
    \centering
    \includegraphics[width=\linewidth]{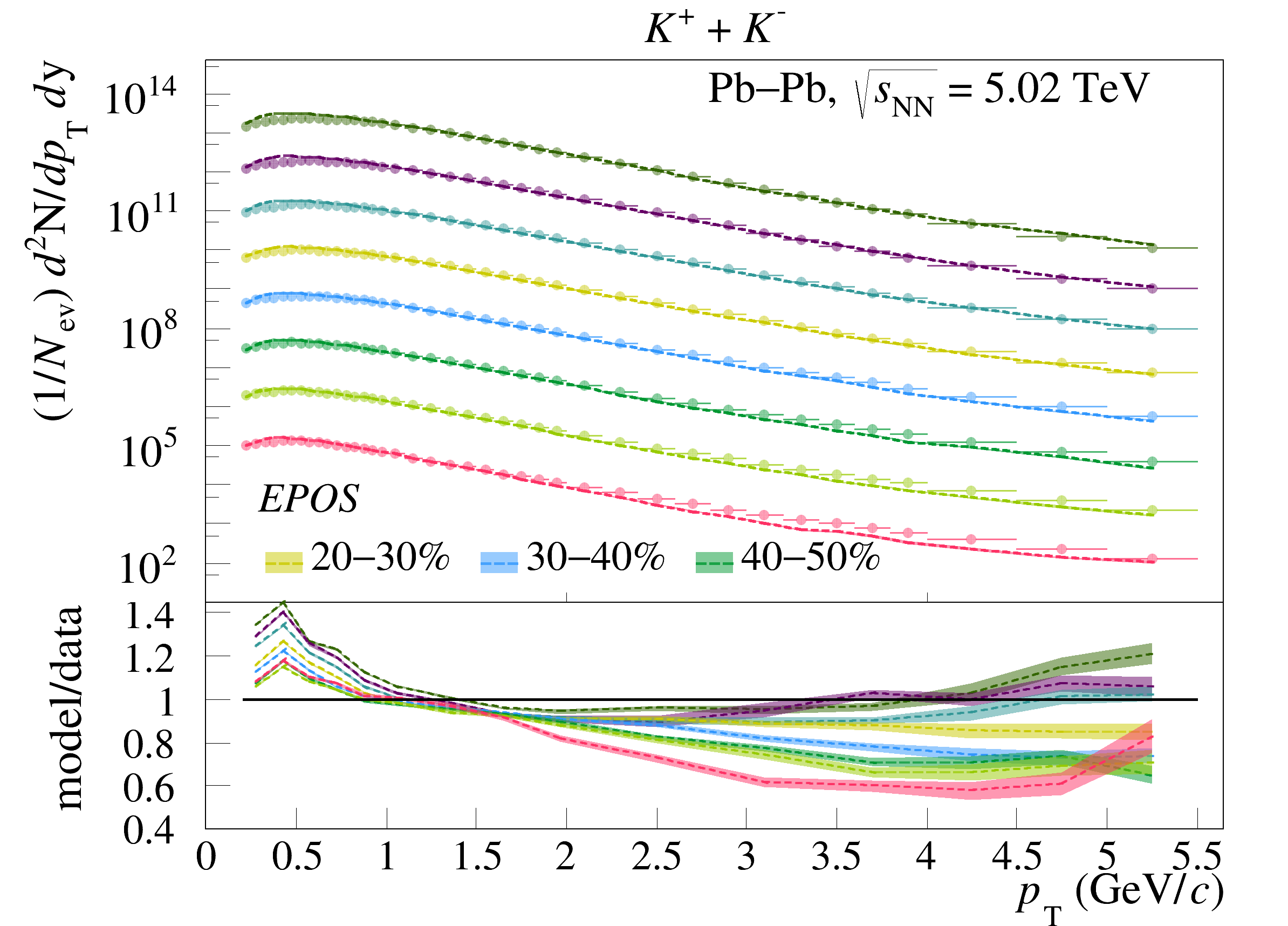}
    % \subcaption{Pb--Pb K}
  \end{minipage}
  \hfill
  \begin{minipage}{0.325\textwidth}
    \centering
    \includegraphics[width=\linewidth]{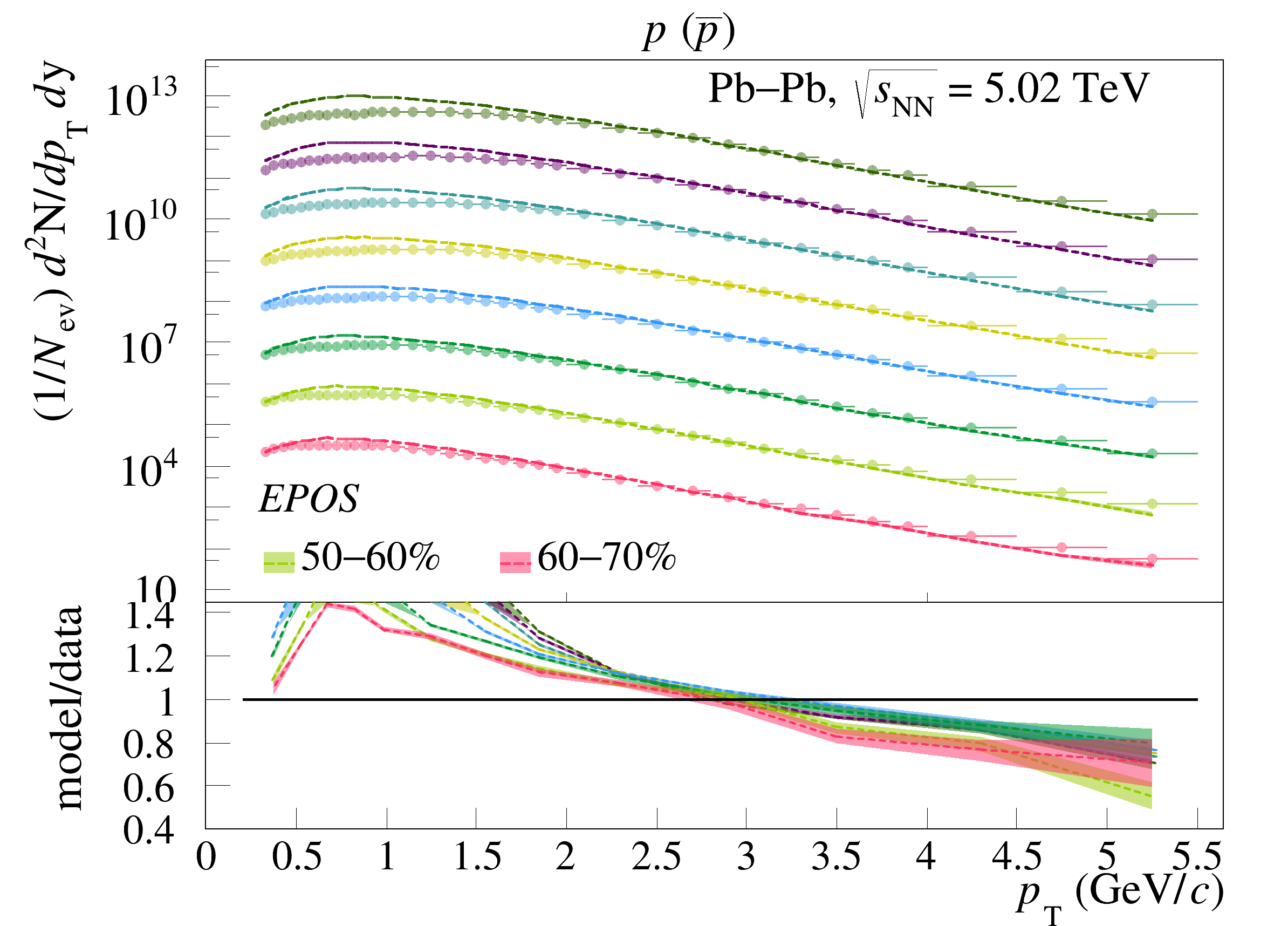}
    % \subcaption{Pb--Pb p}
  \end{minipage}

  \caption{
    Identified particle yields as a function of transverse momentum (\tpt) for pions (left), kaons (middle), and protons (right) at midrapidity in Pb--Pb collisions at (\snn = 5.02) TeV. The top row shows EPOS4 predictions including the UrQMD hadronic afterburner and the bottom row shows corresponding results without hadronic rescattering. Curves of different colours represent several centrality classes, as indicated in the legends, and the markers represent ALICE measurements~\cite{ALICE:2019hno}. For visual clarity, the curves are scaled by successive factors of 10, from $10^{11}$ for the most central (0--5\%) to $10^4$ for the most peripheral class (60--70\%).
  }
  \label{fig:yields_pbpb}
\end{figure*}

\begin{figure*}[htbp!]
  \centering

  % --------- with UrQMD ---------
  \begin{minipage}{0.325\textwidth}
    \centering
    \includegraphics[width=\linewidth]{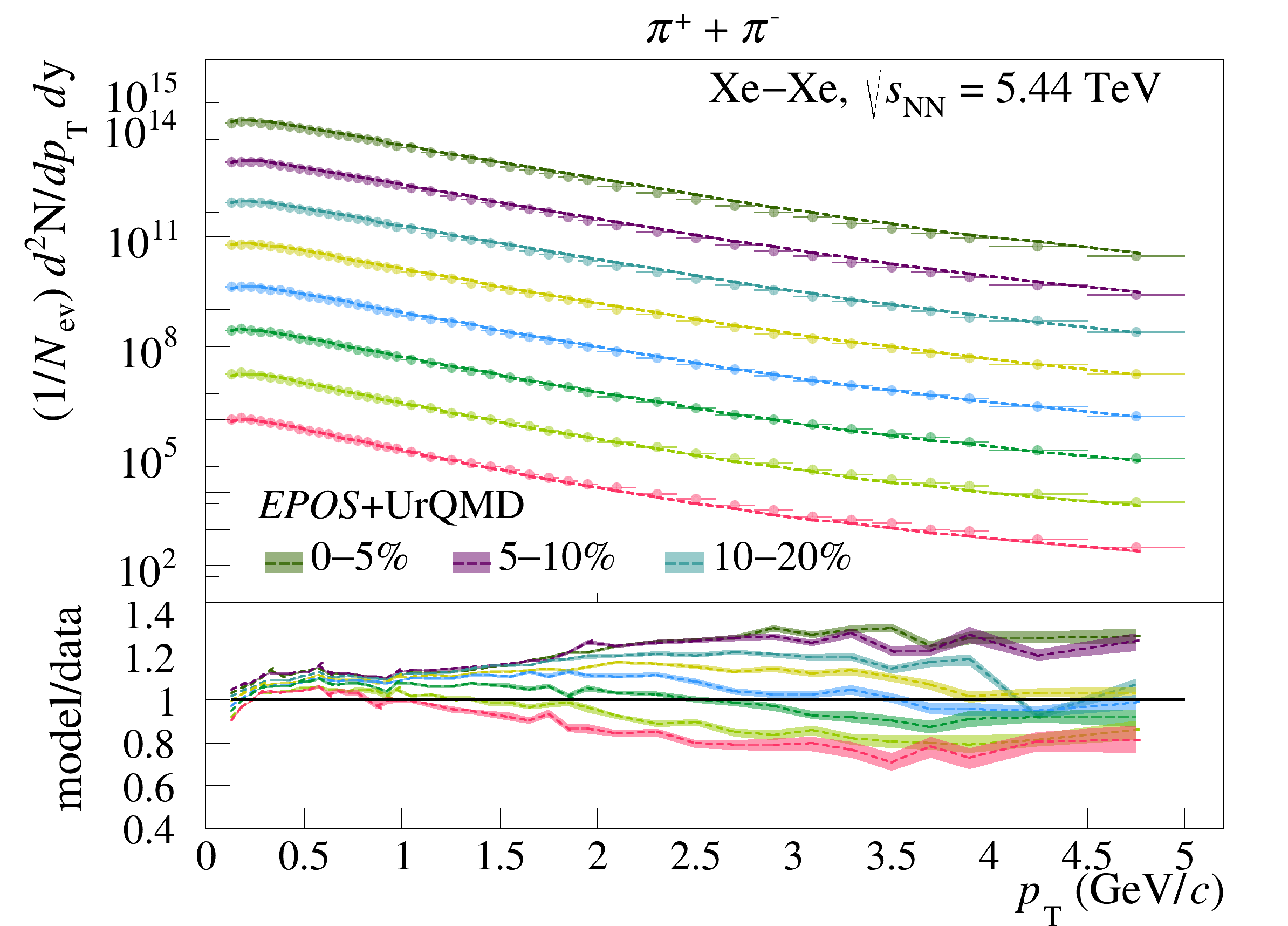}
    % \subcaption{Xe--Xe $\pi$}   % optional, if using subcaption
  \end{minipage}
  \hfill
  \begin{minipage}{0.325\textwidth}
    \centering
    \includegraphics[width=\linewidth]{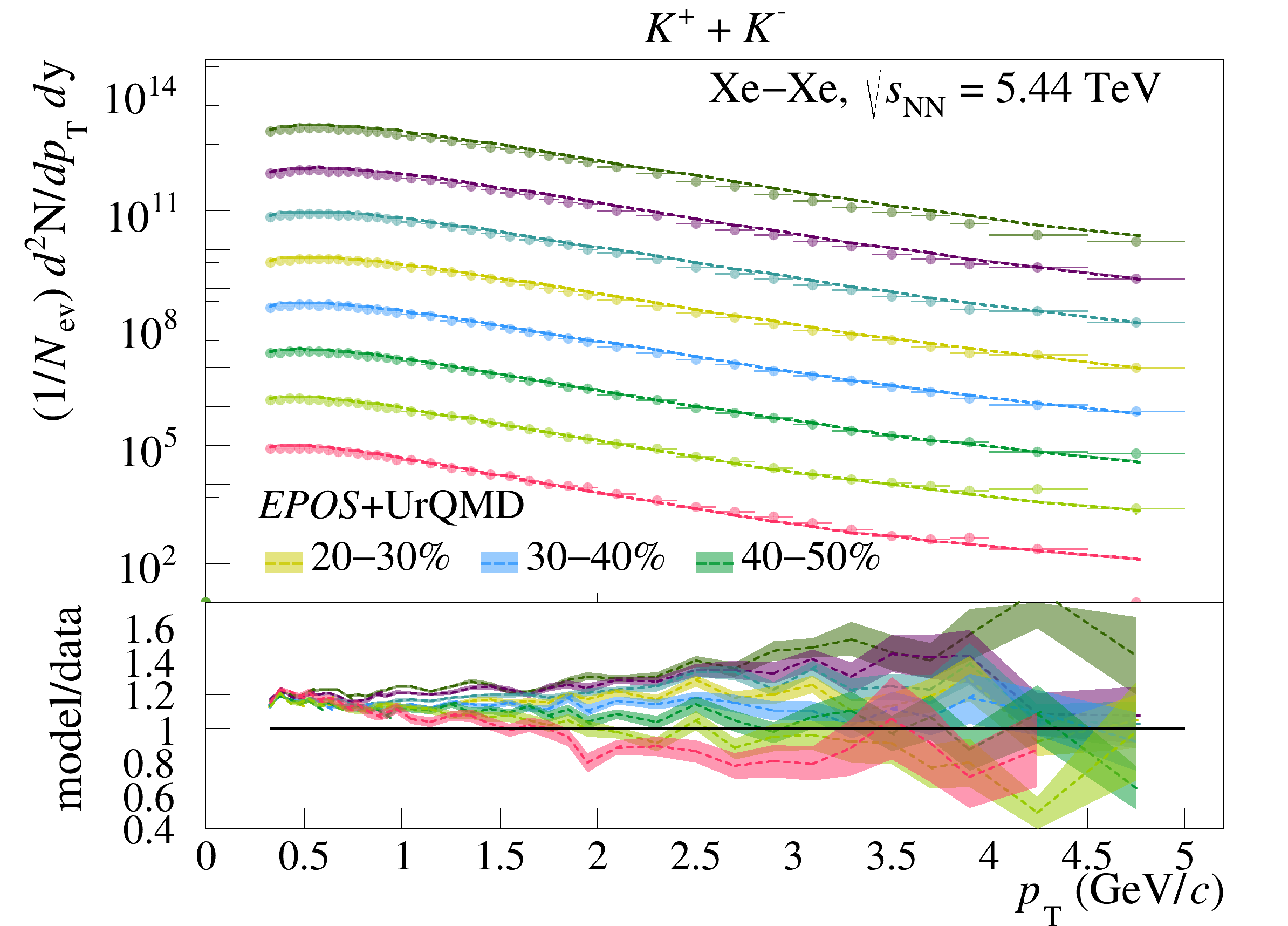}
    % \subcaption{Xe--Xe K}
  \end{minipage}
  \hfill
  \begin{minipage}{0.325\textwidth}
    \centering
    \includegraphics[width=\linewidth]{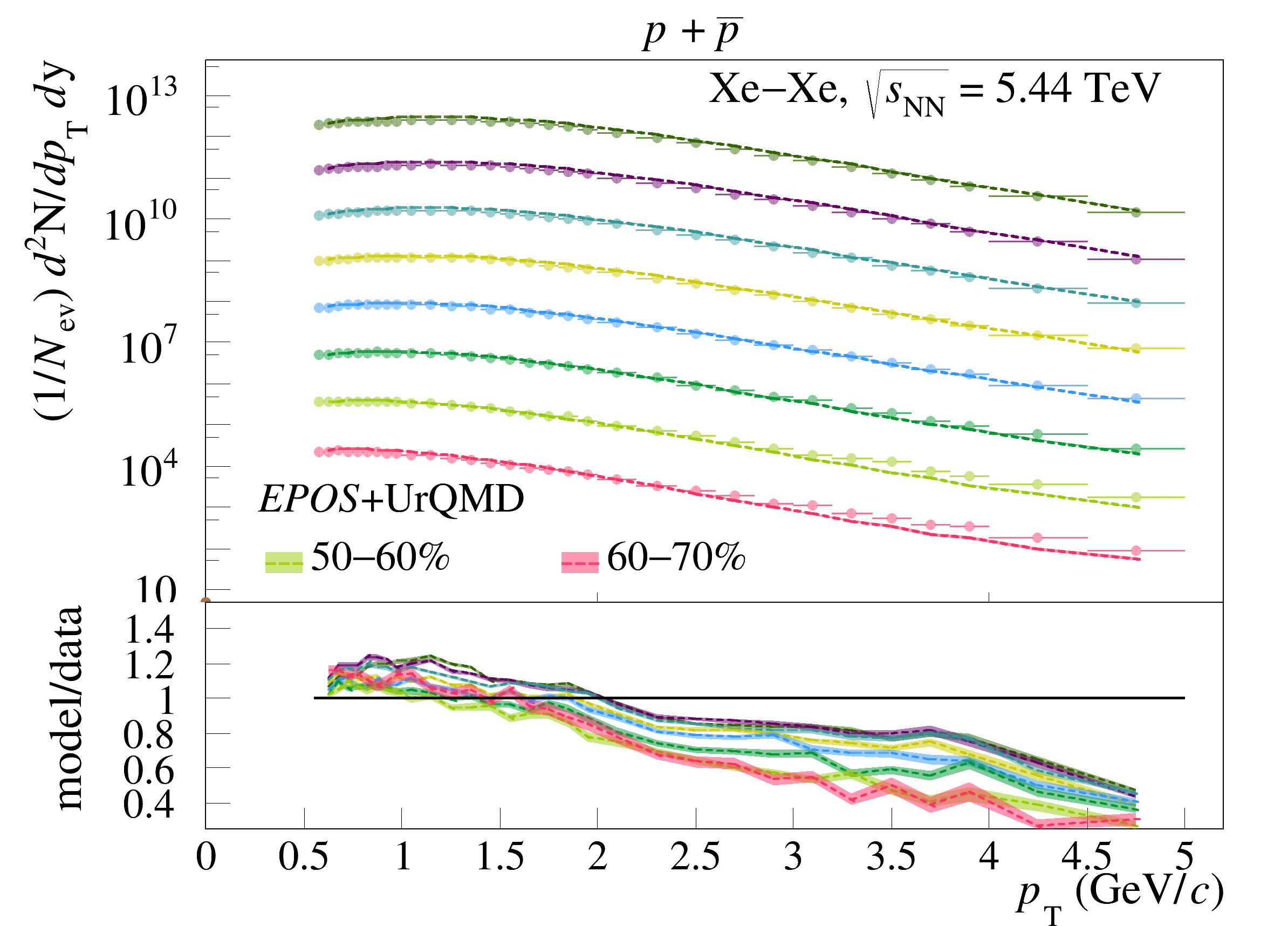}
    % \subcaption{Xe--Xe p}
  \end{minipage}

  %\vspace{-1.5em} % vertical gap between rows

  % --------- without UrQMD ---------
  \begin{minipage}{0.325\textwidth}
    \centering
    \includegraphics[width=\linewidth]{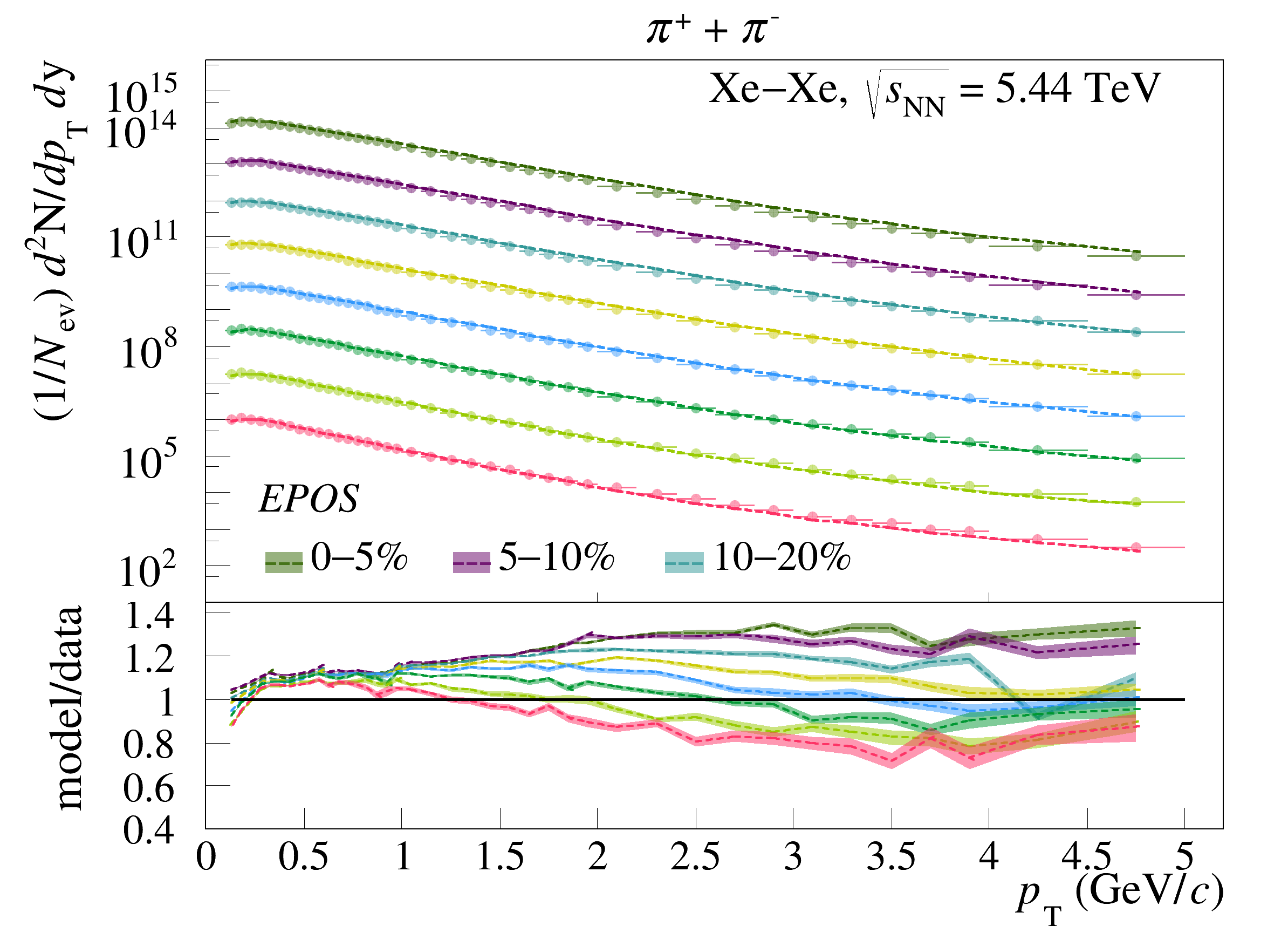}
    % \subcaption{Xe--Xe $\pi$}
  \end{minipage}
  \hfill
  \begin{minipage}{0.325\textwidth}
    \centering
    \includegraphics[width=\linewidth]{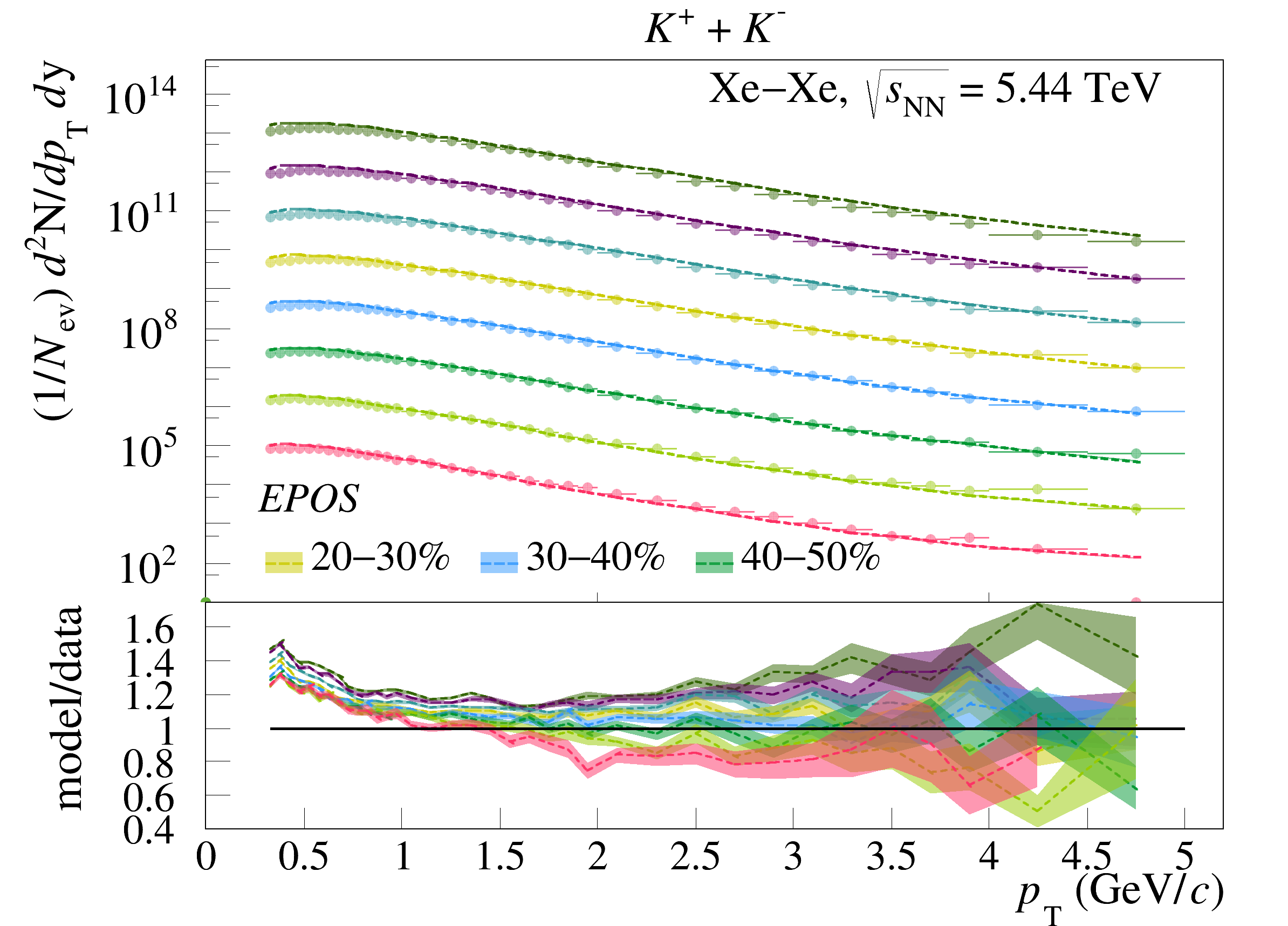}
    % \subcaption{Xe--Xe K}
  \end{minipage}
  \hfill
  \begin{minipage}{0.325\textwidth}
    \centering
    \includegraphics[width=\linewidth]{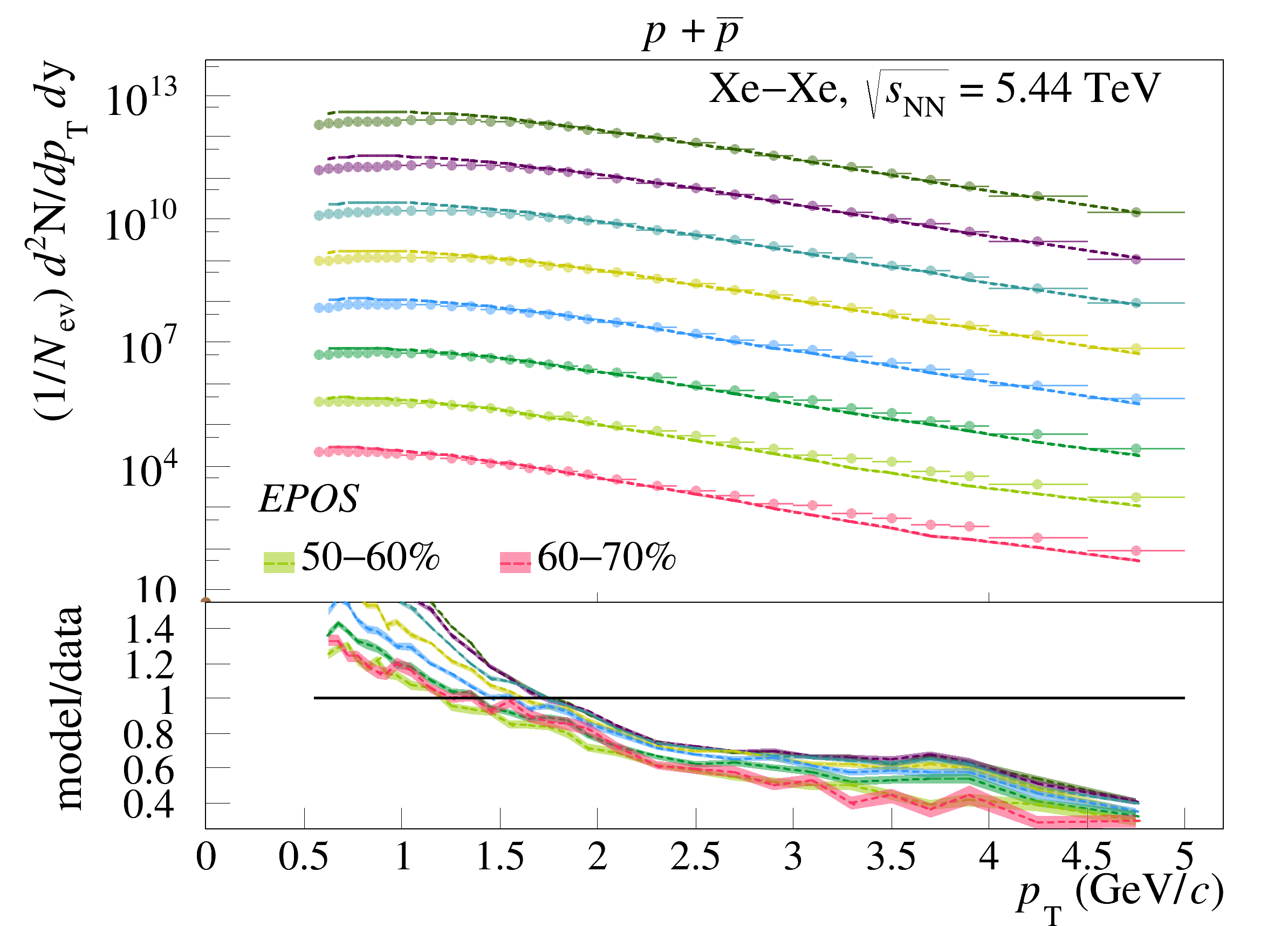}
    % \subcaption{Xe--Xe p}
  \end{minipage}

  \caption{
    Identified particle yields as a function of transverse momentum (\tpt) for pions (left), kaons (middle), and protons (right) at midrapidity in Xe--Xe collisions at (\snn = 5.44) TeV. The top row shows EPOS4 predictions including the UrQMD hadronic afterburner and the bottom row shows corresponding results without hadronic rescattering. Curves of different colours represent several centrality classes, as indicated in the legends, and the markers represent ALICE measurements~\cite{ALICE:2021lsv}.  For visual clarity, the curves are scaled by successive factors of 10, from $10^{11}$ for the most central (0--5\%) to $10^4$ for the most peripheral class (60--70\%).
    }
  \label{fig:yields_xexe}
\end{figure*}

%%%%%%%%%%%%%%%%%%%%%%%%%%%%%%%%%%%%%%%%%%%%%%%%%%%%%%%%%%%%%%%%%%%%%%%%%%%%%%%%%%%%%%%%
%%%%%%%%%%%%%%%%%%%%%%%%%%%%%%%%%%%%%%%%%%%%%%%%%%%%%%%%%%%%%%%%%%%%%%%%%%%%%%%%%%%%%%%%
%%%%%%%%%%%%%%%%%%%%%%%%%%%%%%%%%%%%%%%%%%%%%%%%%%%%%%%%%%%%%%%%%%%%%%%%%%%%%%%%%%%%%%%%
%%%%%%%%%%%%%%%%%%%%%%%%%%  CHARGED PARTICLE PRO %%%%%%%%%%%%%%%%%%%%%%%%%%%%%%%%%%%%%%%
%%%%%%%%%%%%%%%%%%%%%%%%%%%%%%%%%%%%%%%%%%%%%%%%%%%%%%%%%%%%%%%%%%%%%%%%%%%%%%%%%%%%%%%%
%%%%%%%%%%%%%%%%%%%%%%%%%%%%%%%%%%%%%%%%%%%%%%%%%%%%%%%%%%%%%%%%%%%%%%%%%%%%%%%%%%%%%%%%
%%%%%%%%%%%%%%%%%%%%%%%%%%%%%%%%%%%%%%%%%%%%%%%%%%%%%%%%%%%%%%%%%%%%%%%%%%%%%%%%%%%%%%%%
\subsection{Transverse Momentum, \tpt ~Spectra}
The transverse momentum (\tpt) distributions of identified hadrons serve as a primary probe of the collective dynamics and the thermal properties of the system created in high-energy nuclear collisions. The spectral shapes encode information regarding the radial flow velocity and the freeze-out temperature, while the particle species dependence reveals the interplay between thermal motion and collective expansion.

Figure~\ref{fig:yields_pbpb} and Fig.~\ref{fig:yields_xexe} present the invariant yields for pions (\partpm{\pi}), kaons (\partpm{K}), and protons (\partpr) at midrapidity in Pb--Pb collisions at \snn = 5.02 TeV and Xe--Xe collisions at \snn = 5.44 TeV, respectively. The EPOS4 calculations including the UrQMD hadronic stage (dashed lines) are compared with experimental data from the ALICE (markers)~\cite{ALICE:2019hno,ALICE:2021lsv}. The spectra exhibit a characteristic hardening with increasing centrality, consistent with radial flow driven by pressure gradients in the hydrodynamic core. This effect is mass-dependent, protons, being heavier, receive a larger momentum boost from the collective velocity field than pions, leading to a flatter spectral shape at low and intermediate \tpt. The model provides a consistent description of the experimental data for both Pb--Pb and Xe--Xe systems across the measured centrality classes. 
The role of the microscopic hadronic afterburner is illustrated by the comparison between calculations with (dashed lines) and without (solid lines) UrQMD. For pions, the effect of the hadronic phase is small, as their low mass makes them less sensitive to additional radial flow and their yields remain largely unchanged. For protons, the hadronic stage introduces two competing effects. At low \tpt ($\leq 2$ \gev), the inclusion of UrQMD leads to a noticeable suppression of the yield, which is also visible for kaons. This reduction is attributed to baryon--antibaryon annihilation processes in the dense hadronic phase. At intermediate \tpt, hadronic rescattering generates additional radial flow, shifting protons to higher transverse momenta and hardening the spectra. This mechanism plays an important role in reproducing the shape of the proton spectra observed in the data.

\begin{figure*}[htbp!]
  \centering

  % --------- with and without UrQMD ---------
  \begin{minipage}{0.325\textwidth}
    \centering
    \includegraphics[width=\linewidth]{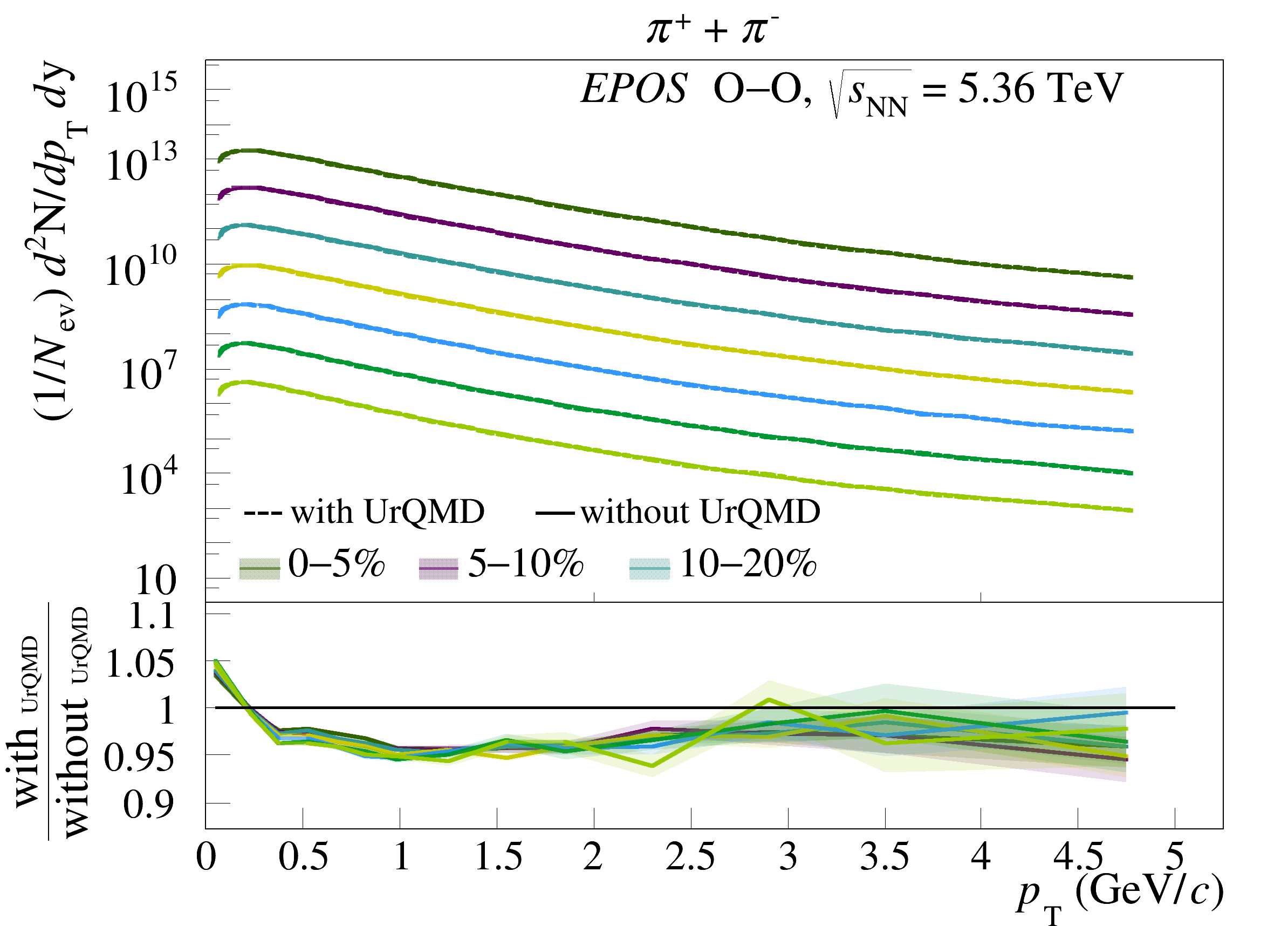}
    % \subcaption{O--O $\pi$}   % optional, if using subcaption
  \end{minipage}
  \hfill
  \begin{minipage}{0.325\textwidth}
    \centering
    \includegraphics[width=\linewidth]{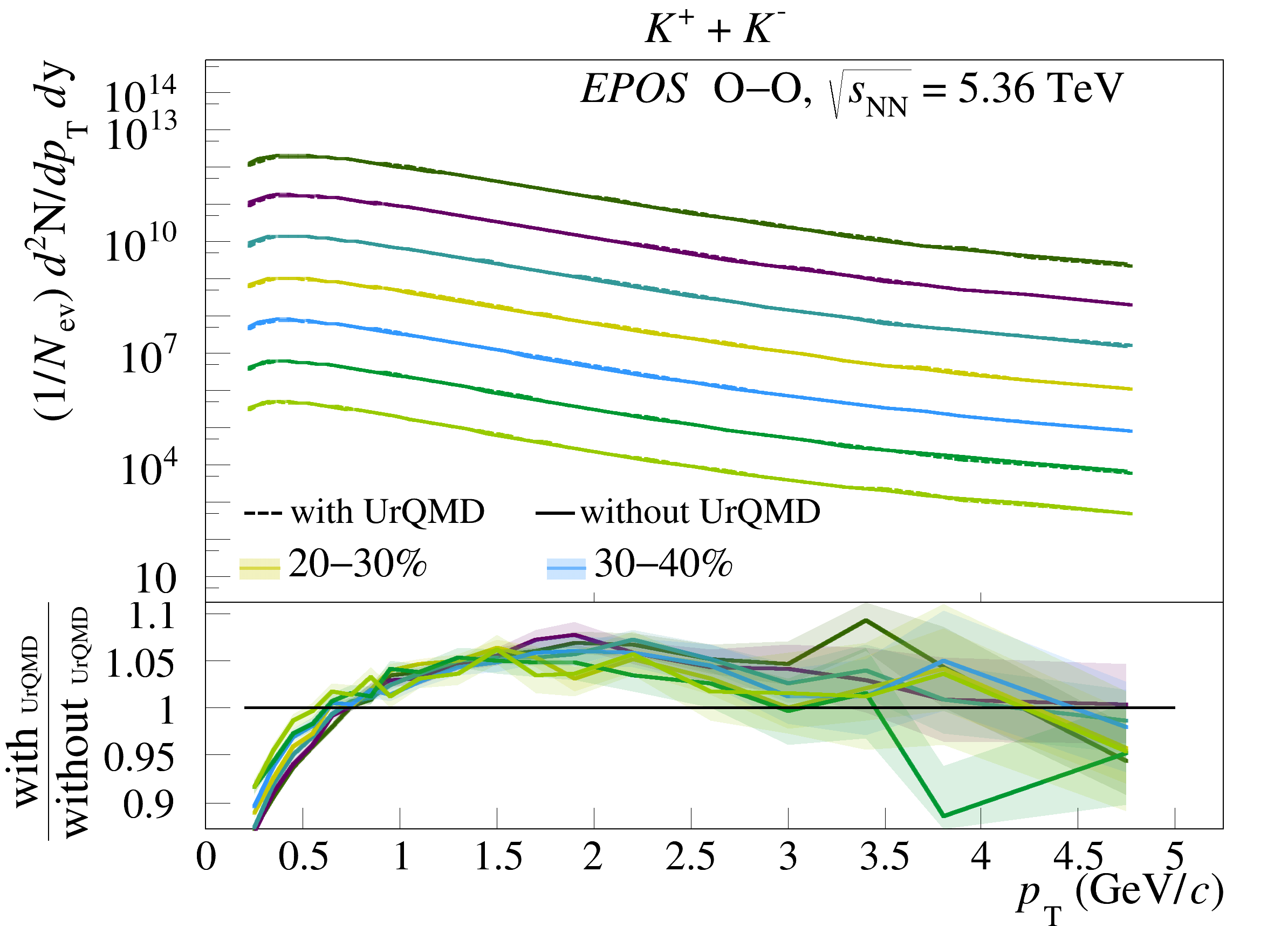}
    % \subcaption{O--O K}
  \end{minipage}
  \hfill
  \begin{minipage}{0.325\textwidth}
    \centering
    \includegraphics[width=\linewidth]{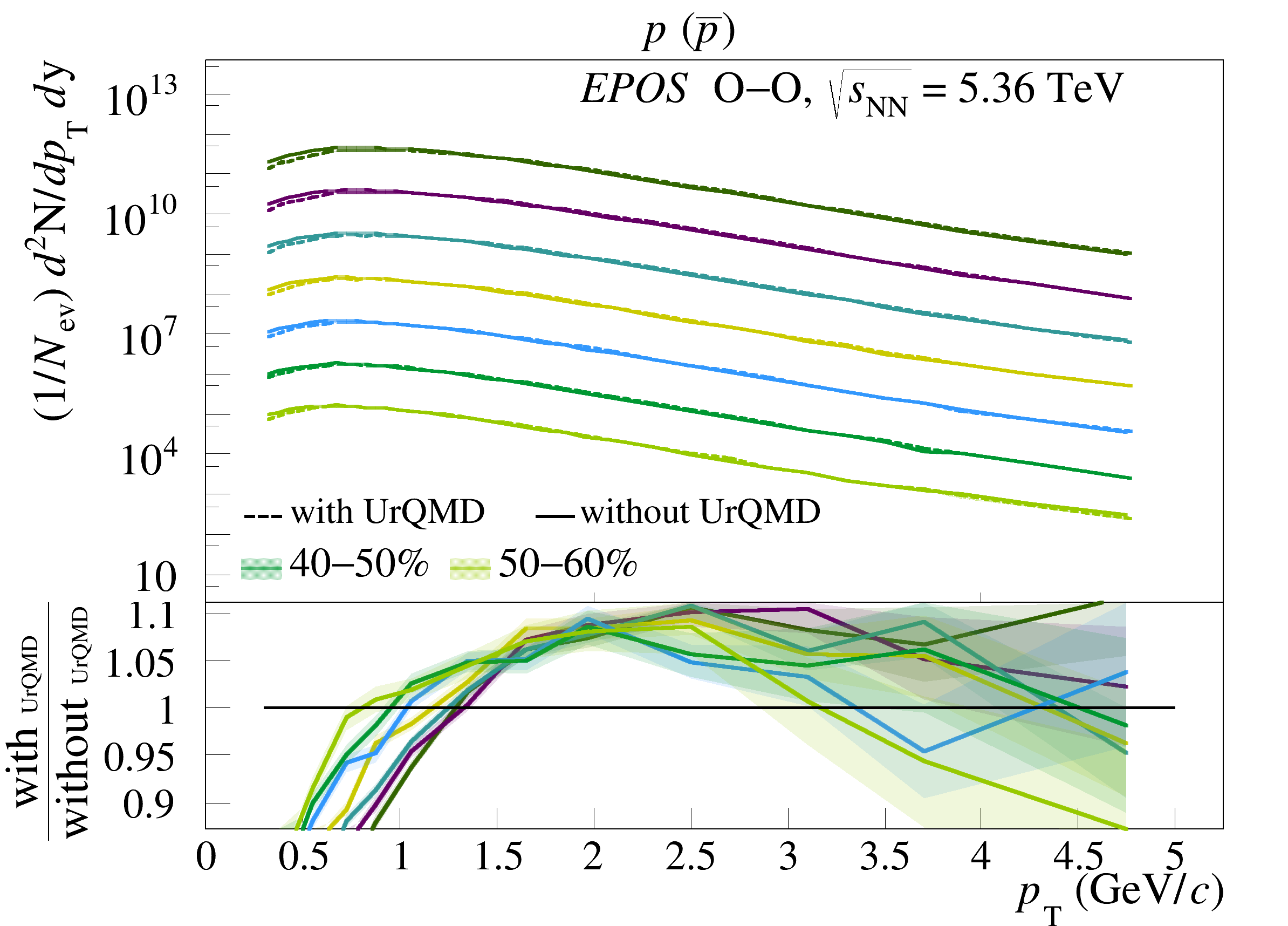}
    % \subcaption{O--O p}
  \end{minipage}
  
  \caption{
    Identified particle yields as a function of transverse momentum (\tpt) for pions (left), kaons (middle), and protons (right) at midrapidity in O--O collisions at (\snn = 5.36) TeV. The results shows EPOS4 predictions including the UrQMD hadronic afterburner (in dashed lines) and without hadronic rescattering (in solid lines). Curves of different colours represent several centrality classes and are scaled by successive factors of 10, from $10^{11}$ for the most central (0--5\%) to $10^5$ for the most peripheral class (50--60\%) for visual clarity.
  }
  \label{fig:yields_oo}
\end{figure*}

Figure~\ref{fig:yields_oo} shows the results for O--O system at \snn = 5.36 TeV. The spectra display similar qualitative features to the larger systems, including the mass-dependent hardening driven by collective expansion. Although the system size is significantly smaller, the EPOS4 framework predicts the formation of a hydrodynamic core even in O--O collisions, resulting in non-negligible radial flow. At low \tpt, the suppression of yields for kaons and protons is $~$10\% when UrQMD is included. The ratio of the calculations with and without UrQMD (bottom panels) indicates these hadronic rescattering effects remain relevant in this smaller system, particularly for modifying the proton yields through annihilation and flow, although the magnitude of the modification is reduced compared to central Pb--Pb collisions.
These spectral features arise from the core–corona separation mechanism implemented in EPOS4~\cite{Werner:2023jps}. The low-\tpt\ region is dominated by particle production from the thermalized core, which expands hydrodynamically and imparts a common flow velocity to the emitted hadrons. In contrast, the high-\tpt\ region is dominated by the corona contribution, originating from the fragmentation of hard string segments that escape the medium. The yield in this power-law tail is governed by the generalized Abramovskii–Gribov–Kancheli (AGK) theorem implemented in EPOS4~\cite{Werner:2023mod}, where event-by-event dynamical saturation scales ensure that factorization and binary scaling are approximately preserved at high \tpt\ despite the complex multiple-scattering environment. The transition from exponential (thermal-like) behaviour to a power-law tail occurs at a \tpt\ scale that depends on the particle mass and the magnitude of the radial flow. 

\begin{figure*}[htbp!]
  \centering
% --------- Pb-Pb ---------
  \begin{minipage}{0.325\textwidth}
    \centering
    \includegraphics[width=\linewidth]{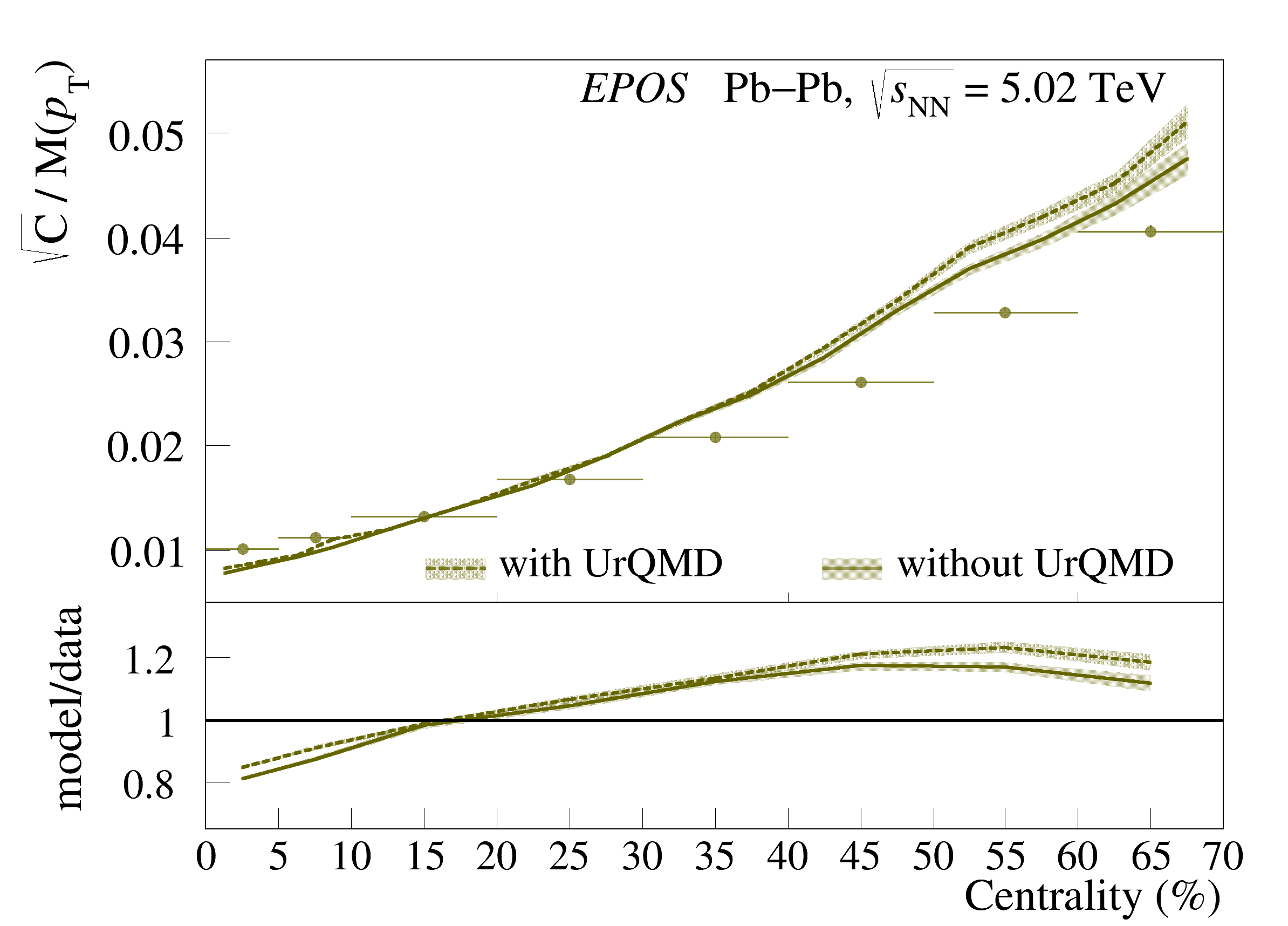}
  \end{minipage}
  \hfill
% --------- Xe-Xe ---------
  \begin{minipage}{0.325\textwidth}
    \centering
    \includegraphics[width=\linewidth]{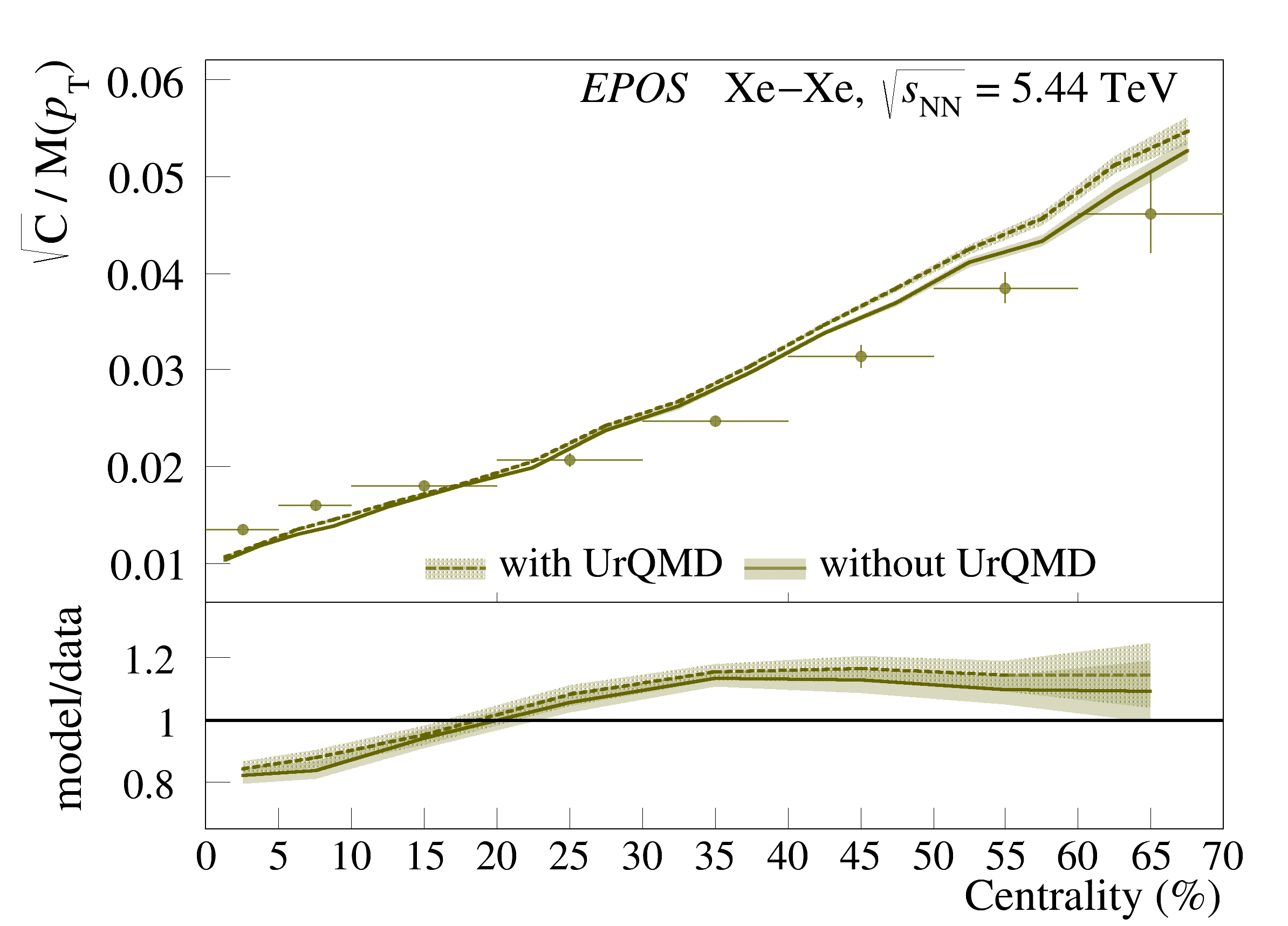}
  \end{minipage}
  \hfill
% --------- O-O ---------
  \begin{minipage}{0.325\textwidth}
    \centering
    \includegraphics[width=\linewidth]{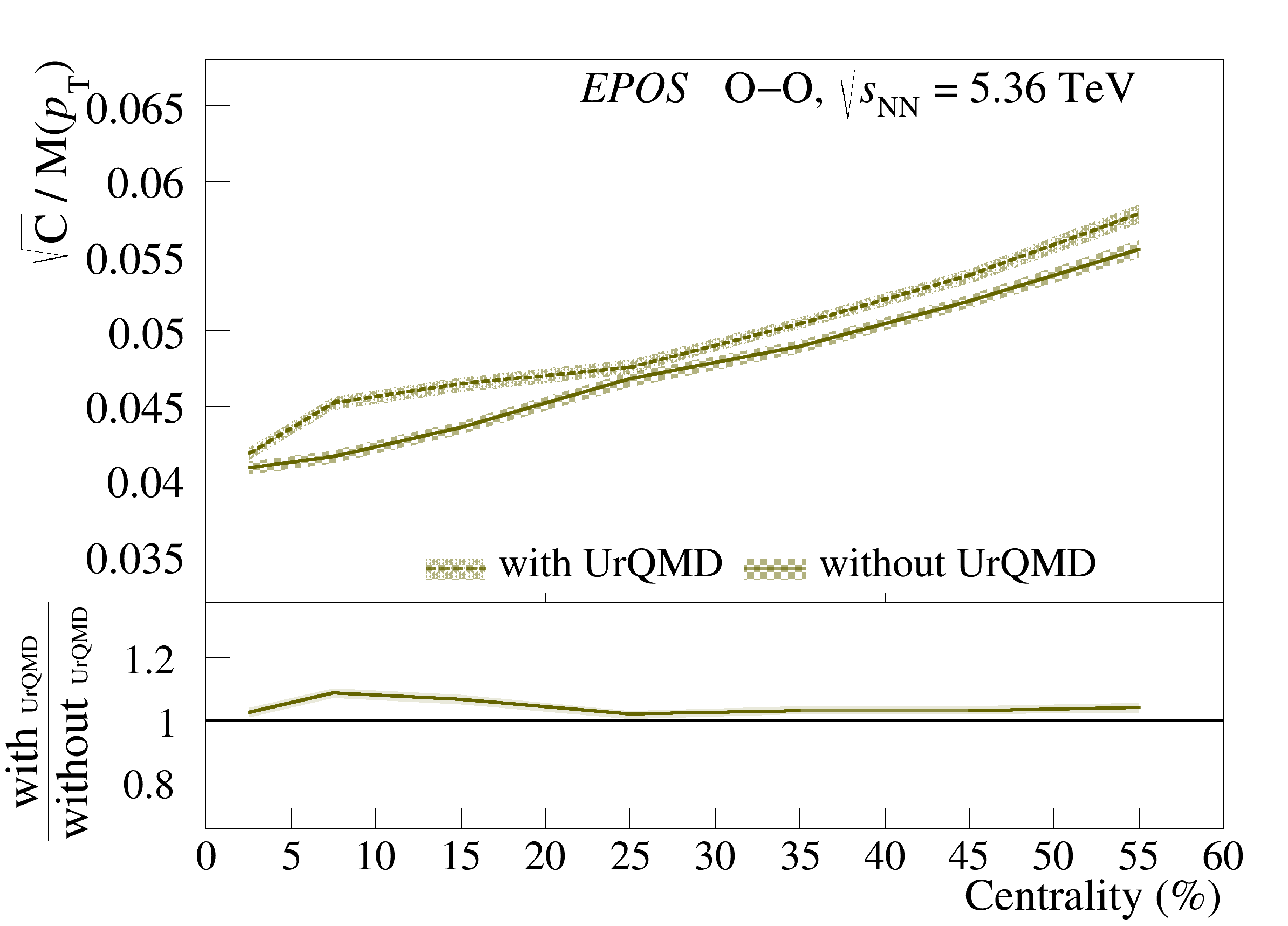}
  \end{minipage}

  \caption{Normalized transverse momentum correlator, (\ptcorr), as a function of centrality in EPOS4 for Pb--Pb collisions at \snn = 5.02 TeV (left), Xe--Xe collisions at \snn = 5.44 TeV (middle), and O--O collisions at \snn = 5.36 TeV (right). Results are shown with (dashed lines) and without (solid lines) the UrQMD hadronic afterburner. For Pb--Pb and Xe--Xe, the EPOS4 predictions are compared to ALICE data points (markers with error bars)~\cite{ALICE:2024apz}.}
  \label{fig:dptoverpt}
\end{figure*}

\begin{figure*}[htbp!]
  \centering
% --------- Pb-Pb ---------
  \begin{minipage}{0.325\textwidth}
    \centering
    \includegraphics[width=\linewidth]{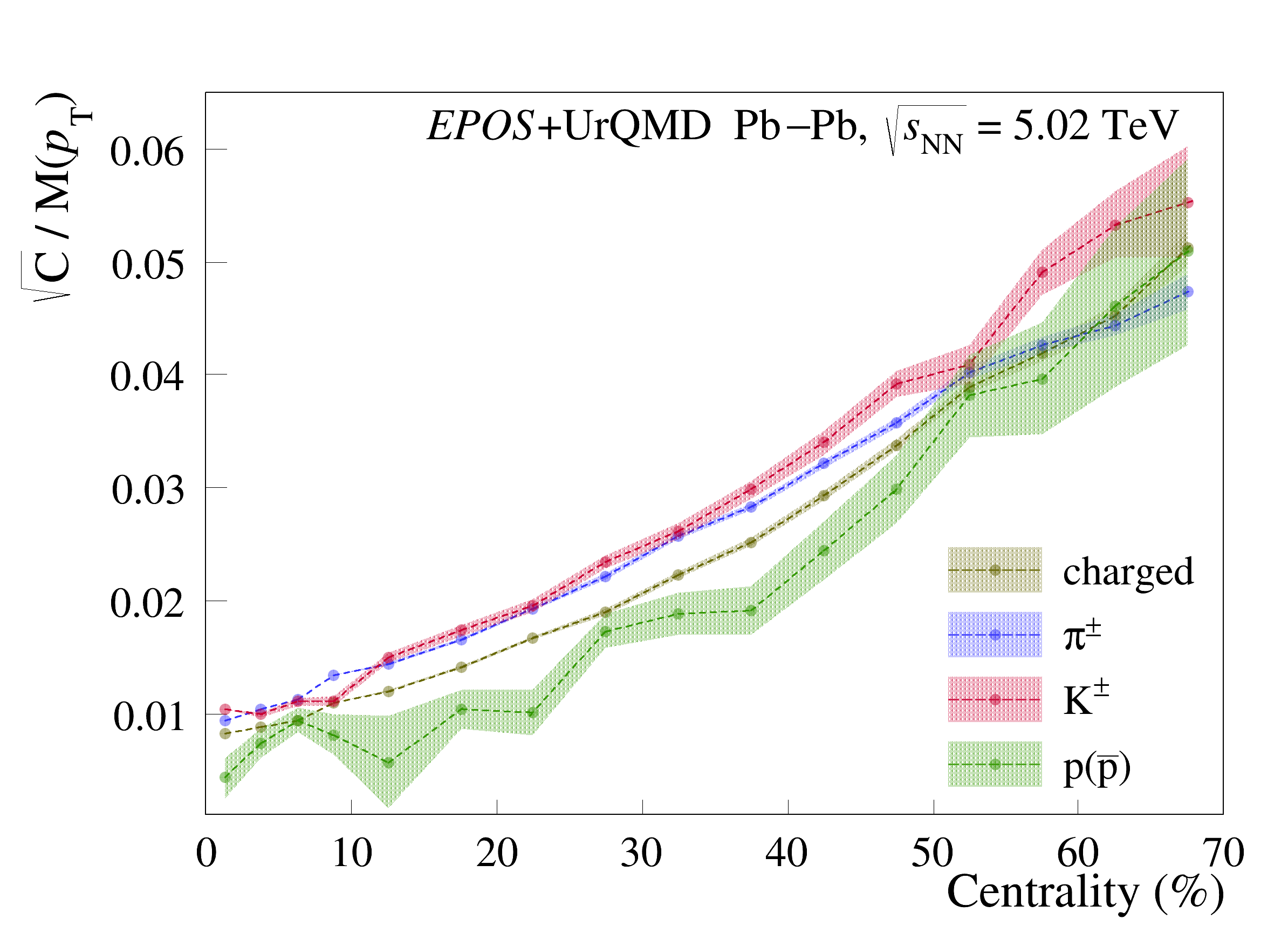}
  \end{minipage}
  \hfill
% --------- Xe-Xe ---------
  \begin{minipage}{0.325\textwidth}
    \centering
    \includegraphics[width=\linewidth]{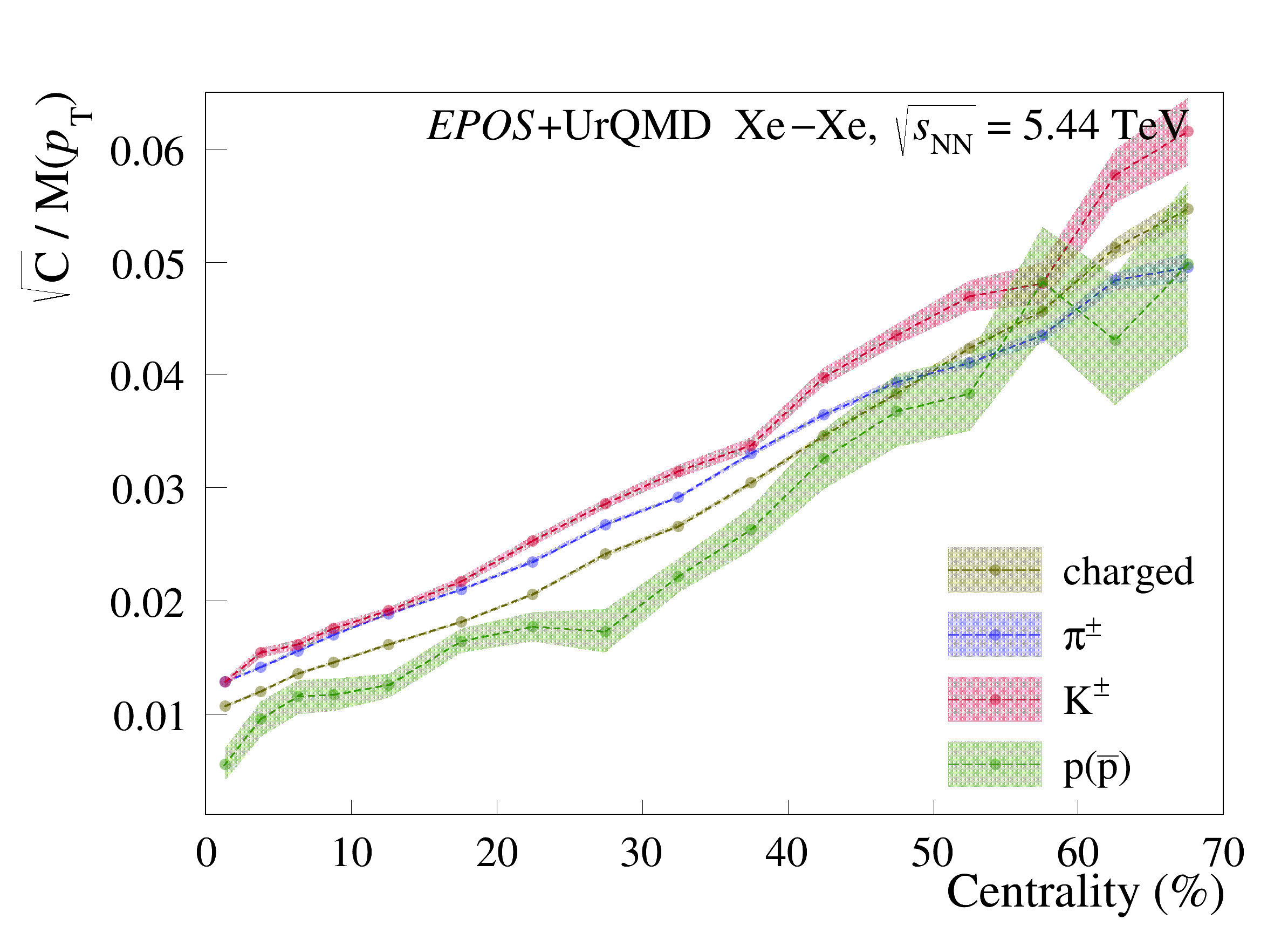}
  \end{minipage}
  \hfill
% --------- O-O ---------
  \begin{minipage}{0.325\textwidth}
    \centering
    \includegraphics[width=\linewidth]{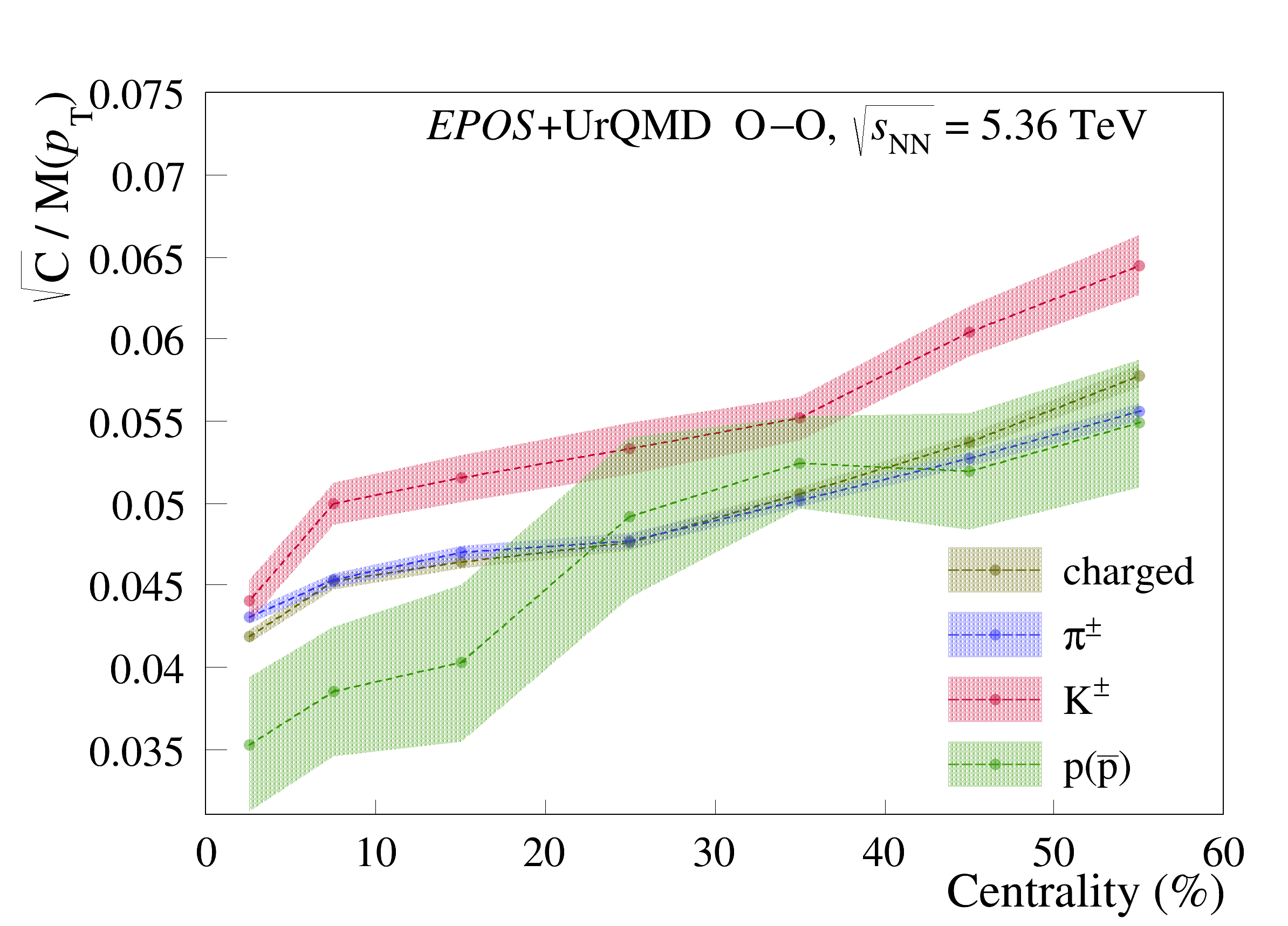}
  \end{minipage}

  \caption{Normalized transverse momentum correlator for charged particles, pions, kaons, and protons as a function of centrality. The results are shown for for Pb--Pb collisions at \snn = 5.02 TeV (left), Xe--Xe collisions at \snn = 5.44 TeV (middle), and O--O collisions at \snn = 5.36 TeV (right).}
  \label{fig:dptoverptall}
\end{figure*}

\begin{figure*}[htbp!]
  \centering

  % --------- with UrQMD ---------
  \begin{minipage}{0.325\textwidth}
    \centering
    \includegraphics[width=\linewidth]{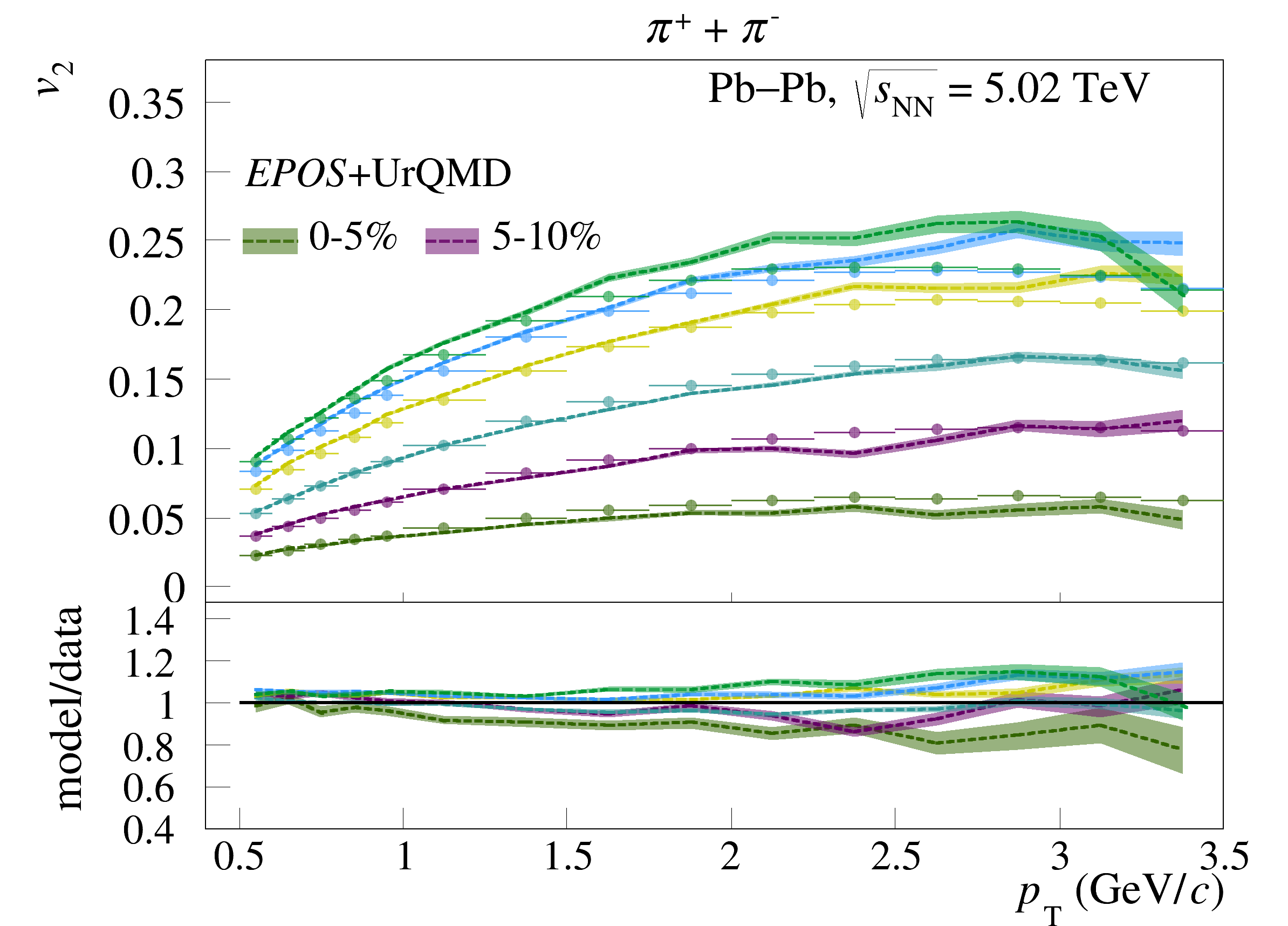}
    % \subcaption{Pb--Pb $\pi$}   % optional, if using subcaption
  \end{minipage}
  \hfill
  \begin{minipage}{0.325\textwidth}
    \centering
    \includegraphics[width=\linewidth]{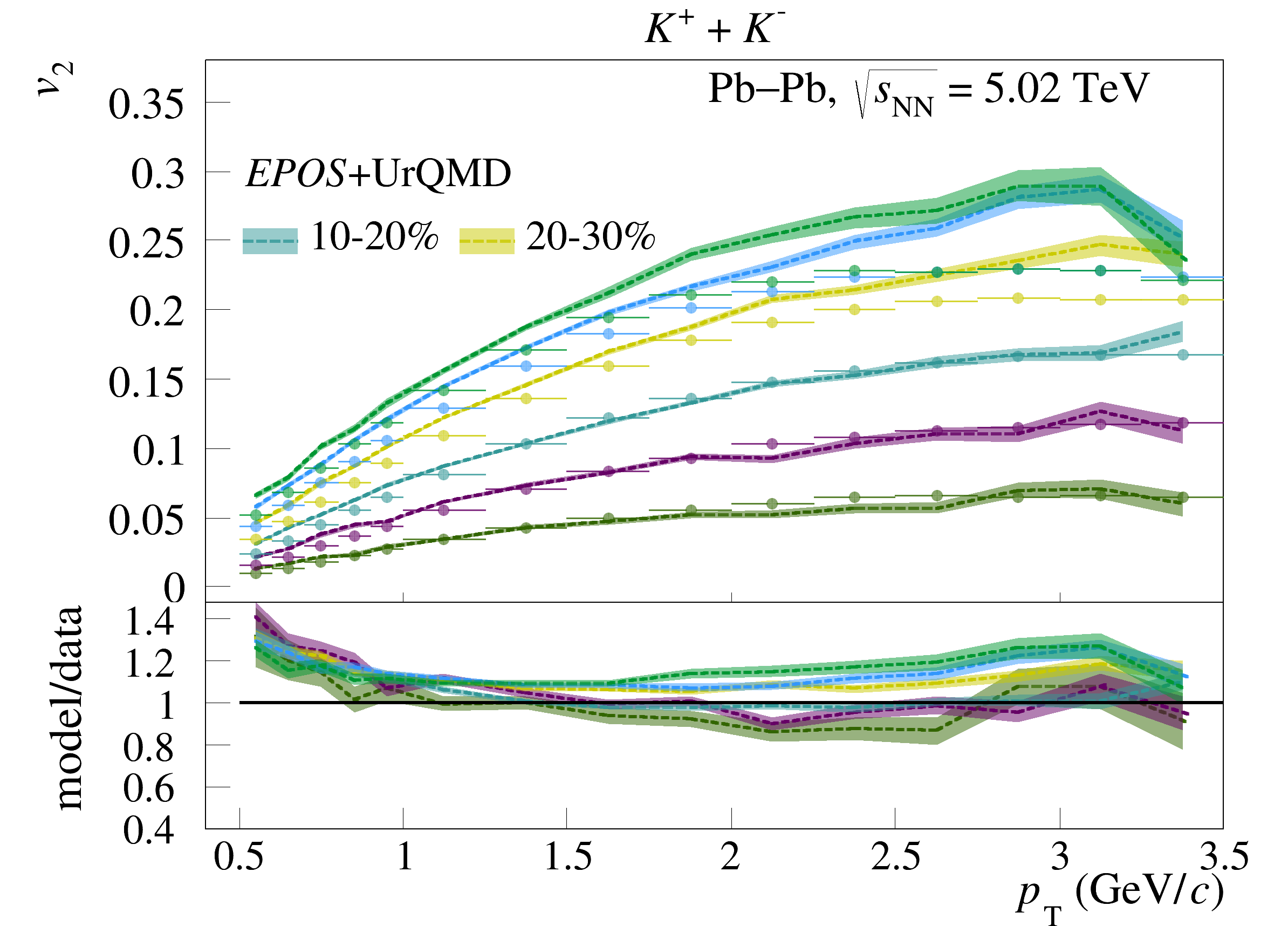}
    % \subcaption{Pb--Pb K}
  \end{minipage}
  \hfill
  \begin{minipage}{0.325\textwidth}
    \centering
    \includegraphics[width=\linewidth]{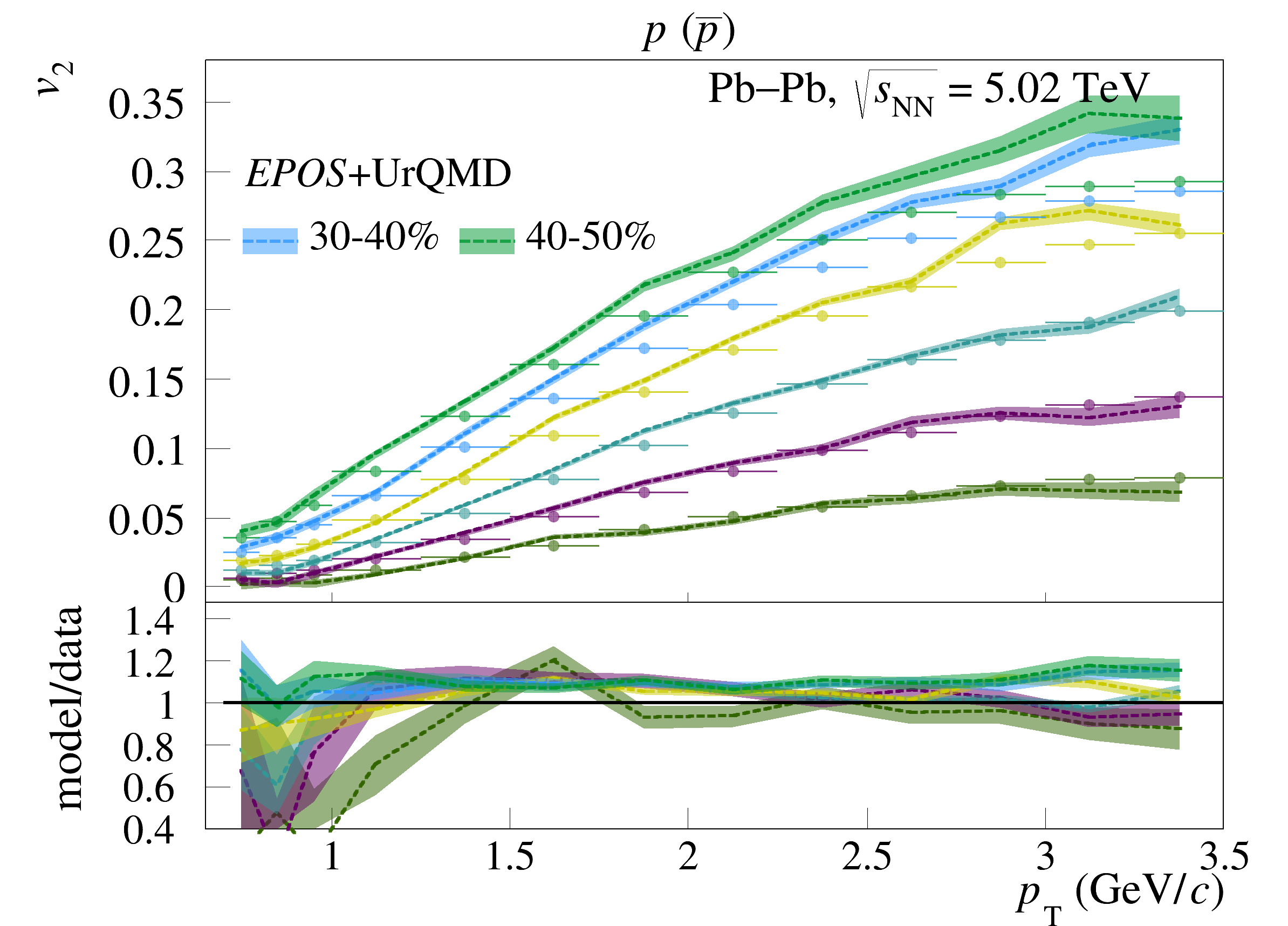}
    % \subcaption{Pb--Pb p}
  \end{minipage}

  %\vspace{-1.5em} % vertical gap between rows

  % --------- without UrQMD ---------
  \begin{minipage}{0.325\textwidth}
    \centering
    \includegraphics[width=\linewidth]{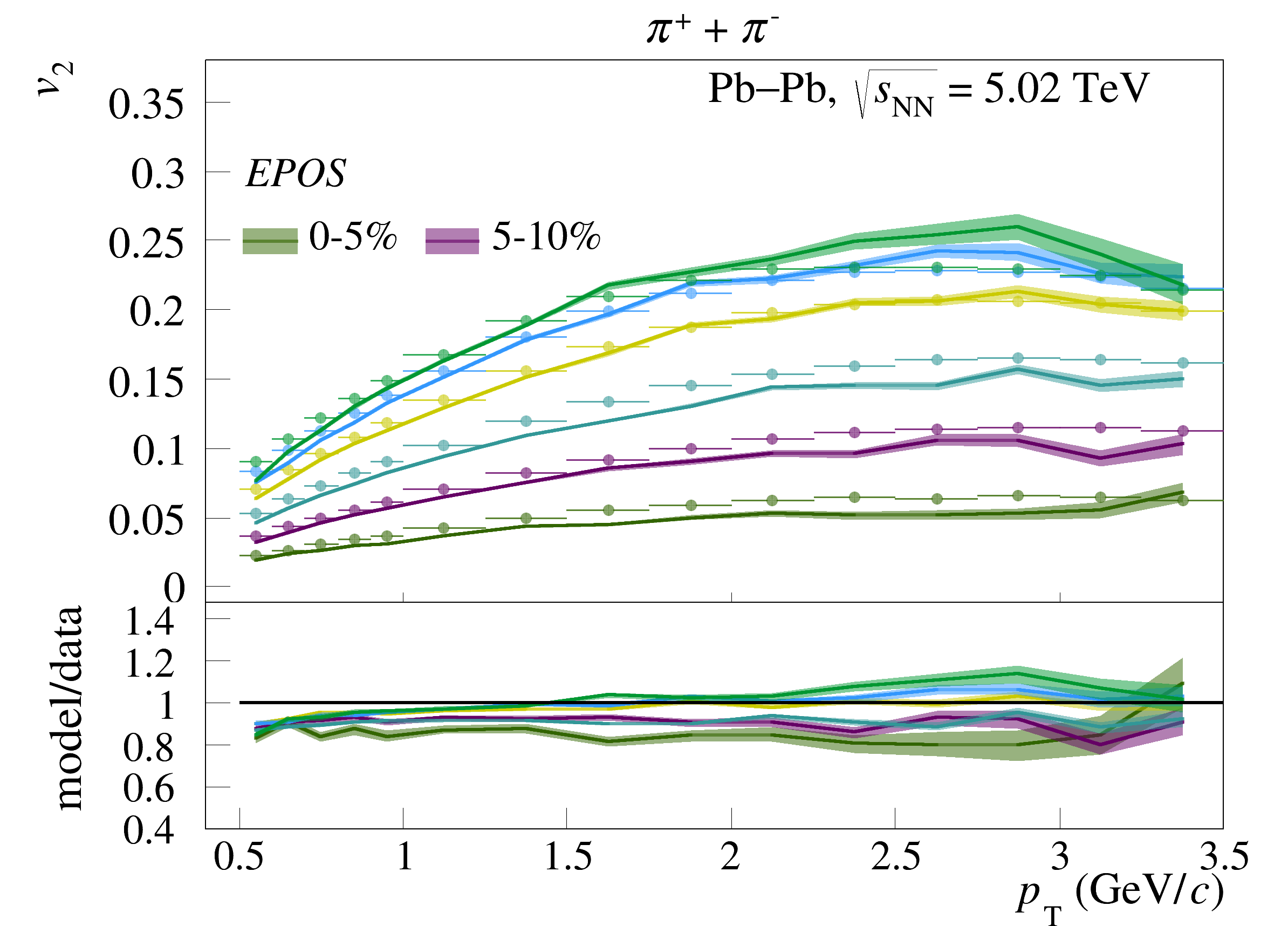}
    % \subcaption{Pb--Pb $\pi$}
  \end{minipage}
  \hfill
  \begin{minipage}{0.325\textwidth}
    \centering
    \includegraphics[width=\linewidth]{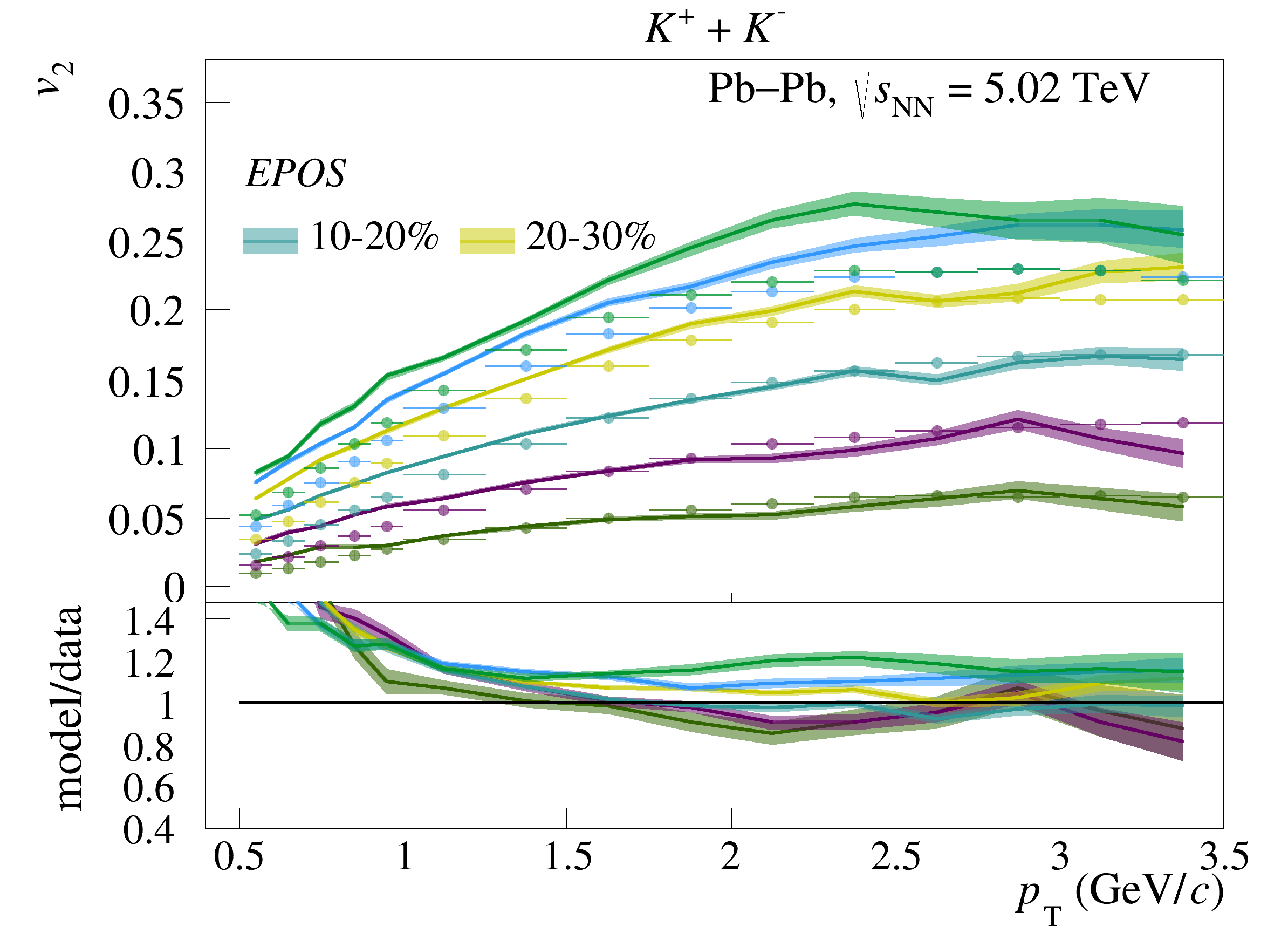}
    % \subcaption{Pb--Pb K}
  \end{minipage}
  \hfill
  \begin{minipage}{0.325\textwidth}
    \centering
    \includegraphics[width=\linewidth]{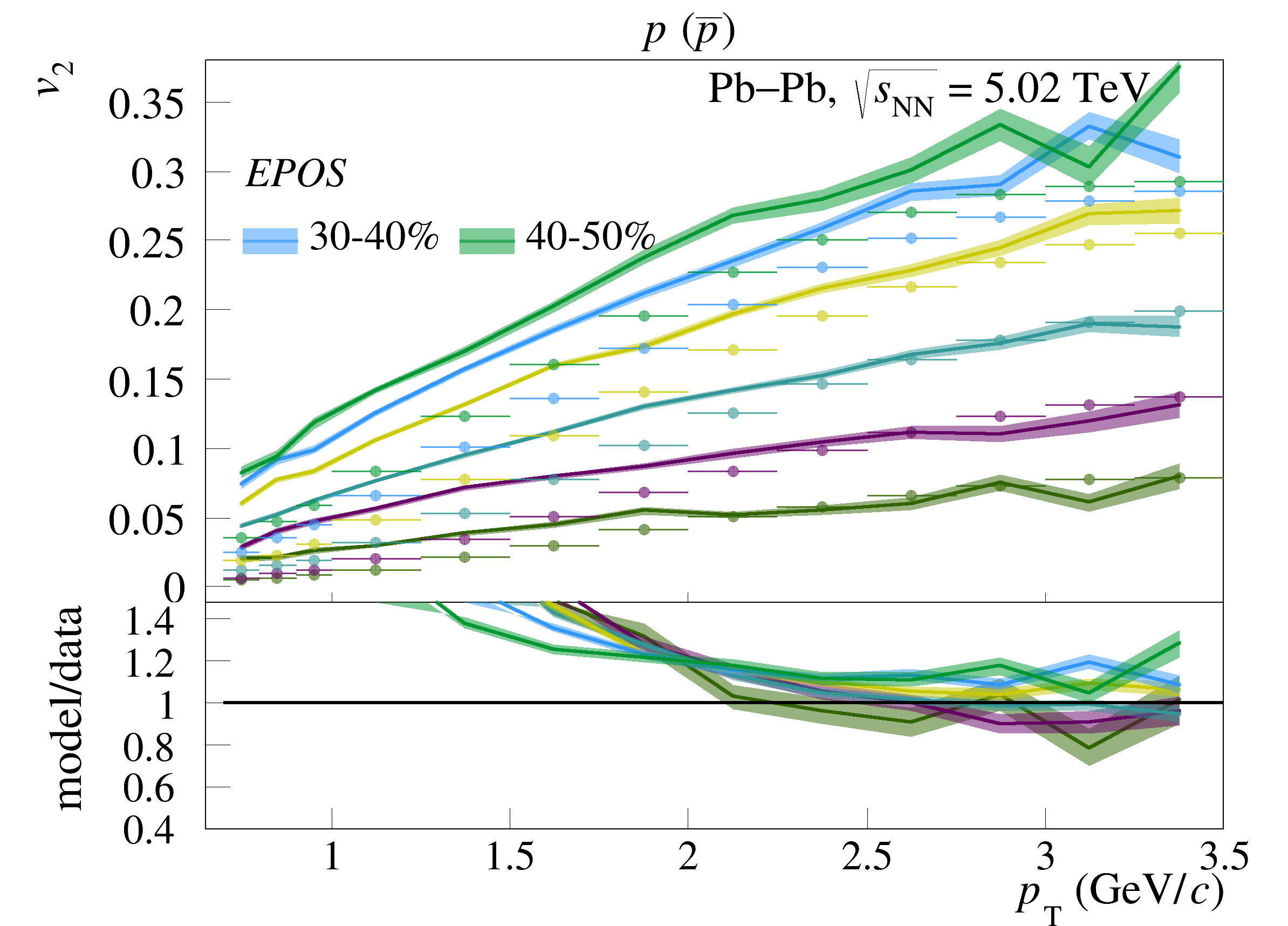}
    % \subcaption{Pb--Pb p}
  \end{minipage}

  \caption{
    Identified \tpt-differential anisotropic flow coefficient \vn{2}\{2\} (\tpt) for pions (left), kaons (middle) and protons (right) in various centrality classes in Pb--Pb collisions at \snn = 5.02 TeV compared with colored ALICE data points~\cite{ALICE:2018yph} (solid markers). The results in the upper row correspond to EPOS4 calculations including the UrQMD hadronic afterburner (dashed lines), while those in the lower row are obtained without hadronic rescattering (solid lines). The colored bands indicate the statistical uncertainties.
  }
  \label{fig:v2vspt_pbpb}
\end{figure*}

\begin{figure*}[htbp!]
  \centering

  % --------- with UrQMD ---------
  \begin{minipage}{0.325\textwidth}
    \centering
    \includegraphics[width=\linewidth]{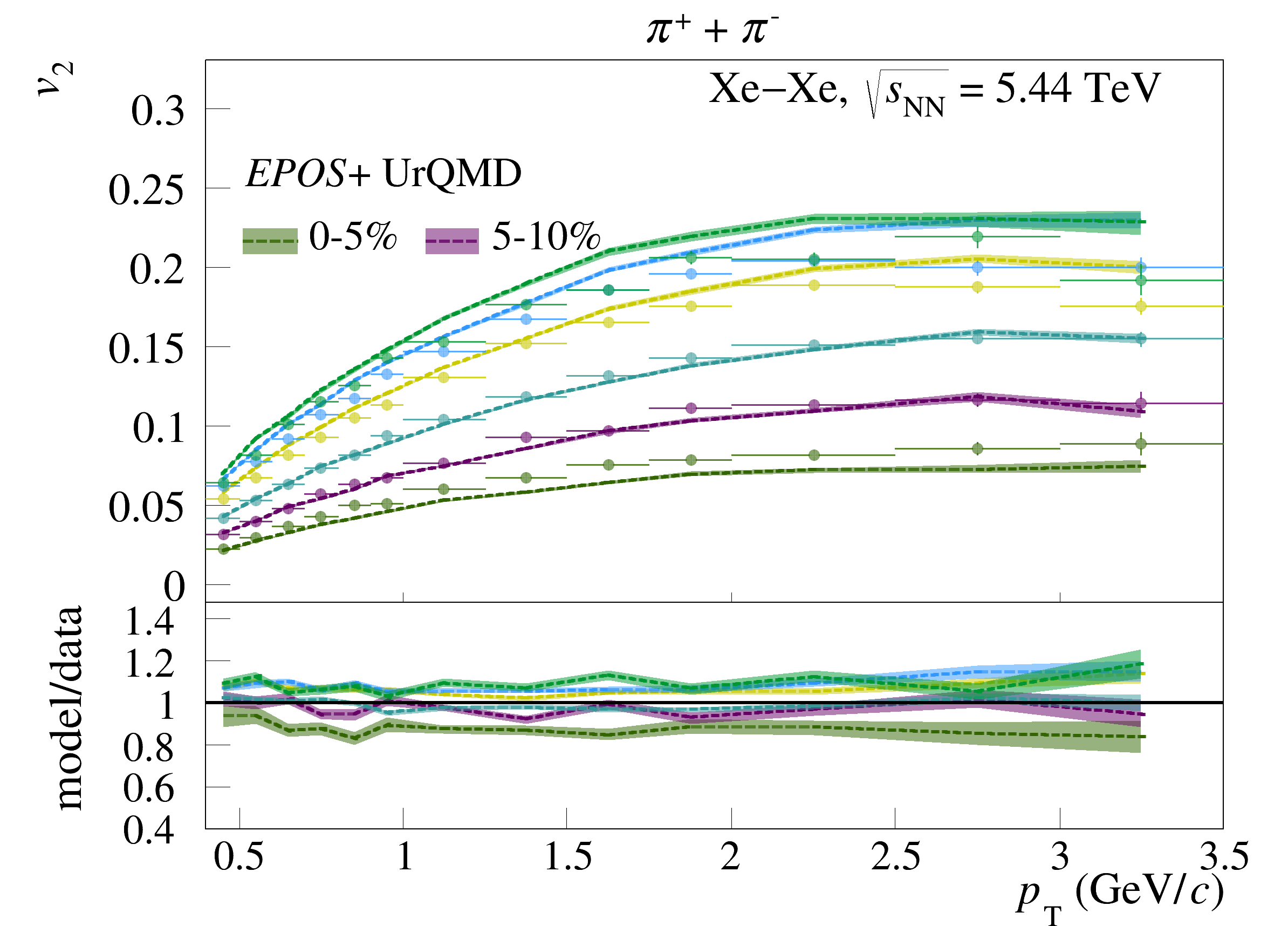}
    % \subcaption{Xe--Xe $\pi$}   % optional, if using subcaption
  \end{minipage}
  \hfill
  \begin{minipage}{0.325\textwidth}
    \centering
    \includegraphics[width=\linewidth]{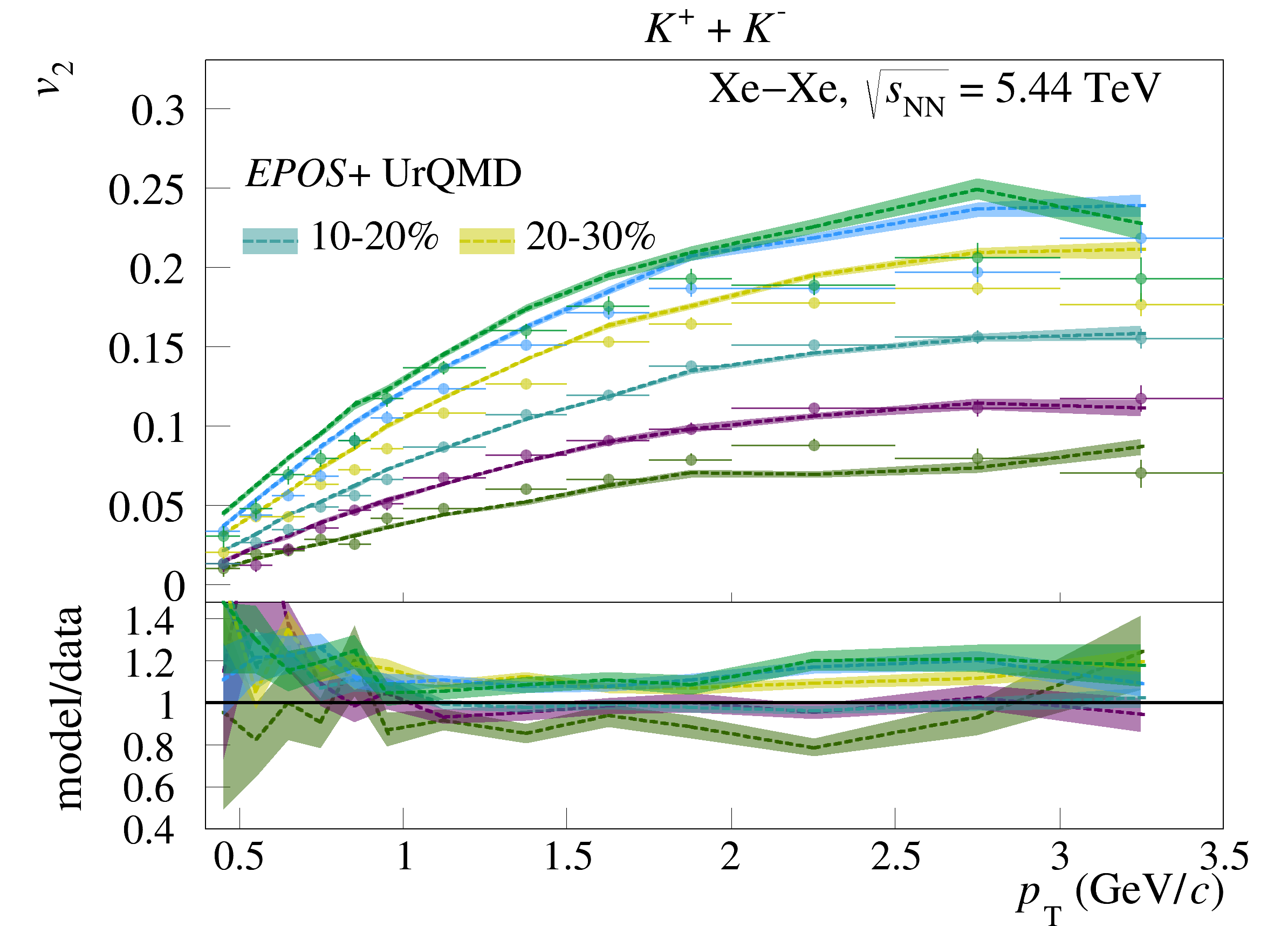}
    % \subcaption{Xe--Xe K}
  \end{minipage}
  \hfill
  \begin{minipage}{0.325\textwidth}
    \centering
    \includegraphics[width=\linewidth]{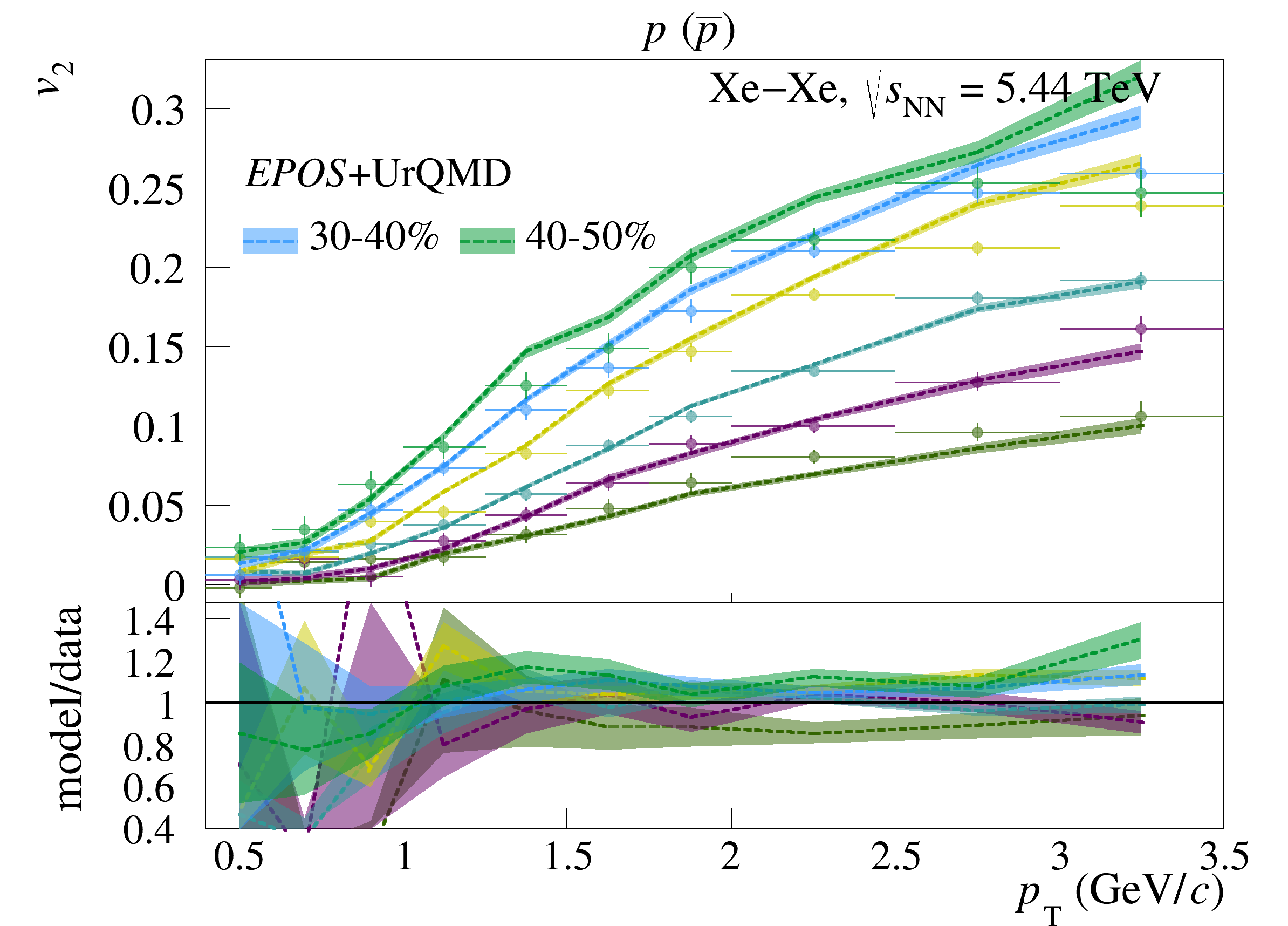}
    % \subcaption{Xe--Xe p}
  \end{minipage}

  %\vspace{-1.5em} % vertical gap between rows

  % --------- without UrQMD ---------
  \begin{minipage}{0.32\textwidth}
    \centering
    \includegraphics[width=\linewidth]{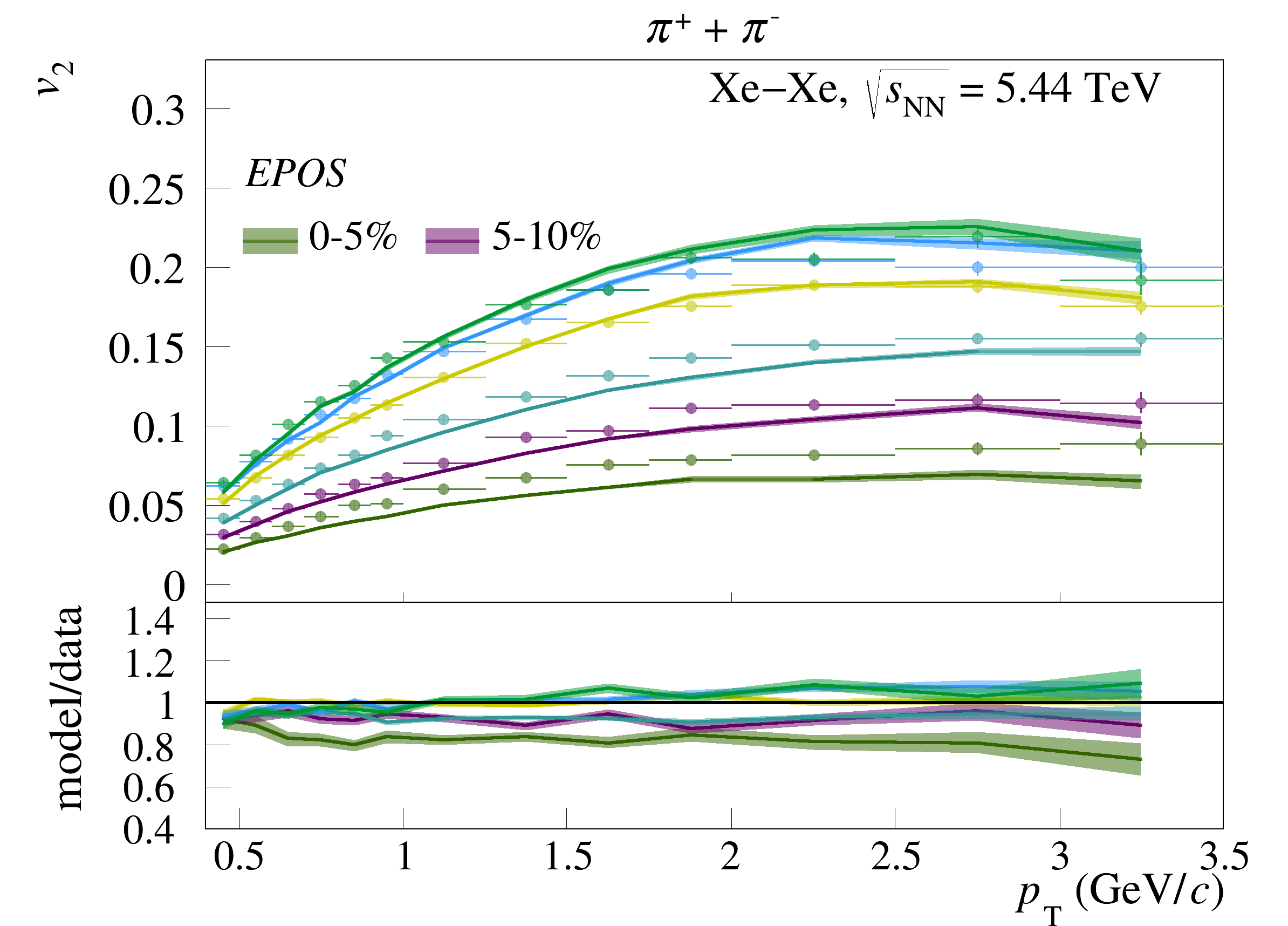}
    % \subcaption{Xe--Xe $\pi$}
  \end{minipage}
  \hfill
  \begin{minipage}{0.32\textwidth}
    \centering
    \includegraphics[width=\linewidth]{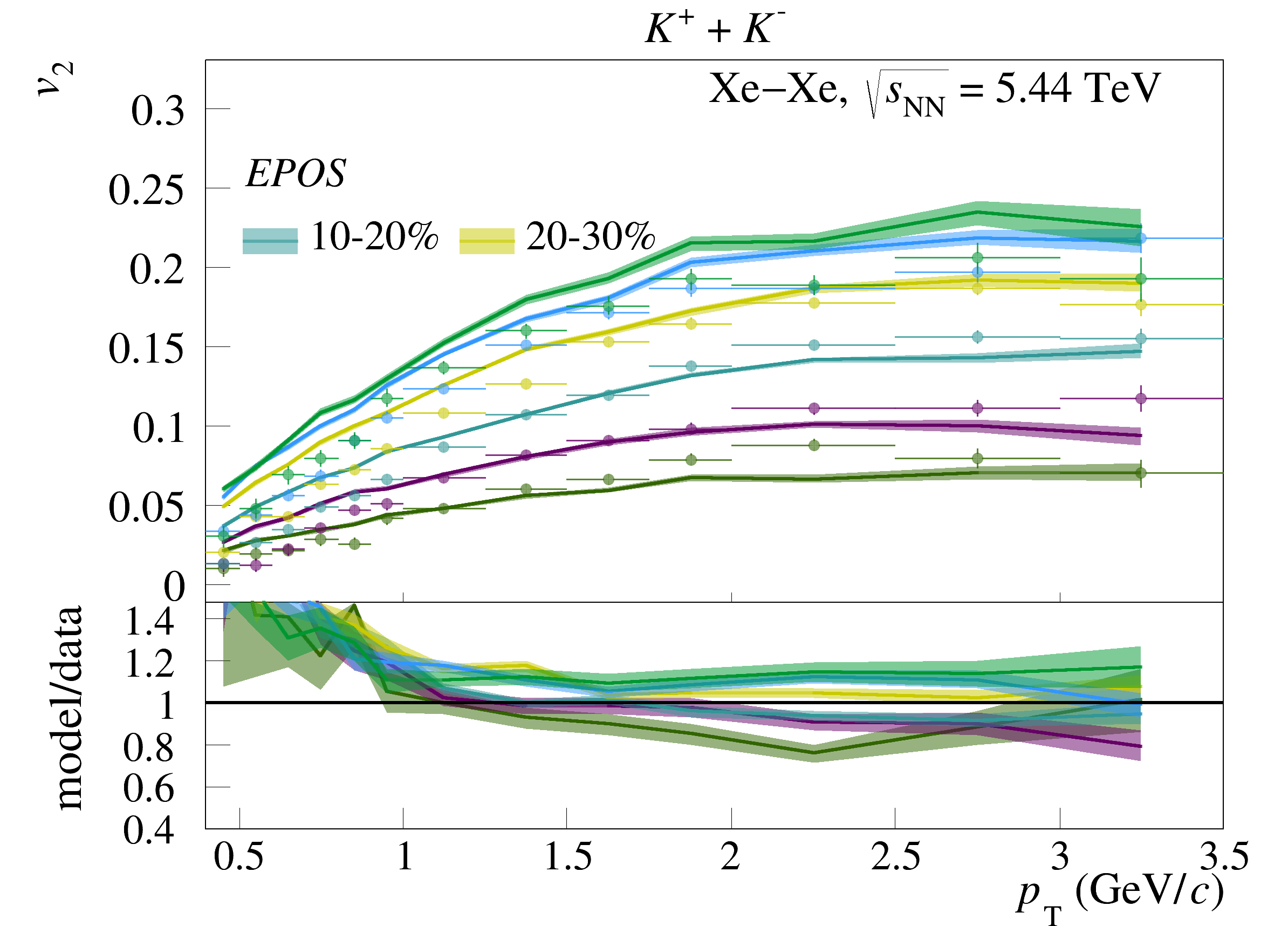}
    % \subcaption{Xe--Xe K}
  \end{minipage}
  \hfill
  \begin{minipage}{0.32\textwidth}
    \centering
    \includegraphics[width=\linewidth]{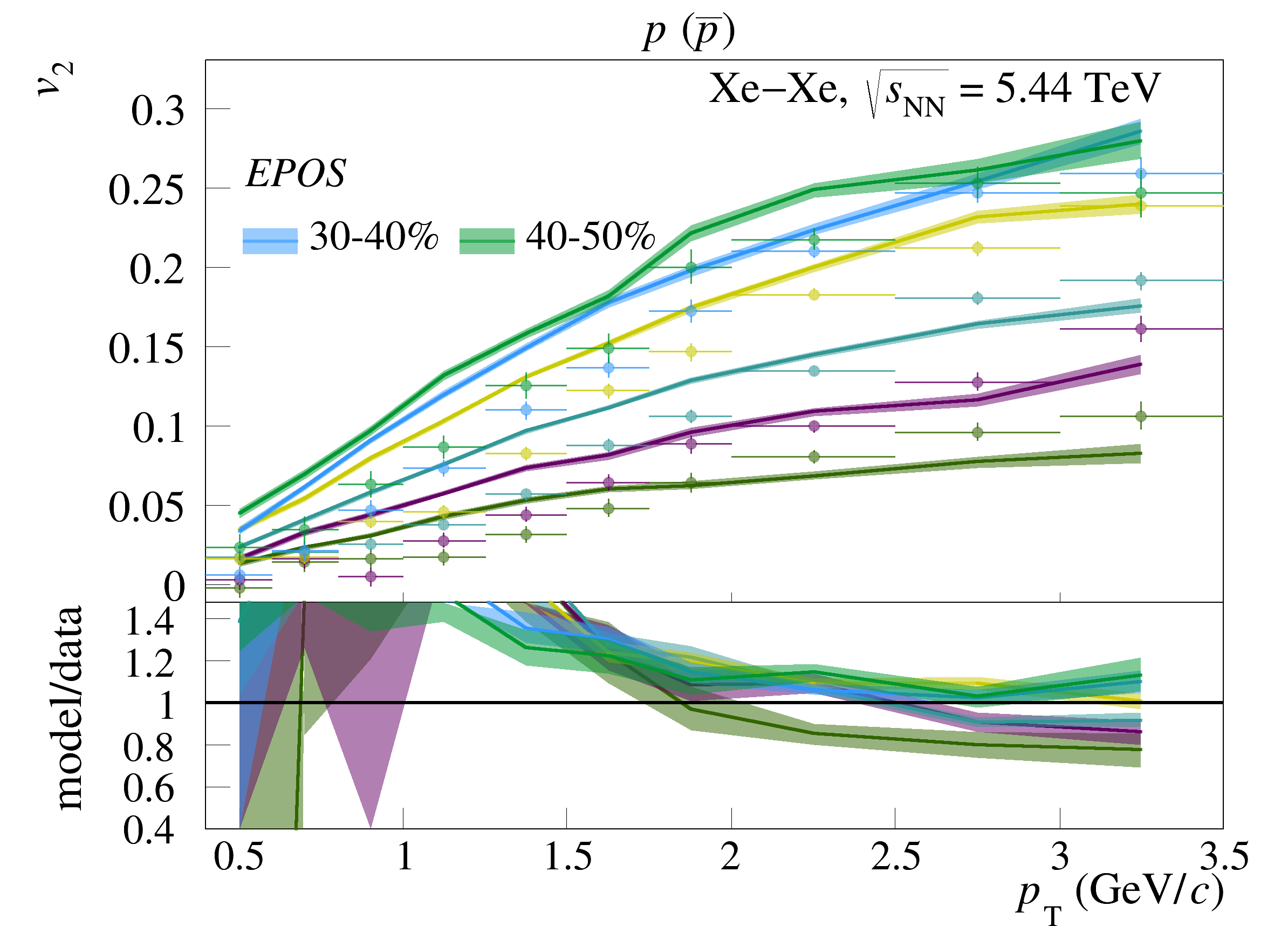}
    % \subcaption{Xe--Xe p}
  \end{minipage}

  \caption{
    Identified \tpt-differential anisotropic flow coefficient \vn{2}\{2\} (\tpt) for pions (left), kaons (middle), and protons (right) in various centrality classes in Xe--Xe collisions at \snn = 5.44 TeV, compared with ALICE data points~\cite{ALICE:2021ibz} (solid markers). The upper row shows EPOS4 calculations including the UrQMD hadronic afterburner (dashed lines), while the lower row corresponds to results without hadronic rescattering (solid lines). The colored bands indicate the statistical uncertainties.
    }
  \label{fig:v2vspt_xexe}
\end{figure*}

\begin{figure*}[htbp!]
  \centering

  \begin{minipage}{0.99\textwidth}
    \centering
    \includegraphics[width=\linewidth]{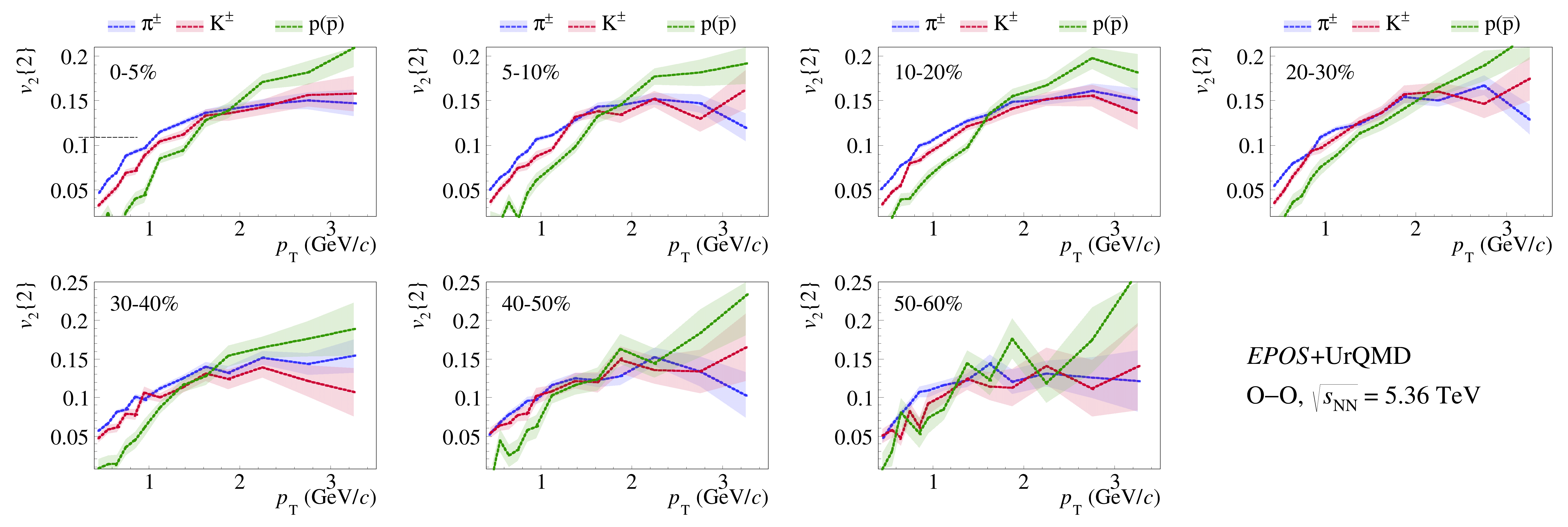}
  \end{minipage}

  \caption{Identified \tpt-differential anisotropic flow coefficient \vn{2}\{2\} (\tpt) for pions, kaons, and protons in various centrality classes in O--O collisions at \snn = 5.36 TeV. The results show EPOS4 predictions including UrQMD (dashed lines) with the colored bands indicating statistical uncertainties.}
  \label{fig:v2vspt_oo}
\end{figure*}

%%%%%%%%%%%%%%%%%%%%%%%%%%%%%%%%%%%%%%%%%%%%%%%%%%%%%%%%%%%%%%%%%%%%%%%%%%%%%%%%%%%%%%%%
%%%%%%%%%%%%%%%%%%%%%%%%%%%%%%%%%%%%%%%%%%%%%%%%%%%%%%%%%%%%%%%%%%%%%%%%%%%%%%%%%%%%%%%%
%%%%%%%%%%%%%%%%%%%%%%%%%%%%%%%%%%%%%%%%%%%%%%%%%%%%%%%%%%%%%%%%%%%%%%%%%%%%%%%%%%%%%%%%
%%%%%%%%%%%%%%%%%%%%%%%%%%  CH PART PT CORRELATOR %%%%%%%%%%%%%%%%%%%%%%%%%%%%%%%%%%%%%%
%%%%%%%%%%%%%%%%%%%%%%%%%%%%%%%%%%%%%%%%%%%%%%%%%%%%%%%%%%%%%%%%%%%%%%%%%%%%%%%%%%%%%%%%
%%%%%%%%%%%%%%%%%%%%%%%%%%%%%%%%%%%%%%%%%%%%%%%%%%%%%%%%%%%%%%%%%%%%%%%%%%%%%%%%%%%%%%%%
%%%%%%%%%%%%%%%%%%%%%%%%%%%%%%%%%%%%%%%%%%%%%%%%%%%%%%%%%%%%%%%%%%%%%%%%%%%%%%%%%%%%%%%%
\subsection{Transverse Momentum, \tpt ~fluctuations}

Event-by-event fluctuations of the mean transverse momentum $\langle \tpt \rangle$, provide a sensitive probe of the dynamics and correlations within the created system. These fluctuations are quantified using the normalized transverse momentum correlator which measures the non-statistical deviation of particle momenta relative to the event ensemble average given as~\cite{ALICE:2024apz}:
\begin{equation}
    \ptcorr = \sqrt{C_2}\,\big/\,\langle p_{\mathrm{T}}\rangle \,
\end{equation}
where $\Delta p_{\mathrm{T},i} = p_{\mathrm{T},i} - \langle p_{\mathrm{T}}\rangle$ is the deviation of the $i$-th particle's transverse momentum from the event-ensemble average, and $C_2 = \langle p_{\mathrm{T},i}\,p_{\mathrm{T},j}\rangle_{i\neq j} - \langle p_{\mathrm{T}}\rangle^{2}$ is the two-particle covariance. The averages $\langle\cdot\rangle$ are taken first over distinct particle pairs within an event and then over all events in a given centrality class, while the double brackets $\langle\langle\cdot\rangle\rangle$ denote the combined event-and-particle average. By construction, the correlator vanishes for uncorrelated particle emission and becomes positive when event-by-event $\langle p_{\mathrm{T}}\rangle$ fluctuations exceed statistical expectations~\cite{ALICE:2024apz}. In EPOS4, such fluctuations arise from the interplay between initial-state variations, particularly event-by-event fluctuations of the saturation scale, and the subsequent hydrodynamic response.

Figure~\ref{fig:dptoverpt} presents the centrality dependence of the normalized \tpt~correlator at midrapidity ($|\eta| < 0.8$) within 0.15 $< \tpt <$ 2.0 \gev~for Pb--Pb (left), Xe--Xe (middle) and O--O (right) collision systems. For the large systems, Pb--Pb and Xe--Xe, the EPOS4 calculations including the UrQMD afterburner (dashed lines) show good agreement with the experimental data from ALICE~\cite{ALICE:2024apz}. The correlator decreases when going from peripheral to central collisions, consistent with the dilution of fluctuations as the number of particle sources increases. An underestimation of 20\% at more central and a similar overestimation towards peripheral region. %However, the dependence deviates from a simple $N_{\text{ch}}^{-1/2}$ power-law scaling, a feature that in hydrodynamic models is attributed to the onset of transverse radial flow which correlates the momenta of particles emitted from the same fluid elements.%
Result for O--O system follow a similar qualitative trend, with the magnitude of the fluctuations being comparable to those in peripheral Pb--Pb and Xe--Xe collisions at similar multiplicities. One can see that the centrality dependence is weaker in O--O than in Pb--Pb. This scaling behaviour suggests that the mechanisms driving \tpt~fluctuations primarily the fluctuations in the core-corona fraction and the saturation scale operate continuously from small to large systems. The saturation scale $Q_{\text{sat}}^2$ depends on the local connection number \substy{N}{conn}~\cite{Werner:2023mod}. Event-by-event fluctuations in the spatial configuration of nucleons lead to variations in \substy{N}{conn}, which in turn modulate the hardness of the initial parton production. These initial momentum correlations are subsequently processed by the collective expansion of the core.

The impact of the hadronic afterburner is visible in the comparison between calculations with (dashed lines) and without (solid lines) UrQMD. In all three systems, the inclusion of hadronic rescattering leads to a modest modification of the correlator strength. This stability indicates that the primary source of the \tpt~correlations is established during the initial state and the hydrodynamic phase, rather than being generated dynamically during the late hadronic stage. However, the fine-tuning provided by the afterburner, likely through resonance decays and elastic scattering, is beneficial for achieving the precise description of the experimental data observed in the Pb–Pb and Xe–Xe systems. Figure~\ref{fig:dptoverptall} decomposes the correlator into contributions from identified charged particles, pions, kaons, and protons. A clear species dependence is observed, particularly in the O–O and peripheral Pb–Pb/Xe–Xe collisions. This mass splitting is a further signature of radial flow, as the common collective velocity field imparts stronger momentum correlations to heavier particles. The identified transverse momentum fluctuations do not seem to be ordered according to mass, but rather the kaon transverse momentum fluctuations are the largest. The consistent description of these fluctuation observables across system sizes reinforces the interpretation that collective flow, driven by initial-state density fluctuations, plays a dominant role even in light-ion collisions.

\begin{figure*}[htbp!]
  \centering

  \begin{minipage}{0.325\textwidth}
    \centering
    \includegraphics[width=\linewidth]{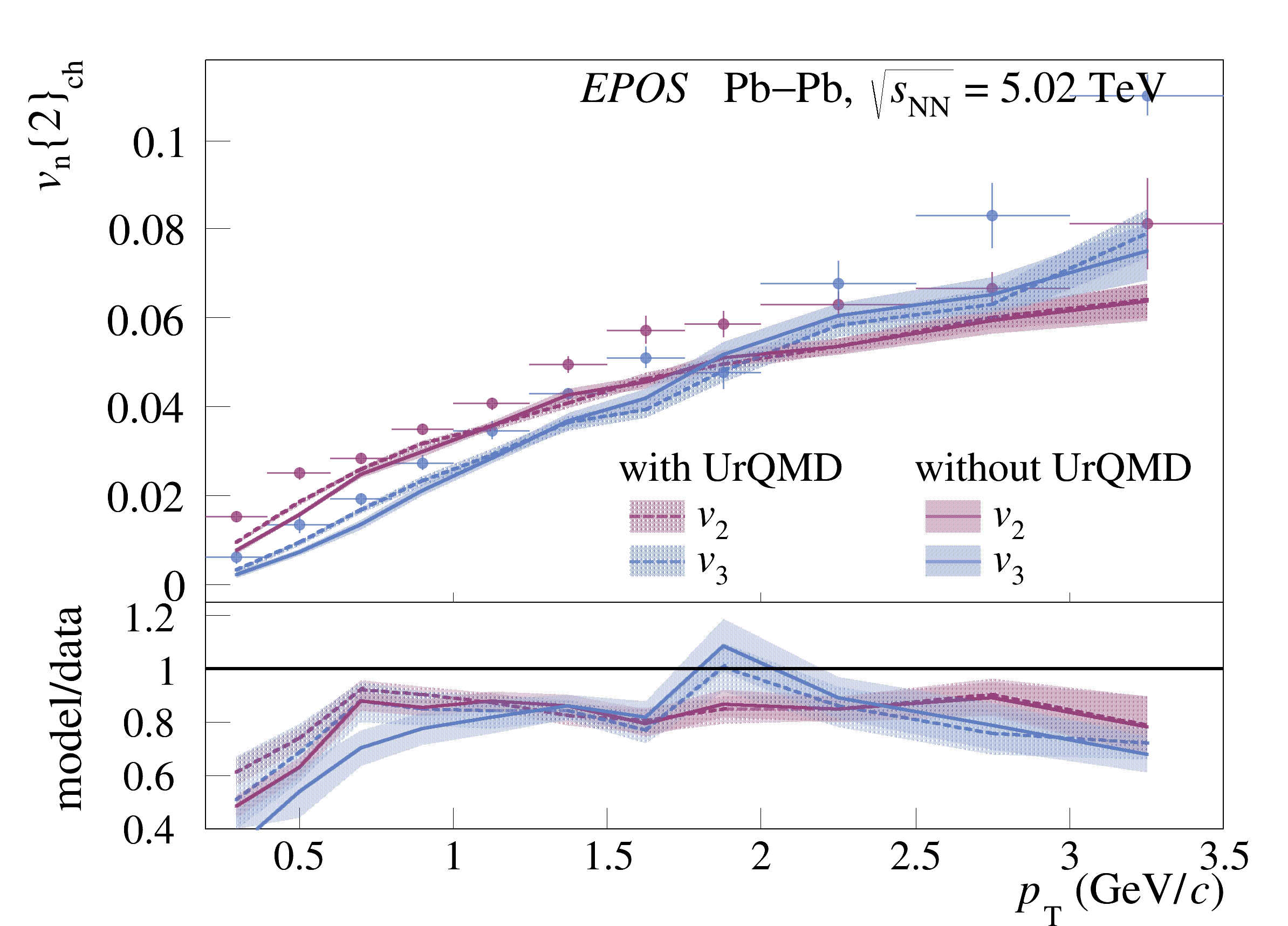}
  \end{minipage}
  \hfill
  \begin{minipage}{0.325\textwidth}
    \centering
    \includegraphics[width=\linewidth]{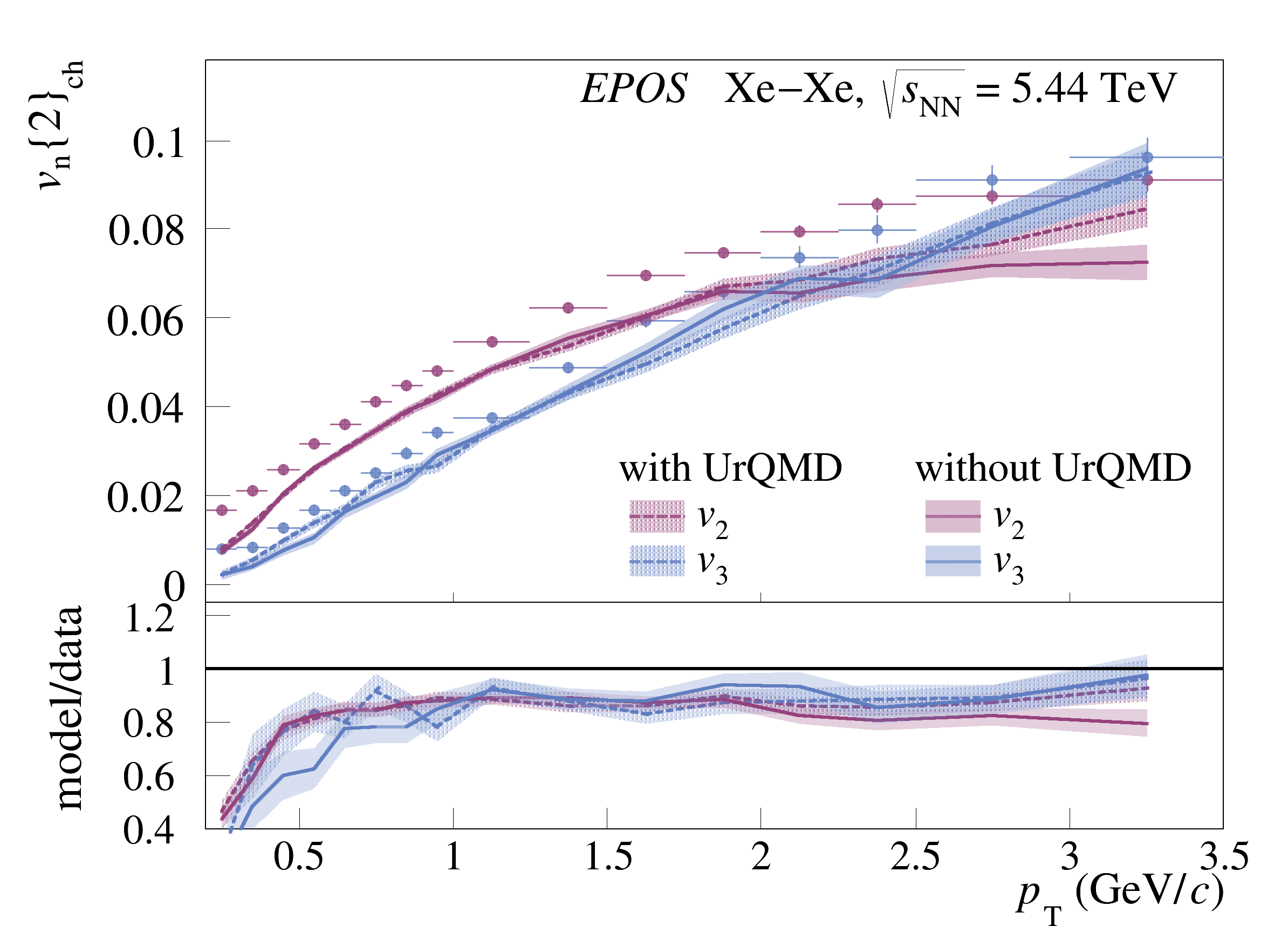}
  \end{minipage}
  \hfill
  \begin{minipage}{0.325\textwidth}
    \centering
    \includegraphics[width=\linewidth]{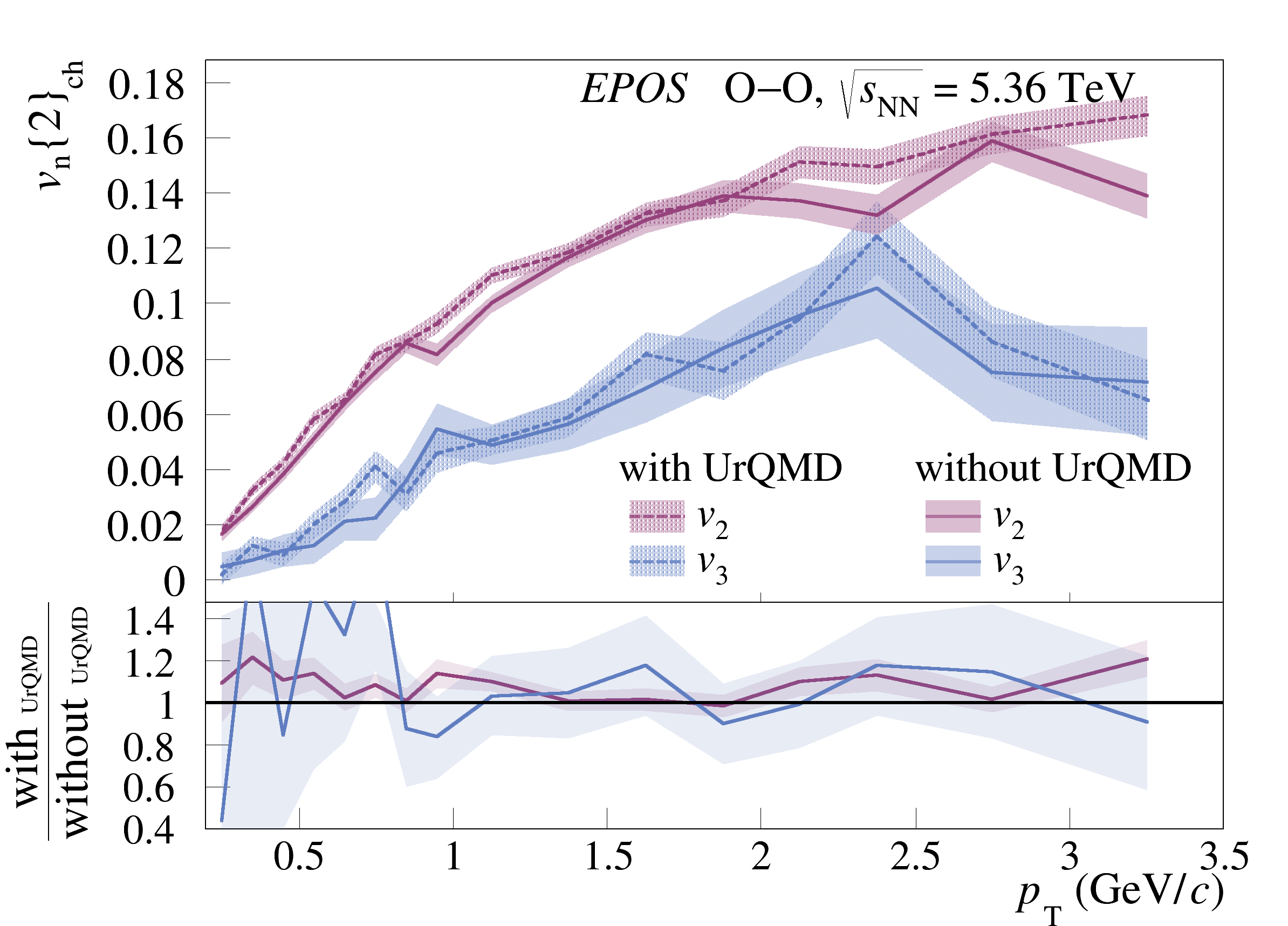}
  \end{minipage}

  \caption{Integrated anisotropic flow coefficients, $v_n \{2\}$ of charged particles as a function of transverse momentum, \tpt~ for Pb--Pb collisions at \snn = 5.02 TeV (left), Xe--Xe collisions at \snn = 5.44 TeV (middle) and predictions for O--O collisions at \snn = 5.36 TeV (right). EPOS4 calculations with (dashed lines) and without (solid lines) the UrQMD hadronic afterburner are compared to ALICE data points (solid markers) for Pb--Pb~\cite{ALICE:2016ccg} and Xe--Xe~\cite{ALICE:2018lao}.}
  \label{fig:vn_pt}
\end{figure*}

\begin{figure*}[htbp!]
  \centering

  \begin{minipage}{0.325\textwidth}
    \centering
    \includegraphics[width=\linewidth]{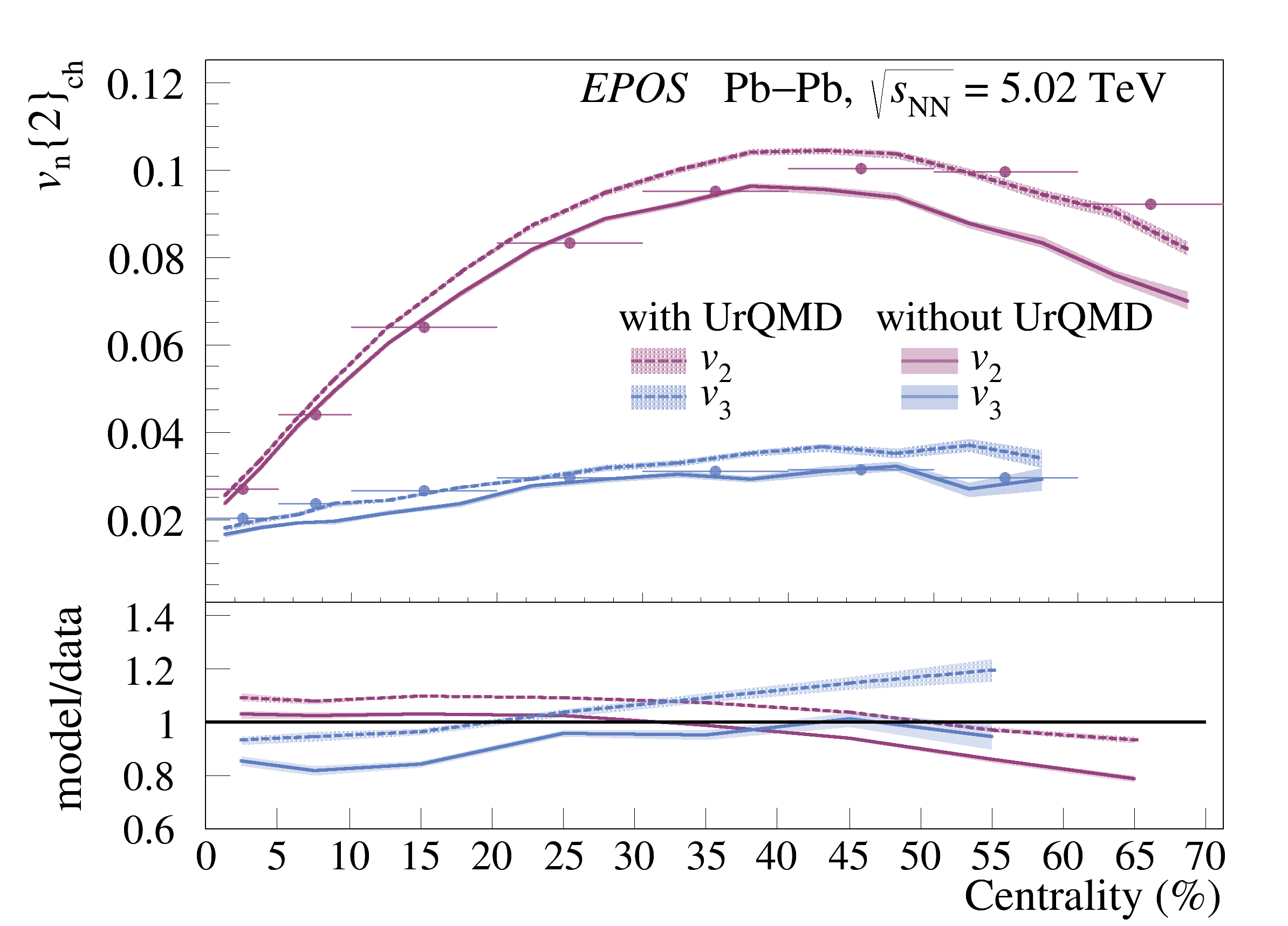}
  \end{minipage}
  \hfill
  \begin{minipage}{0.325\textwidth}
    \centering
    \includegraphics[width=\linewidth]{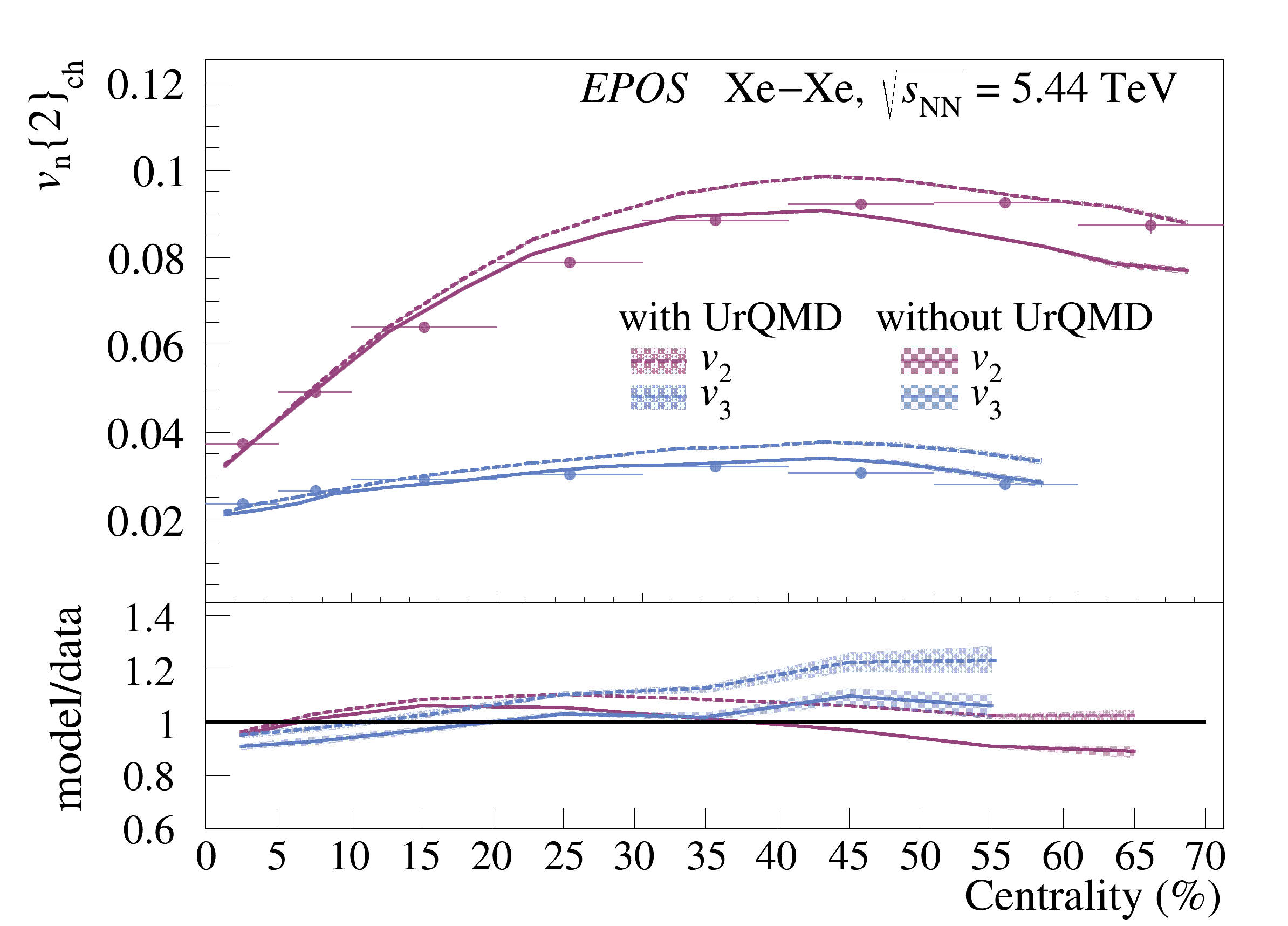}
  \end{minipage}
  \hfill
  \begin{minipage}{0.325\textwidth}
    \centering
    \includegraphics[width=\linewidth]{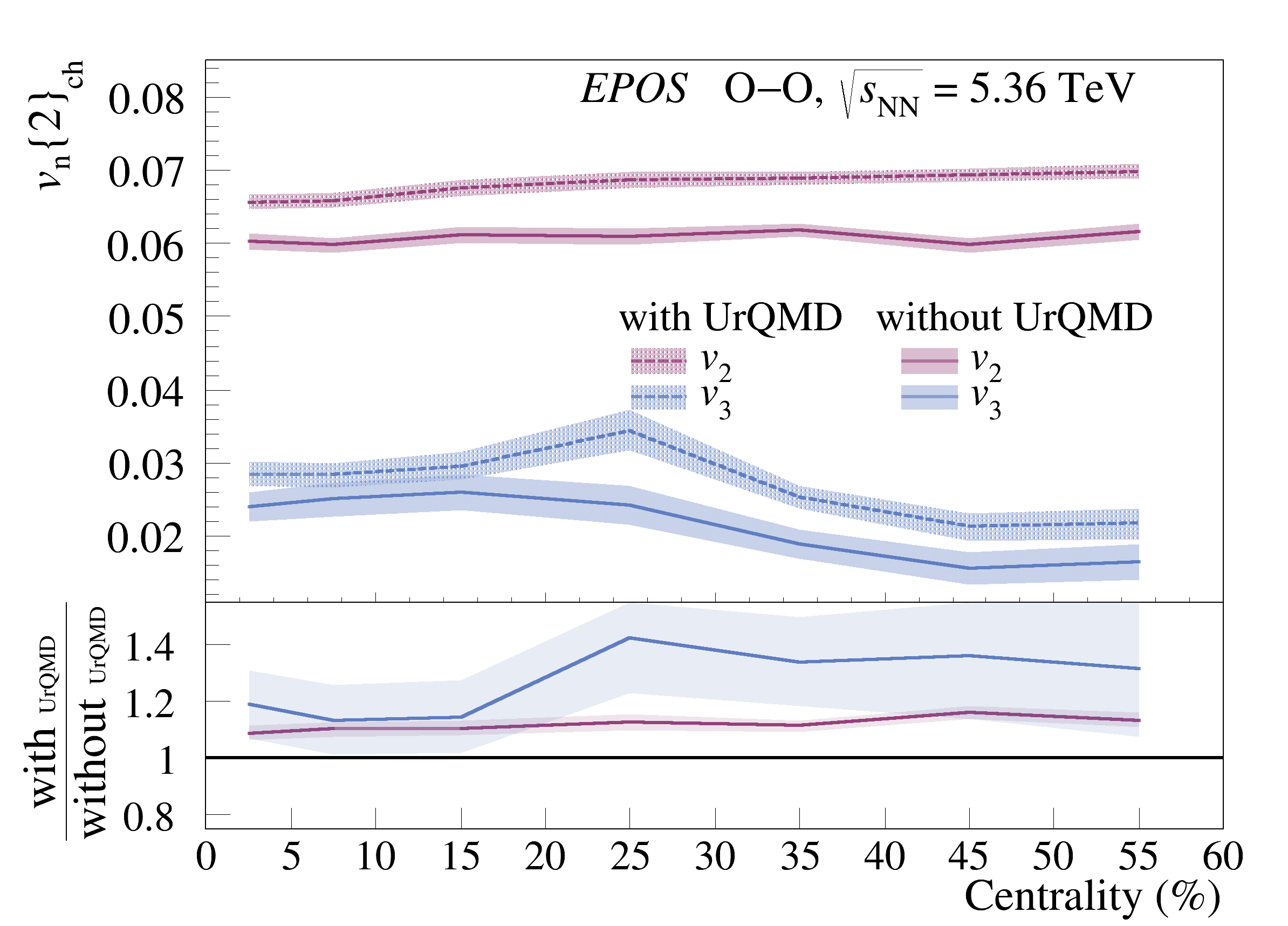}
  \end{minipage}

  \caption{Integrated anisotropic flow coefficients, $v_n \{2\}$ of charged particles as a function of centrality for Pb–Pb collisions at \snn = 5.02 TeV (left), Xe–Xe collisions at \snn = 5.44 TeV (middle), and predictions for O–O collisions at \snn = 5.36 TeV (right). EPOS4 calculations with (dashed lines) and without (solid lines) the UrQMD hadronic afterburner are compared to ALICE data points (solid markers) for Pb–Pb and Xe–Xe~\cite{ALICE:2018lao}.}
  \label{fig:vn_cent}
\end{figure*}

%%%%%%%%%%%%%%%%%%%%%%%%%%%%%%%%%%%%%%%%%%%%%%%%%%%%%%%%%%%%%%%%%%%%%%%%%%%%%%%%%%%%%%%%
%%%%%%%%%%%%%%%%%%%%%%%%%%%%%%%%%%%%%%%%%%%%%%%%%%%%%%%%%%%%%%%%%%%%%%%%%%%%%%%%%%%%%%%%
%%%%%%%%%%%%%%%%%%%%%%%%%%%%%%%%%%%%%%%%%%%%%%%%%%%%%%%%%%%%%%%%%%%%%%%%%%%%%%%%%%%%%%%%
%%%%%%%%%%%%%%%%%%%%%%%%%%%%  FLOW COEFFICIENTS %%%%%%%%%%%%%%%%%%%%%%%%%%%%%%%%%%%%%%%%
%%%%%%%%%%%%%%%%%%%%%%%%%%%%%%%%%%%%%%%%%%%%%%%%%%%%%%%%%%%%%%%%%%%%%%%%%%%%%%%%%%%%%%%%
%%%%%%%%%%%%%%%%%%%%%%%%%%%%%%%%%%%%%%%%%%%%%%%%%%%%%%%%%%%%%%%%%%%%%%%%%%%%%%%%%%%%%%%%
%%%%%%%%%%%%%%%%%%%%%%%%%%%%%%%%%%%%%%%%%%%%%%%%%%%%%%%%%%%%%%%%%%%%%%%%%%%%%%%%%%%%%%%%
\subsection{Anisotropic Flow}
The anisotropic flow coefficients \vn{n}, quantify the azimuthal asymmetry of particle production and provide direct insight into the collective expansion of the created medium. These anisotropies in EPOS originate from the hydrodynamic response of the core to initial spatial asymmetries, which are subsequently modified by the hadronic afterburner.

The $v_n$ coefficients are obtained using the scalar product method~\cite{Voloshin:2008dg, STAR:2002hbo, ALICE:2018lao}. Particles of interest (POI) are selected at midrapidity ($|\eta|<0.8$ for charged particles, $|y|<0.5$ for identified hadrons), while the reference $Q$-vectors are constructed from particles in the forward and backward pseudorapidity regions, analogous to the ALICE V0 detector acceptance. A pseudorapidity gap of $|\Delta\eta|>2.0$ between the POI and reference regions is used in the ALICE results suppresses short-range non-flow correlations~\cite{ALICE:2018yph}. The flow coefficient for the $n$-th harmonic is computed as
\begin{equation}
v_n\{\mathrm{SP}\} = \frac{\langle\boldsymbol{u}_{n,\mathrm{POI}} \cdot \boldsymbol{Q}_{n,\mathrm{ref}}^{\,*}\rangle} {\sqrt{\langle \boldsymbol{Q}_{n}^{\,\mathrm{fwd}} \cdot \boldsymbol{Q}_{n}^{\,\mathrm{bwd}\,*}\rangle}}\,,
\label{eq:vnsp}
\end{equation}
where $\boldsymbol{Q}_{n} = \sum_k e^{in\varphi_k}$ is the flow vector, $\boldsymbol{u}_{n,\mathrm{POI}}$ denotes the unit flow vector of the particles of interest, and $\boldsymbol{Q}_{n}^{\mathrm{fwd}}$ ($\boldsymbol{Q}_{n}^{\mathrm{bwd}}$) is the reference flow vector from the forward (backward) $\eta$ region. Both the numerator and the denominator, $Q_n$ vectors are normalized by the respective multiplicities and averaged over all events in a given centrality class. The denominator provides the event-plane resolution correction by correlating the two independent reference sub-events. The $p_{\mathrm{T}}$-differential flow, $v_n(p_{\mathrm{T}})$, is obtained by evaluating Eq.~\eqref{eq:vnsp} separately in each $p_{\mathrm{T}}$ interval, using the same reference $Q$-vectors.

% \begin{equation}
% v_n\{\mathrm{SP}\} = \frac{\langle\langle \boldsymbol{q}_{n,\mathrm{POI}}
% \cdot \boldsymbol{Q}_{n,\mathrm{ref}}^{\,*}\rangle\rangle}
% {\sqrt{\langle\langle \boldsymbol{Q}_{n}^{\,\mathrm{fwd}}
% \cdot \boldsymbol{Q}_{n}^{\,\mathrm{bwd}\,*}\rangle\rangle}}\,,
% \label{eq:vnsp}
% \end{equation}
% where $\boldsymbol{Q}_{n} = \sum_k e^{in\varphi_k}$ is the flow vector
% and $\boldsymbol{q}_{n} = \boldsymbol{Q}_{n}/M$ is the corresponding
% multiplicity-normalized flow vector.
% The numerator is evaluated as
% $\mathrm{Re}(\boldsymbol{Q}_{n,\mathrm{POI}}\cdot\boldsymbol{Q}_{n,\mathrm{ref}}^{\,*})
% /(M_{\mathrm{POI}}\,M_{\mathrm{ref}})$,
% where $M_{\mathrm{ref}} = M_{\mathrm{fwd}}+M_{\mathrm{bwd}}$ is the total reference
% multiplicity, and the resolution denominator as
% $\mathrm{Re}(\boldsymbol{Q}_{n}^{\mathrm{fwd}}\cdot\boldsymbol{Q}_{n}^{\mathrm{bwd}\,*})
% /(M_{\mathrm{fwd}}\,M_{\mathrm{bwd}})$.
% Both quantities are averaged over events within each centrality class using
% multiplicity-pair weights ($M_{\mathrm{POI}}\times M_{\mathrm{ref}}$ for the numerator,
% $M_{\mathrm{fwd}}\times M_{\mathrm{bwd}}$ for the denominator), which is equivalent to
% averaging over all distinct particle pairs.
% The $p_{\mathrm{T}}$-differential flow, $v_n(p_{\mathrm{T}})$, is obtained by
% evaluating Eq.~\eqref{eq:vnsp} separately in each $p_{\mathrm{T}}$ interval using the same
% reference flow vectors.

Figure~\ref{fig:v2vspt_pbpb} displays the elliptic flow coefficient, \vn{2}\{2\} (\tpt), for pions (\partpm{\pi}), kaons (\partpm{K}), and protons (\partpr) in Pb--Pb collisions at \snn = 5.02 TeV. The results are shown for various centrality classes, comparing calculations with UrQMD (upper panel) and without UrQMD (lower panel) to ALICE data~\cite{ALICE:2016ccg}. A mass ordering of \vn{2} is observed in the low-\tpt ~region (\tpt $\leq$ 3 \gev), where \vn{2}($\pi$) $>$ \vn{2}(K) $>$ \vn{2}(p). This splitting arises from the interplay between radial flow and elliptic flow during the hydrodynamic expansion, the common radial flow velocity field pushes heavier particles to higher transverse momenta, thereby depleting their anisotropy at low \tpt. The full EPOS4 simulation (EPOS+UrQMD) reproduces the experimental data across the presented centrality classes. Comparing the upper and lower panels reveals that the hadronic afterburner suppresses the magnitude of \vn{2} at low \tpt, particularly for kaons and protons. This enhancement is attributed to the continued buildup of flow and the specific cross-sections in the hadronic phase. The results for the intermediate Xe--Xe system at \snn =5.44 TeV are shown in Fig.~\ref{fig:v2vspt_xexe}. The differential elliptic flow exhibits the same qualitative features as observed in Pb--Pb collisions, including the characteristic mass ordering indicative of hydrodynamic behavior. The model, using the same parameter set as for Pb--Pb, successfully describes the centrality and \tpt~dependence of the ALICE data.

Differential \vn{2}\{2\}(\tpt) for O--O at \snn = 5.36 TeV are shown in Fig.~\ref{fig:v2vspt_oo}. Even in this light-ion system, EPOS4 predicts a significant non-zero elliptic flow for identified particles. The mass splitting between pions, kaons, and protons persists, suggesting that a core with hydrodynamic properties is formed and develops radial flow similar to larger systems. The hadronic phase continues to play a non-negligible role in modifying the flow harmonics in O--O collisions. The observation of such collective signatures in O--O within the model can hint at the unified picture of a smooth transition in soft physics dynamics from small to large collision systems.

Figure~\ref{fig:vn_pt} presents the differential anisotropic flow coefficients, \vnn{2}{2} (\tpt) and \vnn{3}{2} (\tpt), for charged particles in Pb--Pb collisions at \snn = 5.02 TeV (left), Xe--Xe at \snn = 5.44 TeV (middle), and O--O at  \snn = 5.36 TeV (right). The model predictions including the UrQMD afterburner (dashed lines) and without it (solid lines) are compared with experimental data from ALICE for the heavy-ion systems~\cite{ALICE:2016ccg,ALICE:2018lao}. EPOS4 reproduces the trend of both elliptic (\vn{2}) and triangular (\vn{3}) flow harmonics as a function of transverse momentum with good accuracy.  The characteristic rise of \vn{n} at low \tpt~is driven by the hydrodynamic expansion of the core. At higher \tpt, the system transitions into the corona regime. Here, the implementation of event-by-event dynamical saturation scales in EPOS4 ensures that particle production recovers the perturbative QCD limit (binary scaling), leading to a diminishing collective flow signal relative to the hydrodynamic region~\cite{Werner:2023mod}.

The influence of the hadronic phase on the differential flow is quantified in the bottom panels of Fig.~\ref{fig:vn_pt}, which show the ratio of model values and data for Pb--Pb and Xe--Xe system. The afterburner seems to have very little effect as the ratios show the same trends for both systems. For \tpt~$\lesssim 0.6$, a suppression in model values of flow harmonics is seen, while the effect is smaller for \tpt~$\gtrsim 0.6$ \gev~where the difference is \~ 20\%. This modification is likely due to the interplay between the randomization of momenta during hadronic rescattering and the additional flow built up for heavier species which dominate different \tpt~regions. In the O--O system, the ratio shows minimal effect of afterburner on the flow harmonics.

%calculations with UrQMD to those without. For the larger Pb--Pb system, the hadronic afterburner tends to slightly suppress the flow harmonics at intermediate \tpt (\tpt $\geq$ 2 \gev), while the effect is smaller or even reversed at lower \tpt. This modification is likely due to the interplay between the randomization of momenta during hadronic rescattering and the additional flow built up for heavier species which dominate different \tpt regions. In the O--O system, the ratio fluctuates around unity, suggesting that while hadronic rescattering is crucial for particle yields (as seen in the spectra), its net impact on the \tpt-differential flow shape of inclusive charged particles is more subtle in this centrality range.

Figure~\ref{fig:vn_cent} displays the \tpt-integrated flow coefficients \vnn{2}{2} and \vnn{3}{2} at midrapidity ($|\eta| < 0.8$) within 0.2 $< \tpt <$ 3.0 \gev~ as a function of centrality for the three systems. The elliptic flow, \vn{2} exhibits a strong dependence on centrality in Pb--Pb and Xe--Xe, increasing from central to mid-peripheral collisions, driven by the initial elliptic geometry of the overlap region~\cite{Werner:2025yse}. In contrast, the triangular flow \vn{3} shows a much weaker dependence on centrality, consistent with its origin in event-by-event fluctuations of the initial nucleon positions rather than the global collision geometry. EPOS4 calculations with UrQMD (dashed lines) show a small enhancement of the integrated flow compared to the case without UrQMD (solid lines) for \vn{2} and \vn{3} across centralities. This hints that the late-stage hadronic interactions do contribute to the total momentum anisotropy of the system. %The results for the O--O system are shown in the right panel of Fig.~\ref{fig:vn_cent}. The model predicts a significant elliptic flow signal (\vnn{2}{2} $\approx$ 0.06−0.07) that persists across a wide centrality range. The magnitude of \vn{2} in O--O is comparable to that in central Pb--Pb collisions. The centrality dependence is flatter compared to the larger systems. This reflects the fact that in light-ion collisions, the anisotropy is driven predominantly by event-by-event fluctuations of the initial nucleon positions, analogous to the origin of triangular flow (\vn{3}) in heavy-ion collisions rather than by the static average elliptic geometry of the overlap region which dominates \vn{2} in semi-central Pb--Pb collisions. The presence of a non-zero \vn{3} in O--O further supports the fluctuation-driven nature of the anisotropy. The comparison between the solid and dashed lines highlights that the hadronic afterburner provides a significant contribution to the development of collective flow in O--O collisions, enhancing integrated \vn{2} by approximately 10\%-15\% and \vn{3} by approximately 20\%-40\%.
The results for the O--O system are shown in the right panel of Fig.~\ref{fig:vn_cent}. The model predicts a finite elliptic flow signal (\vnn{2}{2} $\approx 0.06$--0.07) across the studied centrality range, with a weaker centrality dependence than in the larger systems. This behavior suggests that the anisotropy in O--O collisions is largely driven by fluctuations of the initial geometry rather than by the average elliptic shape of the overlap region. A non-zero \vn{3} is also observed, consistent with the fluctuation-driven origin of triangular flow. The comparison between the two calculations shows that the hadronic transport stage increases the integrated flow coefficients, enhancing \vnn{2}{2} by about $10$--$15\%$ and \vnn{3}{2} by about $20$--$40\%$.

%%%%%%%%%%%%%%%%%%%%%%%%%%%%%%%%%%%%%%%%%%%%%%%%%%%%%%%%%%%%%%%%%%%%%%%%%%%%%%%%%%%%%%%%
%%%%%%%%%%%%%%%%%%%%%%%%%%%%%%%%%%%%%%%%%%%%%%%%%%%%%%%%%%%%%%%%%%%%%%%%%%%%%%%%%%%%%%%%
%%%%%%%%%%%%%%%%%%%%%%%%%%%%%%%%%%%%%%%%%%%%%%%%%%%%%%%%%%%%%%%%%%%%%%%%%%%%%%%%%%%%%%%%
%%%%%%%%%%%%%%%%%%%%%%%%%%%%%%%%%%%%%%%%%%%%%%%%%%%%%%%%%%%%%%%%%%%%%%%%%%%%%%%%%%%%%%%%
%%%%%%%%%%%%%%%%%%%%%%%%%%%%%%%%%%%%%%%%%%%%%%%%%%%%%%%%%%%%%%%%%%%%%%%%%%%%%%%%%%%%%%%%
%%%%%%%%%%%%%%%%%%%%%%%%%%%%%%%%%%%%%%%%%%%%%%%%%%%%%%%%%%%%%%%%%%%%%%%%%%%%%%%%%%%%%%%%
%%%%%%%%%%%%%%%%%%%%%%%%%%%%%%%%%%%%%%%%%%%%%%%%%%%%%%%%%%%%%%%%%%%%%%%%%%%%%%%%%%%%%%%%
%%%%%%%%%%%%%%%%%%%%%%%%%%%%%%%%%%%%%%%%%%%%%%%%%%%%%%%%%%%%%%%%%%%%%%%%%%%%%%%%%%%%%%%%
%%%%%%%%%%%%%%%%%%%%%%%%%%%%       SUMMARY       %%%%%%%%%%%%%%%%%%%%%%%%%%%%%%%%%%%%%%%
%%%%%%%%%%%%%%%%%%%%%%%%%%%%%%%%%%%%%%%%%%%%%%%%%%%%%%%%%%%%%%%%%%%%%%%%%%%%%%%%%%%%%%%%
%%%%%%%%%%%%%%%%%%%%%%%%%%%%%%%%%%%%%%%%%%%%%%%%%%%%%%%%%%%%%%%%%%%%%%%%%%%%%%%%%%%%%%%%
%%%%%%%%%%%%%%%%%%%%%%%%%%%%%%%%%%%%%%%%%%%%%%%%%%%%%%%%%%%%%%%%%%%%%%%%%%%%%%%%%%%%%%%%
%%%%%%%%%%%%%%%%%%%%%%%%%%%%%%%%%%%%%%%%%%%%%%%%%%%%%%%%%%%%%%%%%%%%%%%%%%%%%%%%%%%%%%%%
%%%%%%%%%%%%%%%%%%%%%%%%%%%%%%%%%%%%%%%%%%%%%%%%%%%%%%%%%%%%%%%%%%%%%%%%%%%%%%%%%%%%%%%%
%%%%%%%%%%%%%%%%%%%%%%%%%%%%%%%%%%%%%%%%%%%%%%%%%%%%%%%%%%%%%%%%%%%%%%%%%%%%%%%%%%%%%%%%
%%%%%%%%%%%%%%%%%%%%%%%%%%%%%%%%%%%%%%%%%%%%%%%%%%%%%%%%%%%%%%%%%%%%%%%%%%%%%%%%%%%%%%%%
%%%%%%%%%%%%%%%%%%%%%%%%%%%%%%%%%%%%%%%%%%%%%%%%%%%%%%%%%%%%%%%%%%%%%%%%%%%%%%%%%%%%%%%%
\section{Summary}
This work reports a detailed study of soft global observables in relativistic high energy collisions for Pb--Pb and Xe--Xe data along with providing predictions for O--O collisions using EPOS4. Primary interactions in this model are described by a rigorous parallel scattering scenario based on S-matrix theory, incorporating energy-momentum sharing and a core-corona separation mechanism. An update in this framework is the implementation of event-by-event dynamical saturation scales, which ensures the validity of a generalized AGK theorem and preserves factorization for hard processes while generating the necessary initial conditions for collective behaviour in the soft sector. The secondary stage features a viscous hydrodynamic evolution of the core, followed by microcanonical hadronization and a microscopic hadronic afterburner (UrQMD)~\cite{Werner:2023mod, Werner:2023fne}.

The analysis of bulk production reveals that the centrality dependence of charged-particle pseudorapidity densities and identified particle yields in Pb--Pb and Xe--Xe collisions is well reproduced by the model. The hadronic afterburner is found to be essential for describing the yields of protons and other baryons, where baryon-antibaryon annihilation significantly modifies the composition established at hadronization. The transverse momentum spectra exhibit a distinct mass ordering at low \tpt, a signature of radial flow driven by the hydrodynamic expansion of the core. The predictions for O--O collisions at \snn = 5.36 TeV show particle spectral shapes and radial flow signatures comparable to those observed in semi-peripheral heavy-ion collisions.

This study also reports the transverse momentum (\tpt) fluctuations via \tpt~correlator, \ptcorr. The normalized \tpt correlator in the Pb--Pb and Xe--Xe systems shows a monotonic decrease with increasing multiplicity, consistent with the experimental data. This observable is particularly sensitive to the interplay between the fluctuating initial state and the subsequent hydrodynamic evolution. The model predicts that O--O collisions follow a continuous system-size evolution consistent with that of the larger systems, suggesting that the mechanism of converting initial-state density fluctuations into final-state momentum correlations is universal across these system sizes.

EPOS4 successfully describes the differential elliptic (\vn{2}) and triangular (\vn{3}) flow harmonics for identified particles in Pb--Pb and Xe--Xe collisions, reproducing the mass splitting characteristic of fluid dynamics. For O--O collisions, the model predicts a substantial integrated elliptic flow (\vn{2} $\approx$ 0.06), comparable in magnitude to mid-central Pb--Pb collisions. Unlike in semi-central heavy-ion collisions, where $\vn{2}$ is driven by the average elliptic geometry of the overlap region, the anisotropy in O--O is driven predominantly by event-by-event geometric fluctuations of nucleon positions. The hadronic afterburner is shown to contribute to the flow harmonics, very slightly enhancing the momentum anisotropy developed during the hydrodynamic phase.

In conclusion, EPOS4 provides a consistent description of soft observables from the largest (Pb--Pb) to the smallest (O--O) nuclear systems at the LHC. The results suggest that the mechanisms driving collective radial and anisotropic flow in macroscopic heavy-ion collisions may scale continuously down to smaller systems.

%%%%%%%%%%%%%%%%%%%%%%%%%%%%%%%%%%%%%%%%%%%%%%%%%%%%%%%%%%%%%%%%%%%%%%%%%%%%%%%%%%%%%%%%
%%%%%%%%%%%%%%%%%%%%%%%%%%%%%%%%%%%%%%%%%%%%%%%%%%%%%%%%%%%%%%%%%%%%%%%%%%%%%%%%%%%%%%%%
%%%%%%%%%%%%%%%%%%%%%%%%%%%%%%%%%%%%%%%%%%%%%%%%%%%%%%%%%%%%%%%%%%%%%%%%%%%%%%%%%%%%%%%%
%%%%%%%%%%%%%%%%%%%%%%%%%%%%%%%%%%%%%%%%%%%%%%%%%%%%%%%%%%%%%%%%%%%%%%%%%%%%%%%%%%%%%%%%
%%%%%%%%%%%%%%%%%%%%%%%%%%%%%%%%%%%%%%%%%%%%%%%%%%%%%%%%%%%%%%%%%%%%%%%%%%%%%%%%%%%%%%%%
%%%%%%%%%%%%%%%%%%%%%%%%%%%%%%%%%%%%%%%%%%%%%%%%%%%%%%%%%%%%%%%%%%%%%%%%%%%%%%%%%%%%%%%%
%%%%%%%%%%%%%%%%%%%%%%%%%%%%%%%%%%%%%%%%%%%%%%%%%%%%%%%%%%%%%%%%%%%%%%%%%%%%%%%%%%%%%%%%
%%%%%%%%%%%%%%%%%%%%%%%%%%%%%%%%%%%%%%%%%%%%%%%%%%%%%%%%%%%%%%%%%%%%%%%%%%%%%%%%%%%%%%%%
%%%%%%%%%%%%%%%%%%%%%%%%%%%%   Acknowledgements  %%%%%%%%%%%%%%%%%%%%%%%%%%%%%%%%%%%%%%%
%%%%%%%%%%%%%%%%%%%%%%%%%%%%%%%%%%%%%%%%%%%%%%%%%%%%%%%%%%%%%%%%%%%%%%%%%%%%%%%%%%%%%%%%
%%%%%%%%%%%%%%%%%%%%%%%%%%%%%%%%%%%%%%%%%%%%%%%%%%%%%%%%%%%%%%%%%%%%%%%%%%%%%%%%%%%%%%%%
%%%%%%%%%%%%%%%%%%%%%%%%%%%%%%%%%%%%%%%%%%%%%%%%%%%%%%%%%%%%%%%%%%%%%%%%%%%%%%%%%%%%%%%%
%%%%%%%%%%%%%%%%%%%%%%%%%%%%%%%%%%%%%%%%%%%%%%%%%%%%%%%%%%%%%%%%%%%%%%%%%%%%%%%%%%%%%%%%
%%%%%%%%%%%%%%%%%%%%%%%%%%%%%%%%%%%%%%%%%%%%%%%%%%%%%%%%%%%%%%%%%%%%%%%%%%%%%%%%%%%%%%%%
%%%%%%%%%%%%%%%%%%%%%%%%%%%%%%%%%%%%%%%%%%%%%%%%%%%%%%%%%%%%%%%%%%%%%%%%%%%%%%%%%%%%%%%%
%%%%%%%%%%%%%%%%%%%%%%%%%%%%%%%%%%%%%%%%%%%%%%%%%%%%%%%%%%%%%%%%%%%%%%%%%%%%%%%%%%%%%%%%
%%%%%%%%%%%%%%%%%%%%%%%%%%%%%%%%%%%%%%%%%%%%%%%%%%%%%%%%%%%%%%%%%%%%%%%%%%%%%%%%%%%%%%%%
%\footnotesize
\section*{Acknowledgements}
We sincerely thank the authors of EPOS4. We gratefully acknowledge the project ``Indian participation in ALICE experiment at CERN,’’ funded by the Department of Science and Technology (DST), Government of India, for the financial support of this work under the sanction order No.~3015/1/2021/Gen/R\&D-I/13283. The second author acknowledges the Council of Scientific and Industrial Research (CSIR), Government of India, for financial support through a research fellowship. We also thank the Research and Seed Grant under the Quality Assurance Fund (DIQA), University of Jammu, sanction No.~RA/23/1293-1300dl-7/8/2023, for computing facilities used in the simulations presented in this work.

\end{document}